\documentclass[12pt]{article}
\usepackage[colorlinks=false,
   linkcolor=red, 
   citecolor=blue,
    filecolor=red,
    urlcolor=red,
    linktoc=all, 
    pdfstartview=FitV,
    bookmarksopen=true]{hyperref}
\usepackage[left=2cm,top=1.5cm,right=3cm,nohead]{geometry}

\usepackage{amsfonts}
\usepackage{bm}
\usepackage{amsmath}
\usepackage{epsfig}
\usepackage[mathscr]{euscript}

\usepackage{slashed}

\newcommand{\bmat}{\left(\begin{array}}
\newcommand{\emat}{\end{array}\right)}

\newcommand{\rO}{{\mathrm{O}}}
\newcommand{\hog}{{\mathrm{SO}(32)}}
\newcommand{\mre}{{\mathrm{E}}}
\newcommand{\rL}{{\mathrm{L}}}
\newcommand{\rR}{{\mathrm{R}}}
\newcommand{\rB}{{\mathrm{B}}}
\newcommand{\rF}{{\mathrm{F}}}
\def\gtrsim{\mathrel{\raise.3ex\hbox{$>$\kern-.75em\lower1ex\hbox{$\sim$}}}}

\def\ap{\alpha^{\prime}}

\def\-{\hphantom{-}}
\def\ov{\overline}
\def\s2{\frac{1}{\sqrt2}}

\def\wt{\widetilde}

\def\mg{m_{3/2}}
\def\mg2{m^2_{3/2}}

\def\Dsl{\,\raise.15ex\hbox{/}\mkern-13.5mu D} 

\def\be{\begin{equation}}
\def\ee{\end{equation}}
\def\bea{\begin{eqnarray}}
\def\eea{\end{eqnarray}}
\def\ov{\overline}

\newcommand{\nn}{\nonumber}

\begin{document}

\pagestyle{plain}

\makeatletter
\@addtoreset{equation}{section}
\makeatother
\renewcommand{\theequation}{\thesection.\arabic{equation}}
\pagestyle{empty}
\begin{center}
\ \

\vskip .5cm
\LARGE{\LARGE\bf Supersymmetry, T-duality and \\  Heterotic
$ \alpha'$-corrections \\[10mm]}
\vskip 0.3cm

\large{Eric Lescano$^\dag$, Carmen A. N\'u\~nez$^{\dag\, *}$ and Jes\'us A. Rodr\' iguez$^*$
 \\[6mm]}

{\small  $^\dag$ Instituto de Astronom\'ia y F\'isica del Espacio (IAFE-CONICET-UBA)\\ [.01 cm]}
{\small\it Ciudad Universitaria, Pabell\'on IAFE, 1428 Buenos Aires, Argentina\\ [.3 cm]}
{\small  $^*$ Departamento de F\'isica, FCEyN, Universidad de Buenos Aires (UBA) \\ [.01 cm]}
{\small\it Ciudad Universitaria, Pabell\'on 1, 1428 Buenos Aires, Argentina\\ [.5 cm]}

{\small \verb"elescano@iafe.uba.ar,  carmen@iafe.uba.ar, jarodriguez@df.uba.ar"}\\[1cm]

\small{\bf Abstract} \\[0.5cm]\end{center}

Higher-derivative interactions and transformation rules of the fields in the effective field theories of the massless string states
are strongly constrained by  space-time symmetries  and  dualities.  Here we use an exact    formulation   of ten dimensional ${\cal N}=1$ supergravity coupled to Yang-Mills with manifest T-duality symmetry to
construct the first order $\alpha'$-corrections of the  heterotic string effective action.  The theory contains a supersymmetric and  T-duality covariant generalization of the Green-Schwarz mechanism that determines the
 modifications to the leading order
supersymmetry transformation rules  of the  fields. We compute the resulting field-dependent deformations of the coefficients in the supersymmetry algebra and construct the invariant action, with up to and including four-derivative terms of all the massless bosonic and fermionic  fields of the heterotic string spectrum.

\newpage
\setcounter{page}{1}
\pagestyle{plain}
\renewcommand{\thefootnote}{\arabic{footnote}}
\setcounter{footnote}{0}

\tableofcontents
\newpage

\section{Introduction}

At low energy, or small curvature,  heterotic string theory reduces to  ten dimensional ${\cal N}= 1$ supergravity coupled to super Yang-Mills \cite{gw}. 
Successive terms in the $\alpha'$-expansion may be expressed as higher-derivative interactions  that are strongly constrained by the symmetries of string theory.  There are several reasons to  study the higher-order terms in the effective field theories of the massless string modes. They  are needed to evaluate the stringy effects  on solutions to the supergravity
equations of motion \cite{ort,ortin}, they play
 a central role in the tests of duality conjectures \cite{minasian}, in the microstate
counting of black hole entropy \cite{blackh} and in moduli stabilization \cite{louis}.The swampland  program \cite{swamp} has revealed that the effective field theories of low energy physics and cosmology  are limited by their couplings to quantum gravity \cite{swampdesitter}, and together with the string lamppost principle \cite{Kim:2019ths},  reinforces the interest in the restrictions imposed by  string theory on the higher-derivative corrections to General Relativity.

The first few orders of  the heterotic string $\alpha'$-expansion are known explicitly. The  interactions of the bosonic fields up to ${\cal O}(\alpha'^3)$ were originally determined from the computation of  scattering amplitudes of the massless string states at tree \cite{gw,higherder} and one loop \cite{oneloop} levels in the string coupling and from  conformal anomaly cancellations  \cite{foakes}. The  contributions of the fermionic fields  have been computed using supersymmetry and superspace methods  \cite{Chapline:1982ww}-\cite{howe}. Supersymmetry completely fixes the leading order terms
 \cite{Chapline:1982ww}  and it often provides an elegant underlying explanation of  the higher-derivative corrections. But it holds iteratively in powers of $\alpha'$ and the transformation rules of the  fields demand order by order modifications that are further restricted by other string symmetries and dualities. 

In particular, the effective field theories for the massless string fields exhibit a  global $\rO(n,n;\mathbb R)$ symmetry   when the fields are independent of $n$ spatial coordinates. This continuous T-duality symmetry holds to all orders in $\alpha'$  \cite{sen} (see also \cite{ms}-\cite{garu}) and it  has been  explicitly displayed  recently  for the quadratic and some of the quartic interactions of the bosonic fields   in \cite{ehs,chm}. 
This feature  motivated the construction of field  theories with T-duality covariant structures, such as 
double field theory (DFT)  \cite{Siegel:1993xq,dft} and generalized geometry \cite{hitchin}, which provide    reformulations of the string (super)gravities  in which the global duality invariance is made manifest. 

In the duality covariant frameworks, the standard  local symmetries are generalized to larger groups:
 diffeomorphism invariance is extended to 
also include the gauge transformations of the two-form and the tangent space is enhanced with an
extended  Lorentz symmetry.  Interestingly, the duality covariant gauge transformations completely determine the lowest order field interactions in  string (super)gravities even before dimensional reduction  (for reviews see \cite{reviews} and references therein). Moreover,  extensions of the duality group  \cite{bmn,otros} as well as enhancings of the gauge structure of DFT \cite{mn,bfmn} allowed to reproduce the four-derivative interactions of the   massless bosonic  heterotic string fields.

 Supersymmetry can be naturally  incorporated in the duality covariant formulations \cite{waldram}-\cite{diego2018}.  A supersymmetric and manifestly $\rO(10,10+n_g)$  covariant  DFT reformulation of ten dimensional  ${\cal N}=1$ supergravity coupled to $n_g$ abelian vector multiplets  was introduced in \cite{corfer,Hohm:2012}.  Although it is formally constructed on a  $20+n_g$ dimensional space-time,  the apparent inconsistency of supergravity  beyond eleven dimensions is  avoided through a strong constraint that admits solutions removing the field dependence  on $10+n_g$ coordinates, and fermions transform as spinors under the $\rO(9,1)_\rL$ factor of the  local $\rO(9,1)_\rL\times \rO(1,9+n_g)_\rR$   double Lorentz symmetry.  

More recently, an exact  supersymmetric and manifestly duality covariant 
mechanism  was  introduced in \cite{diego2018}, in which the global symmetry of the theory is taken to be \mbox{$\rO(D,D+k)$},  $k$ being the dimension of the  $\rO(1,D+k-1)$ Lorentz group. To preserve  duality covariance, the $\rO(D,D+k)$ multiplets are parameterized with elements of $\rO(D,D)$. Additionally identifying the $\rO(D,D)$ vector with the generalized spin connection of  $\rO(D,D+k)$, the construction produces an exact supersymmetric and duality covariant  generalization of the Green-Schwarz transformation, which requires an infinite tower of $\rO(D,D)$
 covariant higher-derivative terms in the gauge invariant action.

With the motivation to further understand the structure of the heterotic string $\alpha'$-expansion, in this paper we  perform a perturbative expansion of the formal exact construction of \cite{diego2018} and obtain  the first order corrections to ${\cal N}=1$ supersymmetric DFT. Further parameterizing the duality multiplets in terms of supergravity and super Yang-Mills multiplets, we show that the   supersymmetric duality covariant generalized Green-Schwarz transformation completely  fixes the first order deformations of the transformation rules of the  fields. 
 We also construct the invariant action with up to and including four-derivative terms  of all the massless bosonic and fermionic fields of the heterotic string  and up to bilinear terms in fermions.

The paper is organized as follows. In section \ref{sec:leading}  we review the basic features of the  $\mathcal{N}=1$ supersymmetric  DFT introduced in \cite{Hohm:2012} and we trivially extend it to incorporate non-abelian gauge vectors.   In section \ref{sec:DFTalpha}, after briefly recalling the relevant aspects of the duality covariant mechanism proposed in \cite{diego2018}, we extract the first order corrections to the transformation rules of the $\rO(10,10+n_g)$  generalized fields from those of the $\rO(10,10+k)$ multiplets, and  obtain the manifestly duality covariant and gauge invariant ${\cal N}=1$ supersymmetric DFT action to ${\cal O}(\alpha')$. We then parameterize the $\rO(10,10+n_g)$ fields  in terms of  supergravity and super Yang-Mills multiplets in section \ref{sec:transf} and find the relations between the duality  and the local gauge covariant structures.  We  discuss the deformations  induced from the generalized Green-Schwarz transformation  on the transformation rules of the supergravity fields and compare with previous results in the literature. Finally, in section \ref{sec:hete} we  present the first order $\alpha'$-corrections of the heterotic string effective action including up to bilinear terms in fermions. Conclusions are the subject of section \ref{sec:conclu}. 
The conventions used throughout the paper and some useful gamma function identities
are included in appendix \ref{app:conventions}. Details of the proof of closure of the symmetry algebra on the duality multiplets are contained in appendix \ref{app:closure}. Finally, in appendix  \ref{app:susyac} we compute the deformed supersymmetry algebra on the supergravity multiplets and prove the  supersymmetric invariance of the first order corrections in the heterotic string effective action.

\section{The leading order theory}\label{sec:leading}

In this section we review the basic features of the  DFT reformulation of $\mathcal{N}=1$ supergravity coupled to $n_g$ vector multiplets in ten dimensions that was introduced in \cite{Hohm:2012}, mainly to establish the notation.
The  frame formalism used in  \cite{hohm2011} is most useful to achieve a manifestly $\rO(10,10+n_g)$ covariant rewriting of   heterotic supergravity truncated to the Cartan subalgebra of $\hog$ or  $\mre_8 \times \mre_8$ for $n_g=16$.   Employing  gauged DFT  \cite{md}, we further include the full set of non-abelian gauge fields  and recover the leading order terms of  heterotic supergravity.

\subsection{Review of ${\cal N}=1$ supersymmetric Double Field Theory}

${\cal N}=1$ supersymmetric Double Field Theory is defined on a space with coordinates $X^{\mathbb M}$ belonging to the fundamental representation of G$=\rO(10, 10+n_g|\mathbb R)$, with ${\mathbb M}=(M,i)$, $M=0, \dots, 19; i=1,\dots, n_g$, and $n_g$ is the dimension of the gauge group. The theory has a global G symmetry, a local double Lorentz  H $=\rO(9,1|\mathbb R)_\rL\times \rO(1, 9+n_g|\mathbb R)_\rR$ symmetry, diffeomorphisms generated infinitesimally  by $\xi^{\mathbb M}$  through a generalized Lie derivative $\hat{\cal L}_\xi$ and supersymmetry parameterized by an infinitesimal Majorana fermion $\epsilon$ transforming as a spinor of $\rO(9,1)_\rL$. The
propagating degrees of freedom are:

\begin{itemize}

\item $E^{\mathbb M}{}_{\mathbb A}$:  a generalized vielbein  parameterizing the coset $\frac {\text G}{\text H}=\frac{\rO(10,10+n_g)}{\rO(9,1)_\rL\times \rO(1,9+n_g)_\rR}$,  with tangent space  indices ${\mathbb A}=(\underline{A},\overline A)$  splitting into $\rO(9,1)_\rL$  and $\rO(1,9+n_g)_\rR$ vector indices, $\underline{A}=0,\dots , 9$  and   $\overline A=0,\dots , 9+n_g$,   respectively,

\item $d$: an ${\rO(10,10+n_g)}$ scalar dilaton,

\item $\Psi_{\overline{A}}$:   a Majorana spinor generalized gravitino,   transforming  as a spinor of $\rO(9,1)_{\rL}$, as a vector of $\rO(1,9+n)_{\rR}$, and as a scalar of $\rO(10,10+n_g)$,

\item $\rho$: a Majorana  spinor  `dilatino', transforming as a spinor of $\rO(9,1)_{\rL}$ and as a  scalar of $\rO(10,10+n_g)$.

\end{itemize}

The group invariant symmetric and invertible $\rO(10,10+n_g)$ metric is

\be
{\eta}_{\mathbb{M N}}  = \left(\begin{matrix}\eta^{\mu\nu}&\eta^\mu{}_\nu&\eta^\mu{}_i\\ 
\eta_\mu{}^\nu&\eta_{\mu\nu}&\eta_{\mu i}\\\eta_{i}^\nu&\eta_{i\nu}&\eta_{ij}\end{matrix}\right)= \left(\begin{matrix}0&\delta^\mu{}_\nu&0\\ 
\delta_\mu{}^\nu&0&0\\0&0&\kappa_{ij}\end{matrix}\right) \ , \label{eta}
\ee
with $\mu, \nu=0,\dots, 9$, $i,j=1,\dots, n_g$ and $\kappa_{ij}$ the Killing metric of the gauge group.  There are two constant symmetric and invertible H-invariant metrics $\eta_{\mathbb{AB}}$ and ${H}_{\mathbb{AB}}$. The former is used to raise and lower the indices that are rotated by H and the latter is constrained to satisfy
${H}_{\mathbb A}{}^{\mathbb C}{H}_{\mathbb C}{}^{\mathbb B} = \delta^{\mathbb B}_{\mathbb A}$. The three metrics $\eta_{\mathbb{MN}}, \eta_{\mathbb{AB}}$ and ${H}_{\mathbb{AB}}$ are invariant under the action of  $\hat{\cal L}$, G and H.

 The generalized vielbein $E^{\mathbb M}{}_{\mathbb A}$ is constrained to relate the metrics $\eta_{\mathbb{AB}}$ and $\eta_{\mathbb {MN}}$ and defines a generalized metric ${H}_{\mathbb{MN}}$ from ${H}_{\mathbb {AB}}$
\be
\eta_{\mathbb{AB}}=E^{\mathbb M}{}_{\mathbb A} \eta_{\mathbb {MN}} E^{\mathbb N}{}_{\mathbb B}\, , \quad {H}_{\mathbb{M N}}=E_{\mathbb M}{}^{\mathbb A} {H}_{\mathbb{AB}} E_{\mathbb N}{}^{\mathbb B} \, .
\ee
${H}_{\mathbb{MN}}$  is also an element of $\rO(10,10+n_g)$, constrained as
\be
{H}_{\mathbb{MP}} \eta^{\mathbb{PQ}} {H}_{\mathbb{QN}} = \eta_{\mathbb{MN}} \ ,\qquad {H}_{\mathbb{AC}} \eta^{\mathbb{CD}} {H}_{\mathbb{DB}} = \eta_{\mathbb{AB}}\ . \label{GMconstraint}
\ee

It is convenient to define the projectors
\be
P_{\mathbb{M N}} = \frac 1 2 \left(\eta_{\mathbb{MN}} - H_{\mathbb{MN}}\right)  \ \ {\rm and} \ \
\overline{P}_{\mathbb{MN}} = \frac 1 2 \left(\eta_{\mathbb{MN}} + H_{\mathbb{MN}}\right)\ ,
\ee
satisfying the usual properties 
\be
\overline{P}_{\mathbb{M Q}} \overline{P}^{\mathbb Q}{}_{\mathbb N}=\overline{P}_{\mathbb{M N}}\, , \ \ P_{\mathbb{M Q}} P^{\mathbb Q}{}_{\mathbb N}=P_{\mathbb{M N}}, \ \
P_{\mathbb{M  Q}}\overline{P}^{\mathbb Q}{}_{\mathbb N} = \overline{P}_{\mathbb {M Q}} P^{\mathbb Q}{}_{\mathbb N} = 0\, , \ \  \overline{P}_{\mathbb{MN}} + P_{\mathbb{M N}} = \eta_{\mathbb{M N}}\,,\nn
\ee
and  related with the generalized vielbein in the following way
\be
P_{\underline{A} \underline{B}} = E_{{\mathbb M}\underline{A} } E^{\mathbb M} {}_{\underline{B}}\ , \ \ \overline P_{\overline{AB}}= E_{{\mathbb M}\overline{A} } E^{\mathbb M}{}_{\overline{B}}\, ,\ \
P_{\mathbb{M N}} = E_{{\mathbb M}\underline{A} } E_{\mathbb N}{}^{\underline{A}} , \ \ \overline{P}_{\mathbb{M N}}= E_{{\mathbb M} \overline{A} } E_{\mathbb N}{}^{\overline{A}}  \, .
\ee
We use the convention that $P_{\underline{A} \underline{B}}$ , $\overline P_{\overline{AB}}$ and their inverse lower and raise  projected indices. 

The generalized Lie derivative acts as
\begin{subequations}\label{gendiffeomorphisms}
\begin{align}
\delta_\xi E^{\mathbb M}{}_{\mathbb A}&=\hat{\cal L}_\xi E^{\mathbb M}{}_{\mathbb A}=\xi^{\mathbb N}\partial_{\mathbb N} E^{\mathbb M}{}_{\mathbb A}+(\partial^{\mathbb M}\xi_{\mathbb N}-\partial_{\mathbb N}\xi^{\mathbb M}) E^{\mathbb N}{}_{\mathbb A}+f^{\mathbb M}{}_{\mathbb {N P}}\xi^{\mathbb N}E^{\mathbb P}{}_{\mathbb A} ,\label{dife}\\
\delta_\xi \Psi_{\overline{A}} & = \  \hat{\cal L}_\xi\Psi_{\overline A}\ = \ \xi^{\mathbb M} \partial_{\mathbb M} \Psi_{\overline{A}}  \, \\
\delta_\xi d \  &= \ \hat{\cal L}_\xi d \ \ =\ \xi^{\mathbb M}\partial_{\mathbb M} d-\frac12\partial_{\mathbb  M}\xi^{\mathbb M}\, , \quad
   \qquad \delta_\xi \rho  \ = \  \hat{\cal L}_\xi\rho=\xi^{\mathbb M} \partial_{\mathbb M} \rho\, ,\label{diff}
\end{align}
\end{subequations}
where the partial derivatives $\partial_{\mathbb M}$   belong to the fundamental representation of $\rO(10,10+n_g)$ and  the so-called fluxes or gaugings $f_{\mathbb{MNP}}$  are a set of constants \cite{hohm2011} verifying  linear and quadratic constraints
\bea
f_{\mathbb{ MNP}}=f_{[\mathbb{MNP}]}\, , \qquad f_{[\mathbb {MN}}{}^{\mathbb R}f_{\mathbb{P}]\mathbb R}{}^{\mathbb Q}=0\, . \label{consf}
\eea

Consistency of the construction requires 
constraints which restrict the coordinate dependence of fields and gauge parameters. The strong constraint
\be
\partial_{\mathbb M} \partial^{\mathbb M} \cdots = 0 \ , \ \ \ \ \ \partial_{\mathbb M} \cdots \ \partial^{\mathbb M} \cdots = 0 \, , \ \ \ \ \
f_{\mathbb{MN}}{}^{\mathbb P}\partial_{\mathbb P}\cdots =0\, ,\label{StrongConstraint}
\ee
where $\cdots$ refers to products of fields, will be assumed throughout. This constraint  locally removes   the field dependence  on $10+n_g$ coordinates, so that
fermions can be effectively defined in a $10$-dimensional tangent space\footnote{A supersymmetric DFT without the strong constraint was obtained through a generalized Scherk-Schwarz reduction in \cite{berman}. }.

The   local  $\rO(9,1)_\rL \times O(1,9+n_g)_\rR$ double Lorentz  symmetry is parameterized 
 by an infinitesimal parameter $\Gamma_{\mathbb A\mathbb B}$ satisfying
\be
\Gamma_{\mathbb{AB}} = - \Gamma_{\mathbb{BA}}\, , \label{lorpar}
\ee
 in order to preserve the invariance of $\eta_{\mathbb{AB}}$ and ${H}_{\mathbb{AB}}$.  The two projections of a generic vector $V^{\mathbb A}=V^{\underline{A}}+V^{\overline{A}}$ transform as 
\be
\delta_\Gamma V^{{\underline{A}}} =  V^{\underline{B}}\, \Gamma_{\underline{B}}{}^{\underline{A}} \ , \ \ \ \ \ \delta_\Gamma V^{\overline{A}} =  V^{\overline{B}}\,\Gamma_{\overline{B}}{}^{\overline{A}}\ ,
\ee
where the $\Gamma_{\underline{A}}{}^{\underline{B}}$ and $\Gamma_{\overline{A}}{}^{\overline{B}}$  components   generate the $\rO(9,1)_\rL$ and $O(1,9+n_g)_\rR$ transformations  leaving $ P_{\underline{AB}}$ and $\overline P_{\overline{AB}}$ invariant, respectively, and $\delta_\Lambda H_{\mathbb{AB}}=0$ implies
  $\Gamma_{\underline{A} \overline {B}} = 0$.

 The fields transform under double Lorentz variations as 
\bea
\delta_\Gamma E^{\mathbb M}{}_{\mathbb A}=E^{\mathbb M}{}_{\mathbb B} \Gamma^{\mathbb B}{}_{\mathbb A}\, , \quad \ \
\delta_\Gamma \Psi_{\overline{A}} =  \Psi_{\overline{B}} \Gamma^{\overline{B}}{}_{\overline{A}} +\frac14\Gamma_{\underline{BC}}\gamma^{\underline{BC}}\Psi_{\overline{A}}  \, ,\quad \ \
    \delta_\Gamma \rho =  \frac{1}{4} \Gamma_{\underline{B} \underline{C}} \gamma^{\underline{B} \underline{C}} \rho\, ,\  \label{dltf}
\eea
where the $\rO(9,1)_\rL$  gamma matrices can be chosen to be conventional gamma matrices  in ten dimensions, satisfying
\be
\Big \{ \gamma_{\underline{A}},\gamma_{\underline{B}} \Big \} = -2 P_{\underline{A} \underline{B}}\, . \label{clif}
\ee  
Some useful identities for the product of  gamma matrices  are listed in Appendix \ref{app:gammaids}.

The Lorentz and space-time covariant derivatives act on  generic vectors as
\be
\nabla_{\mathbb A} V_{\mathbb B}= E_{\mathbb A} V_{\mathbb B} +\omega_{\mathbb{A B} }{}^{\mathbb C} V_{\mathbb C}\, ,
\quad \qquad \nabla_{\mathbb M}V_{\mathbb A}=\partial_{\mathbb M}V_{\mathbb A}+\omega_{\mathbb {MA}}{}^{\mathbb B}V_{\mathbb B}\, ,
\ee
with $E_{\mathbb A}\equiv \sqrt2E_{\mathbb A}{}^{\mathbb M}{}\partial_{\mathbb M}$,  implying $
\omega_{[\mathbb{ABC}]}={\sqrt2}\omega_{\mathbb M[\mathbb{AB}}E^{\mathbb M}{}_{\mathbb {C}]}$.

 Only the totally antisymmetric  and trace parts of $\omega_{\mathbb{ABC}}$ can be determined in terms of $E^{\mathbb M}{}_{\mathbb A}$  and $d$,  namely
 \bea
 \omega_{[\mathbb{ABC}]} &=&-E_{[\mathbb A}E^{\mathbb N}{}_{\mathbb B} E_{\mathbb{ NC}]}-\frac{\sqrt2}3f_{\mathbb{MNP}}E^{\mathbb M}{}_{\mathbb{ A}}E^{\mathbb N}{}_{\mathbb{ B}}E^{\mathbb P}{}_{\mathbb{ C}}\equiv -\frac13{ F}_{\mathbb{ABC}}\, ,
\label{gralspinconnectionE}\\
\omega_{\mathbb{B A}}{}^{\mathbb B} &=& -\sqrt{2} e^{2d} \partial_{\mathbb M}\left(E^{\mathbb M}{}_{\mathbb A} e^{-2d}\right)  \equiv -{ F}_{\mathbb A}\, ,
\label{gralspinconnectiontrace}
\eea
the latter arising from partial integration with the dilaton density 
\be
\int e^{-2d}V\nabla_{\mathbb A}V^{\mathbb A}=-\int  e^{-2d}V^{\mathbb A}\nabla_{\mathbb A}V\, ,
\ee
for arbitrary $V$ and $V^{\mathbb A}$. Only the combinations with the same projection on the last two indices are non-vanishing.

The covariant derivatives  of the  (adjoint)  gravitino and dilatino are
\begin{subequations}
\begin{align}
\nabla_{\mathbb A}\Psi_{\overline{B}}&= E_{\mathbb A}\Psi_{\overline{B}} + \omega_{\mathbb{A}\overline{B}}{}^{\overline{C}} \Psi_{\overline{C}} - \frac{1}{4} \omega_{{\mathbb A}\underline{B} \underline{C}} \gamma^{\underline{B} \underline{C}} \Psi_{\overline{B}}\, , \\
 \nabla_{\mathbb A}\overline{\Psi}_{\overline{B}}&= E_{\mathbb A}\overline{\Psi}_{\overline{B}} + \omega_{\mathbb{A} \overline{B}}{}^{\overline{C}} \overline\Psi_{\overline{C}} + \frac{1}{4} \omega_{{\mathbb A}\underline{B} \underline{C}} \overline\Psi_{\overline{B}} \gamma^{\underline{B} \underline{C}} \, , \label{covder} \\
\nabla_{\mathbb A}{\rho}&= E_{\mathbb A}{\rho} - \frac{1}{4} \omega_{\mathbb A \underline{B} \underline{C}}  \gamma^{\underline{B} \underline{C}} \  {\rho}\,  ,\quad
 \nabla_{\mathbb A}\overline{\rho}=E_{\mathbb A}\overline{\rho} + \frac{1}{4} \omega_{\mathbb A \underline{B} \underline{C}}\  \overline{\rho} \gamma^{\underline{B} \underline{C}}  \, . 
\end{align}
\end{subequations}

The supersymmetry transformation rules are parameterized by an  infinitesimal Majorana fermion $\epsilon$ transforming as a spinor of $\rO(1,9)_\rL$
\begin{subequations}\label{gensusytransf}
\begin{align}
   \delta_\epsilon E^{\mathbb M}{}_{\underline{A}} &=  -\frac12 \bar{\epsilon} \gamma_{\underline{A}} \Psi_{\overline B} E^{ {\mathbb M}\overline B}\, , \qquad
 \delta_\epsilon E^{\mathbb M}{}_{\overline A} = \frac12 \bar{\epsilon} \gamma_{\underline{B}} \Psi_{ \overline A} E^{{\mathbb M}\underline{B}}\, , \qquad \delta_\epsilon d  =  -  \frac{1}{4} \bar{\epsilon} \rho \, , \label{gensusytransf0viel}\\
  \delta_\epsilon \Psi_{\overline{A}} &=  \nabla_{\overline{A}} \epsilon \, , \qquad
    \delta_\epsilon \rho  =  \ -\gamma^{\underline{A}} \nabla_{\underline{A}} \epsilon  \, . \qquad\qquad\quad\label{DFTSUSYf}
\end{align}
\end{subequations}

Putting all together, the generalized fields obey the transformation rules
\begin{subequations}\label{gentransf0}
\begin{align}
  \delta E^{\mathbb M} {}_{\underline{A}}&=\hat{\cal L}_\xi E^{\mathbb M} {}_{\underline{A}} + E^{\mathbb M}{}_{\underline{B}} \Gamma^{\underline{B}}{}_{\underline{A}} -\frac12 \bar{\epsilon} \gamma_{\underline{A}} \Psi_{\overline B} E^{{\mathbb M} \overline B}\, ,\label{gentransf0vielund} \\   
 \delta E^{\mathbb M}{}_{\overline A} &=\hat{\cal L}_\xi E^{\mathbb M}{}_{\overline A} + E^{\mathbb M}{}_{\overline B}  \Gamma^{\overline B}{}_{\overline A} +\frac12 \bar{\epsilon} \gamma_{\underline{B}} \Psi_{ \overline A} E^{{\mathbb M}\underline{B} }\, , \label{gentransf0vielov}\\   
  \delta d \ \ &= \xi^{\mathbb P} \partial_{\mathbb P} d - \frac{1}{2} \partial_{\mathbb P} \xi^{\mathbb P} -  \frac{1}{4} \overline{\epsilon} \rho \, , \label{0transf}\\   
\delta \Psi_{\overline{A}} \  &= \xi^{\mathbb M} \partial_{\mathbb M} \Psi_{\overline{A}} + \Gamma^{\overline{B}}{}_{\overline{A}} \Psi_{\overline{B}} + \frac{1}{4} \Gamma_{\underline{B} \underline{C}} \gamma^{\underline{B} \underline{C}} \Psi_{\overline{A}} + \nabla_{\overline{A}} \epsilon \, , \label{gentransf0gravi}\\
    \delta \rho \ \ &= \xi^{\mathbb M} \partial_{\mathbb M} \rho + \frac{1}{4} \Gamma_{\underline{B} \underline{C}} \gamma^{\underline{B} \underline{C}} \rho - \gamma^{\underline{A}} \nabla_{\underline{A}} \epsilon \label{gentransf0rho}\, .
\end{align}
\end{subequations}
In Appendix \ref{app:closure1} we review the  algebra of these  transformations, and show that it closes  up to terms with two fermions,
 with the following parameters 
\begin{subequations}\label{par0}
\begin{align}
\xi^{\mathbb M}_{12} &= [\xi_1, \xi_{2} ]^{\mathbb M}_{(C_f)} - \frac{1}{\sqrt{2}} E^{\mathbb M} {}_{\underline{A}}\overline{\epsilon_1} \gamma^{\underline{A}} \epsilon_{2} ,\\
\Gamma_{12 {\mathbb A} {\mathbb B}} &= 2 \xi_{[1}^{\mathbb P} \partial_{\mathbb P} \Gamma_{2] {\mathbb A} {\mathbb B}} - 2 \Gamma_{[1 \mathbb A}{}^{\mathbb C} \Gamma_{2] \mathbb{C B}}+E_{[\mathbb A}\left(\overline\epsilon_1\gamma_{\mathbb B]}\epsilon_2\right)-\frac1{2}\left(\overline\epsilon_1\gamma^{\underline C}\epsilon_2\right)F_{\mathbb{AB}\underline C} , \\
\epsilon_{12} &= -\frac{1}{2} \Gamma_{[1 \underline{B} \underline{C}} \gamma^{\underline{B} \underline{C}} \epsilon_{2]} + 2 \xi_{[1}^{\mathbb P} \partial_{\mathbb P} \epsilon_{2]} \, ,
\end{align}
\end{subequations}
where the $C_f$-bracket is defined as
\bea
 [\xi_1, \xi_{2} ]^{\mathbb M}_{(C)}=2\xi^{\mathbb P}_{[1}\partial_{\mathbb P}\xi_{2]}^{\mathbb M}-\xi_{[1}^{\mathbb N}\partial^{\mathbb M}\xi_{2]\mathbb N}+f_{\mathbb{ PQ}}{}^{\mathbb M} \xi_{1}^{\mathbb P} \xi_2^{\mathbb Q}\, .
\eea

The transformation rules \eqref{gentransf0}  leave the following action invariant, up to  bilinear  terms in fermions,
\begin{equation}
\label{DFTsusyAction}
{\mathbb S}_{\mathcal{N}=1 \ {\rm DFT}} \ = \ \int d^{20+n_g}X\,e^{-2d}\left(\mathbb{L}_\rB+{\mathbb L}_\rF
 \right)\, ,
\end{equation}
where ${\mathbb{L}}_\rB$
 is the generalized Ricci scalar, which can be written as
\bea
{\mathbb L}_\rB \ \equiv \ {\mathbb R}&=&\frac18{ F}_{\mathbb{ABC}}{ F}_{\mathbb{DEF}}\left({H}^{\mathbb{AD}}\eta^{\mathbb{BE}}\eta^{\mathbb{CF}}-\frac1{3}{H}^{\mathbb{AD}}{H}^{\mathbb{BE}}{H}^{\mathbb{CF}}\right)-{H}^{\mathbb{AB}}\left(\frac12{ F}_{\mathbb A}{ F}_{\mathbb B}+E_{\mathbb A}{ F}_{\mathbb B}\right)\, ,\nn
\eea
up to terms that vanish under the strong constraint, and the fermionic Lagrangian is 
\bea
{\mathbb L}_\rF = \overline{\Psi}^{\overline{A}}\gamma^{\underline{B}}\nabla_{\underline{B}}{\Psi}_{\overline{A}}
 -\bar{\rho}\gamma^{\underline{A}}\nabla_{\underline{A}}\rho+2\overline{\Psi}^{\overline{A}}\nabla_{\overline{A}}\rho\, . \label{lf}
\eea

Using the Bianchi identity
\bea
\frac16{F}_{\mathbb{ABC}}{ F}^{\mathbb{ABC}}=2E_{\mathbb A}{ F}^{\mathbb A}+{ F}_{\mathbb A}{ F}^{\mathbb A}\, ,
\eea
 it is useful to rewrite 
\bea
{\mathbb R}&=& 2E_{\underline{A}}F^{\underline{A}}+F_{ \underline{A}}F^{\underline{A}}-\frac16{ F}_{\underline{A} \underline{B} \underline{C}}{ F}^{\underline{A} \underline{B} \underline{C}}-\frac12{ F}_{\overline A \underline{B} \underline{C}}{ F}^{\overline A \underline{B} \underline{C}}.\label{r}
\eea

The supersymmetry variation of the bosonic piece of the action gives 
\bea
e^{2d}\delta_\epsilon [{e^{-2d}\mathbb{R}(E,d)}]=\frac12\overline\epsilon\rho\mathbb{R}+2\Delta E_{{\underline{A}} \overline B}{\mathbb R}^{\overline B {\underline{A}}}=
\frac12\overline\epsilon\rho\mathbb{R}-\overline\epsilon\gamma^{\underline{A}}\Psi^{\overline B}{\mathbb R}_{\overline B \underline{A}}\, ,\label{varbos}
\eea
where we have used 
\bea
\delta_\epsilon { F}_{\mathbb{ABC}}=-3\left(E_{[\mathbb A}\Delta E_{\mathbb{BC}]}+\Delta E_{[\mathbb{A}}{}^{\mathbb D}{ F}_{\mathbb{BC}]\mathbb D}\right) \label{transff}
\end{eqnarray}
with 
\bea
\Delta E_{\mathbb{AB}}\equiv E^{\mathbb M}{}_{\mathbb A}\delta_\epsilon E_{\mathbb {MB}}=-\Delta E_{\mathbb{BA}}=\left\{\begin{matrix}\Delta E_{\underline{A} \underline{B}}=\Delta E_{\overline{AB}}=0\\
\Delta E_{\underline{A}\overline B}=-\Delta E_{\overline B \underline{A}}=\frac12\overline\epsilon\gamma_{\underline{A}} \Psi_{\overline B}\end{matrix}\right.
\eea
and 
\bea
\delta_\epsilon{\mathbb R}
&=&-\bar\epsilon\gamma_{\underline{A}}\Psi^{\overline B}\left[E_{\overline B}F^{\underline{A} }-E_{\underline{C}}F_{\overline B}{}^{\underline{A} \underline{C}}+F_{\underline{C} \overline {A B}}F^{\overline A \underline{A} \underline{C}}-F_{\underline{D}}F_{\overline B}{}^{\underline{A} \underline{D}}\right] = -\overline\epsilon\gamma^{\underline{A}} \Psi^{\overline B}{\mathbb R}_{\overline B \underline{A}}\, .\nn
\eea

The supersymmetry transformation rules  define the following Lichnerowicz principle
\bea
\left(\gamma^{\underline{ A}}\nabla_{ \underline{A}}\gamma^{ \underline{B}}\nabla_{\underline{ B}}-\nabla^{\overline{A}}\nabla_{\overline{A}}
\right)\epsilon&=&-\frac14{\mathbb R}\epsilon\, ,\label{lich1}\\
\left[\nabla_{\overline{ A}}, \gamma^{ \underline{ B}}\nabla_{\underline{  B}}\right]\epsilon&=&\frac12\gamma^{\underline { B}}{\mathbb R}_{{\overline{ A}} \underline{B}}\epsilon\, , \label{lich2}
\eea
and then, the supersymmetric variation of the fermionic piece of the action
\begin{eqnarray}
e^{2d}\delta_{\epsilon}\left( e^{-2d}\mathbb{L}_{\rF} \right) =  -2\Delta E_{\underline{B} \overline A}{\mathbb R}^{\overline A \underline{B}} -\frac{1}{2}\bar{\epsilon}\rho\mathbb{R} 
  = \overline{\epsilon}\gamma_{\underline{B}}\Psi_{\overline{A}}\mathbb{R}^{\overline{A}{\underline{B}}}-\frac{1}{2}\overline{\epsilon}\rho\mathbb{R}  \, , \label{InvFermAction}
\end{eqnarray}
exactly cancels  (\ref{varbos}).

\subsection{Parameterization and choice of section }
 To make contact with  ten dimensional $\mathcal{N}=1$ supergravity coupled to $n_g$ vector multiplets, 
we split the G and H  indices  as ${\mathbb M}=({}_\mu,{}^\mu,i)$ and ${\mathbb A}=(\underline{A},\overline{A})$,  respectively with $\underline{A}=\underline a, \overline{A}=(\overline{a},\overline{i})$,  $_\mu, ^\mu, \underline a, \overline a=0,\dots, 9$, $i, \overline i=1,\dots, n_g$, and
    parameterize the generalized fields as follows:

\noindent - Generalized frame 
\be
E^{\mathbb M}{}_{\mathbb A}  =\left(\begin{matrix}{ E}_{\mu \underline a}&  { E}^{\mu }{}_{\underline a} & E^i{}_{\underline a}\\ 
E_{\mu \overline  a}& E^\mu{}_{\overline  a}&E^i {}_{\overline a} \\
E_{\mu\overline i} &E^\mu{}_{\overline i} &E^i{}_{\overline i} \end{matrix}\right) \ = \
\frac{1}{\sqrt{2}}\left(\begin{matrix}-{ e}_{\mu \underline a}-C_{ \rho\mu} { e}^{\rho }{}_{\underline a} &  { e}^{\mu }{}_{\underline a} & -A_\rho{}^i { e}^{\rho }{}_{\underline{a}}\, , \\ 
e_{\mu \overline a}-C_{\rho \mu}{} e^{\rho }{}_{\overline{a}}& e^\mu{}_{\overline a}&-A_{\rho}{}^i  e^\rho{}_{\overline a} \\
\sqrt2 A_{\mu i}e^i{}_{\overline i} &0&\sqrt2 e^i{}_{\overline i} \end{matrix}\right)  \, ,
\label{HKparam}
\ee
where   ${e}_{\underline  a}$ and $e_{\overline  a}$ are two vielbein for the same  ten dimensional metric.  To guarantee that the number of DFT and supergravity degrees of freedom agree, we gauge fix
${ e}^{\mu}{}_{\underline a}=e^{\mu}{}_{\overline {a}},  {e}_{\mu\underline  a}= e_{\mu \overline{a}} $, and identify $ e^{\mu}{}_{\overline{a}}, e_{\mu \overline a}$ with the supergravity vielbein   $e^{\mu}{}_{{a}}, e_{\mu a}$, $a,b=0,\dots, 9$, respectively, i.e.  $g_{\mu\nu}={ e}_{\mu}{}^a g_{{ab}}{e}_{\nu}{}^b$, with $g_{ab}$  the Minkowski metric.  $C_{\mu\nu}=b_{\mu\nu}+\frac12 A_{\mu}^i A_{\nu i}$, with $A_\mu^i$ being the gauge connection.
For consistency, we also need to impose 
\bea
P_{\underline{ab}}=-g_{{ab}}\delta_{\underline a}^a\delta_{\underline b}^b, \qquad \overline P_{\overline{ab}}=g_{{ab}}\delta_{\overline a}^a\delta_{\overline b}^b\, , \qquad \overline P_{\overline{ij}}=e^i{}_{\overline i}\eta_{{ij}} e^j{}_{\overline j}\, ,
\eea
with $e^i{}_{\overline i}$  the (inverse) vielbein for the   Killing metric of the $\hog$ or $\mre_8 \times \mre_8$  gauge group, $\eta_{ij}=e_i{}^{\overline i} \eta_{\overline i\overline j}e_j{}^{\overline j}$,  as required for  modular invariance of the heterotic string.
\\

\noindent  - Generalized dilaton and dilatino \bea
d=\phi-\frac{1}{2}\log\sqrt{-g}\, \, \, \qquad {\rm and} \, \, \, \qquad \rho=2\lambda +\gamma^\mu\psi_\mu \, ,\label{gendil}
\eea
where $\phi, \psi_\mu$ and $\lambda$ are the standard dilaton, gravitino and dilatino fields, respectively.\\

\noindent - Generalized gravitino: 
\bea
\Psi_{\mathbb{A}}=(0, e^\mu{}_{ a}\psi_{\mu},\frac1{\sqrt2}e^i{}_{\overline i}\chi_i )\, , \label{gengrav}\eea
$\chi_{ i}$ being the standard gaugino field.

The non-abelian gauge sector is trivially incorporated through the gaugings that deform the generalized Lie derivative \eqref{dife} as
\bea
f_{\mathbb{MN}}{}^{\mathbb P}=\left\{\begin{matrix}f_{ij}{}^k& {\rm for} \  {\mathbb {M}, \mathbb{N}, \mathbb{P}}=i,j,k \\
0 & {\rm otherwise.}\end{matrix}\right.\, \label{fidentification}
\eea

 The $\gamma$-functions  $\gamma^{\underline a}=\gamma^a\delta_a^{\underline a}$ verify the Clifford algebra $\{\gamma^a, \gamma^b\}=2g^{ab}$.

The gauge fixing $e^\mu{}_{\underline a}= e^\mu{}_{\overline  a}$  implies $\delta  e^\mu{}_{\underline a}=\delta e^\mu{}_{\overline  a}$, and  \eqref{dltf}  lead to
\bea
 \Gamma_{\underline{ab}} =\left( -\Lambda_{{ab}}  + \bar{\epsilon} \gamma_{[a} \psi_{b]}\right)\delta^a_{\underline a}\delta^b_{\underline b} \, , \label{lambdagaugefixing}
\eea 
where 
$\Lambda_{ab}$ denotes  the generator of  $\rO(1,9)$ transformations that  parameterizes $\Gamma_{\overline{ab}}$. 

The additional gauge fixings $\delta E^i{}_{\overline i}=0$ and $\delta E^\mu{}_{\overline i}=0$   lead respectively to 
\bea
\Gamma_{\overline{ij}} = \Lambda_{ij}\delta_{\overline i}^i\delta_{\overline j}^j = f_{{ijk}}\xi^{ k} \delta_{\overline i}^i\delta_{\overline j}^j\, \qquad {\rm and}\qquad \Gamma_{\overline {ai}}=\Lambda_{ai}\ \delta^a_{\overline a}\ \delta^i_{\overline i}=\frac1{2\sqrt2} \overline\epsilon\gamma_a \chi_{ i}\ \delta^a_{\overline a}\ \delta^i_{\overline i}\, , \label{lorgau}
\eea
where we have parameterized  $
\xi^{\mathbb M}=(\xi_\mu, \xi^\mu, \xi^i)$ and $\Lambda_{ai}$, $\Lambda_{ij}$ are introduced for convenience, as we will discuss in section \ref{sec:transf}. 

Solving the strong constraint in the supergravity frame, parameterizing  \eqref{gensusytransf} and using  the non-vanishing determined components of the generalized spin connection listed in Appendix \ref{app:fluxes},   we  recover the leading order supersymmetry transformation rules of the coupled ten dimensional ${\cal N}=1$ supergravity and Yang-Mills fields, namely
\begin{subequations}\label{transf0}
\begin{align}
\delta_\epsilon e_{\mu}{}^{a} &=  \frac12 \bar{\epsilon} \gamma^{a} \psi_{\mu}\, ,  \  \ \ \ \ \ \ \ \ \ \ \ \  \ \ \ \ \ \ \ \ \ \ \ \delta_\epsilon \phi =  -  \frac{1}{2} \bar{\epsilon} \lambda =-\frac14\bar\epsilon\rho+\frac14\bar\epsilon\gamma^\mu\psi_\mu\, , \label{transf0dil}\\
\delta_\epsilon b_{\mu \nu} & =  \bar{\epsilon} \gamma_{[\mu} \psi_{ \nu ] }+\frac12\bar\epsilon\gamma_{[\mu}\chi^i A_{\nu]i}\, ,
 \quad \ \ \  \ \delta\rho=\gamma^\mu D_\mu\epsilon-\frac1{24}H_{abc}\gamma^{abc}\epsilon-\gamma^\mu\partial_\mu\epsilon\label{transf0b}\\ 
  \delta_\epsilon \psi_{\mu} &= \partial_{\mu} \epsilon -\frac14w^{(+)}_{\mu ab}\gamma^{ab}\epsilon\,  ,  \  \ \ \ \ \ \ \ \ \ \ \
 \delta_\epsilon \lambda \ = \ - \frac{1}{2}\gamma^{\mu} \partial_{\mu}\phi \epsilon +\frac{1}{24}H_{{a} {b} c}\gamma^{{a} {b} c} \epsilon \, , \ \ \  \label{transf0dilino} \\
\delta_\epsilon A_\mu^i &= \frac12\bar\epsilon\gamma_\mu\chi^i\, ,    \  \ \ \ \ \ \ \ \  \ \ \ \ \ \ \ \ \ \ \ \ \ \ \ \delta_\epsilon\chi^i =-\frac14 F_{\mu\nu}^i\gamma^{\mu\nu}\epsilon \, ,  \label{transf0chi}
\end{align}
\end{subequations}
where $ w_{\mu ab}^{(+)}=w_{\mu ab}+\frac12 H_{\mu ab}$ is the  spin connection with torsion  given by the field strength of the $b$-field 
\be
H_{abc}=e^\mu{}_{[a} e^\nu {}_be^\rho{}_{c]} H_{\mu\nu\rho}=3e^\mu{}_{a} e^\nu{}_b e^\rho{}_{c}\left(\partial_{[\mu}b_{\nu\rho]}-C_{\mu\nu\rho}^{(g)}\right)\, , \label{hcero}
\ee 
with $C_{\mu\nu\rho}^{(g)}$  the Yang-Mills Chern-Simons form
\be
C_{\mu\nu\rho}^{(g)}=A^i_{[\mu}\partial_\nu A_{\rho]i}-\frac13 f_{ijk}A_\mu^i A_\nu^jA_\rho^k \, . \label{CSg}
\ee

The
Lorentz transformations of the supergravity and super Yang-Mills multiplets obtained from \eqref{dltf} are
\be
\delta_\Lambda e_\mu{}^a \ = \ e_\mu{}^b\Lambda_b{}^a \, , \qquad \delta_\Lambda \psi_a \ = \ \psi_b\Lambda^b{}_a-\frac14\gamma^{bc}\Lambda_{bc}\psi_a \, ,\qquad\delta_{\Lambda}\chi=-\frac14\Lambda_{bc}\gamma^{bc}\chi\, , \label{lor0}
\ee
and the gauge transformations derived from \eqref{gendiffeomorphisms} are
\be
\delta_{\xi}A_\mu^i=\partial_\mu\xi^i+f^i{}_{jk}\xi^jA_\mu^k\, ,  \qquad\delta _\xi\chi^i=f^i{}_{jk}\xi^j\chi^k\, ,  \qquad \delta_{\xi} b_{\mu\nu}=2\partial_{[\mu}\xi_{\nu]}-\partial_{[\mu}\xi^i A_{\nu]i}\, , \label{gauge0}\ee
where the second term in the gauge transformation of the $b$-field is the gauge sector of the Green-Schwarz transformation required for anomaly cancellation.

Parameterizing the DFT action \eqref{DFTsusyAction}, using the fluxes listed in Appendix \ref{app:fluxes}, we get 
\bea
S &=& \int d^{10}x \ e \ e^{-2\phi}\left[R(w(e)) -\frac{1}{12}H_{\mu \nu \rho}H^{\mu \nu \rho} + 4\partial_{\mu}\phi\partial^{\mu}\phi-\frac14 F^i_{\mu\nu}F_i^{\mu\nu} \right. \nn\\ 
&  & \ \  - \ov{\psi}{}^{\mu}\slashed{D}\psi_{\mu}  + \ov{\rho}\slashed{D}\rho+ 2\ov{\psi}{}^{\mu}{D}_{\mu}\rho-\frac12\ov\chi^i{\slashed{ D}}\chi_i+ \ \ov\chi_i\left(\gamma^\mu\psi^{\nu}-\frac14\gamma^{\mu\nu}\rho\right)F_{\mu\nu}^i\nn \\
& & \left. \ \ + \frac{1}{24}H_{\rho \sigma \tau}\left(\ov{\psi}{}^{\mu}\gamma^{\rho \sigma\tau }\psi_{\mu} + 12\ov{\psi}{}^{\rho}\gamma^{ \sigma }\psi^\tau - \ov{\rho}\gamma^{\rho \sigma \tau}\rho -6\ov\psi{}^\nu\gamma^{\rho\tau}\rho+\frac12\ov\chi^i\gamma^{\rho\sigma\tau}\chi_i\right) \right] \, .
\label{BdRAction}
\eea

We use standard notation defined in Appendix \ref{app:conventions}.
Both the action and the transformation rules match the corresponding ones   in \cite{Bergshoeff:1988nn}, with the field redefinitions specified in Appendix \ref{app:eom}, where \eqref{BdRAction} is rewritten in terms of  the standard supergravity  dilatino $\lambda$  instead of   $\rho$.

\section{The first order $\alpha'$-corrections}\label{sec:DFTalpha}

In this section we  construct the first order  corrections to  ${\cal N}=1$ supersymmetric DFT, performing a perturbative expansion of the exact formalism developed in \cite{diego2018}.

The duality structure of the first order $\alpha'$-corrections to
heterotic supergravity was
originally considered in \cite{bmn,otros}.
Exploiting a
symmetry
between the gauge and torsionful spin connections that exists in   ten dimensional heterotic supergravity
\cite{bdr1,Bergshoeff:1988nn}, the duality group was extended to $\rO(10,10+n_g+n_l)$, with $n_g(n_l)$ the dimension of the heterotic gauge (Lorentz)  group. In this construction, the gaugings in the  generalized Lie derivative \eqref{dife}  preserve a residual $\rO(10,10)$ global symmetry.  Including one-form  fields
in the GL$(10)$ parameterization of the generalized vielbein,  the formalism reproduces the first order corrections to the interactions of the bosonic fields in the heterotic effective field theory. This construction was supersymmetrized in \cite{lee}.

The  lack of manifest duality covariance  and the difficulties to incorporate higher orders of the $\alpha'$-expansion in  these formulations  motivated  the search of alternative frameworks. A deformation of the gauge structure of DFT was
proposed in \cite{mn}, introducing a  generalized Green-Schwarz  transformation 
that modifies the leading order double Lorentz variations (\ref{dltf}) with two derivative corrections.
 The deformations fix the four derivative  terms of bosonic fields in all T-duality symmetric gravitational theories, including in particular  the bosonic and heterotic string effective actions \cite{bfmn}.

The two formalisms described above were merged in the so-called generalized Bergshoeff-de Roo identification  introduced in \cite{diego2018}. In the first part of this section we briefly review this  exact  supersymmetric and manifestly duality covariant formulation. Then we perform a perturbative expansion and extract the first order corrections to the transformation rules  of the $\rO(10,10+n_g)$ multiplets  \eqref{gentransf0}.   Finally,  we construct the gauge invariant action containing three and four derivatives of the duality multiplets up to bilinear terms in fermions.

\subsection{The generalized Bergshoeff-de Roo identification}\label{sec:gbdr}

The theory has a global $\rO(10,10+k)$ symmetry, where $k$  is the dimension of  the $\rO(1, 9+k)$ group.
 This differs from the construction of the previous section, where the duality group is $\rO(10,10+n_g)$ and $n_g$ denotes the dimension of the  $\hog$ or $\mre_8\times \mre_8$ heterotic gauge group. In the construction of \cite{diego2018} instead the  gauge sector encodes the higher derivatives.

The  vielbein ${\cal{E}}_{\cal {M}}{}^{{\cal A}}$ is an element of $\rO(10,10+k)$,  parameterized  in terms of $\rO(10,10)$ fields as\footnote{Note that this differs from  \eqref{HKparam} and from previous constructions, e.g. \cite{bmn,otros}, where the  generalized vielbein is parameterized with GL(10) multiplets. } 
\bea
\begin{matrix}
{\cal E}_{M}{}^{\overline a} = E_M{}^{\overline a} \, ,&
{\cal E}_{M}{}^{\underline a} = (\triangle^{\frac 1 2})_M{}^P\, E_P{}^{\underline a} \, , &
{\cal E}_{M}{}^{\overline \alpha} = -{\rm A}_M{}^\beta\,  e_\beta{}^{\overline \alpha} \, ,\\
{\cal E}_{\alpha}{}^{\overline a} = 0 \, ,&
{\cal E}_{\alpha}{}^{\underline a} =  E_M{}^{\underline a}\,  {\rm A}^M{}_{\alpha} \, , &
{\cal E}_{\alpha}{}^{\overline \alpha} = (\Box^{\frac 1 2})_\alpha{}^\beta \, e_\beta{}^{\overline \alpha}\ .\end{matrix} \label{frameidentification}
\eea
We use  calligraphic symbols to distinguish the $\rO(D,D+k)$ objects. The indices ${\cal {M}}=(M, \alpha)=(^\mu, _\mu,\alpha)$ and ${\cal A}=(\underline {\cal A}, \overline{\cal A})$ take  values $M=0,\dots, 19$, $\underline {\cal A}\equiv \underline a=0,\dots,9; \overline {\cal A}=(\overline a,\overline\alpha),\overline a=0,\dots,9$ and $\alpha, \overline\alpha=1,\dots, k$.  ${\rm{A}}_{M}{}^{\alpha}$ is a constrained $\rO(10,10)$  vector field satisfying ${\rm {A}}_M{}^\alpha=P_M{}^N {\rm{A}}_N{}^\alpha$ (the projection is fixed by the choice of  $\rO(10,10+k)$ duality group, as opposed to $\rO(10+k,10)$ which would give an equivalent  ${\mathbb Z}_2$ transformed  theory),  and 
\bea
\Box_{\alpha}{}^\beta &=& \kappa_\alpha{}^\beta - {\rm A}_{M\alpha} {\rm A}^{M\beta} \, , \\
\triangle_M{}^N &=& \eta_M{}^N - {\rm A}_{M \alpha} {\rm A}^{N\alpha}\,.
\eea
The gauge freedom is used to set  ${\cal E}_\alpha{}^{\bar a}$ to zero and the bijective map $e_\alpha{}^{\overline \beta}$ relates the Cartan-Killing metrics of $\rO(k)$, $\kappa_{\alpha\beta}$ and $\kappa_{\overline{\alpha\beta}}$, as 
\bea
e_\alpha{}^{\overline \alpha} {\kappa}_{\overline{\alpha\beta} } e_\beta{}^{\overline \beta}  =\kappa_{\alpha\beta}\, .
\eea

 The  parameterization (\ref{frameidentification})  preserves the constraint 
\be
{\cal E}_{\mathcal{M}}{}^{{\cal A}} {\eta}_{\cal {AB}} {\cal E}_{\mathcal{N}}{}^{\cal {B}}= {\eta}_{\mathcal{M N}}\, ,
\ee
 where ${\eta}_{\mathcal{M N}}$  and  ${\eta}_{\cal{AB}}$ are  the invariant metrics  of $\rO(10,10+k)$ and $\rO(9,1)_\rL\times \rO(1, 9+k)_\rR$,
\bea
{\eta}_{\mathcal{M N}}  = \left(\begin{matrix}0&\delta_\mu{}^\nu&0\\ 
\delta^\mu{}_\nu&0&0\\
0&0&\kappa_{\alpha\beta}\end{matrix}\right)\, , \qquad
 {\eta}_{\mathcal{AB}}= \left(\begin{matrix}-g_{\underline {ab}}&0&0\\ 
0&g_{\overline {ab}}&0\\0&0&\kappa_{\overline{\alpha\beta}}\end{matrix}\right) \ .  \label{eta}
\eea

The generalized $\rO(10,10+k)$ gravitino splits as $\Psi_{\cal A}=(0,{\Psi}_{\bar{a}},\Psi_{\overline{\alpha}})$, where $\Psi_{\overline{a}}$ is a generalized $\rO(10,10)$ gravitino and $ \Psi_{\overline{\alpha}}$ is a   gaugino of the $\rO(1, 9+k)_\rR$ gauge group, that will later be  identified with a function of the $\rO(10,10)$ generalized  fields. The  gamma matrices are  $\gamma^{\cal A}=(\gamma^{\underline a},0,0)$, with $\gamma^{\underline a}$  the $\rO(9,1)_\rL$ gamma matrices verifying \eqref{clif}.

The transformation rules of the  $\rO(10,10+k)$ fields  have the same functional form as  \eqref{gentransf0}, namely
\begin{subequations}\label{tranfe}
\begin{align}
\delta {\cal E}_{\cal M}{}^{\cal A}&= \xi^{\cal P} \partial_{\cal P} {\cal E}_{\cal M}{}^{\cal A} + (\partial_{\cal M} \xi^{\cal P} - \partial^{\cal P} \xi_{\cal M}) {\cal E}_{\cal P}{}^{\cal A} +g f_{{\cal M}{\cal N}}{}^{\cal P} \xi^{\cal N} {\cal E}_{\cal P}{}^{\cal A} \nn\\
& \ \ \ + \ {\cal E}_{\cal M}{}^{\cal B} {\mathscr T}_{\cal B}{}^{\cal A}-\overline{\epsilon} \gamma^{[{\cal A}} \Psi^{{\cal B}]} {\cal E}_{{\cal M} \cal B}\, , \label{transfeviel}\\   
   \delta d &= \xi^{\cal P} \partial_{\cal P} d - \frac{1}{2} \partial_{\cal P} \xi^{\cal P} -  \frac{1}{4} \bar{\epsilon} \rho  \\
  \delta \Psi_{\overline{\cal{ A}}} &= \xi^{\cal M} \partial_{\cal M} \Psi_{\overline{\cal{ A}} }+ {\mathscr T}^{\overline{\cal{B}}}{}_{\overline{\cal{A}}} \Psi_{\overline{\cal{ B}}} + \frac{1}{4} {\mathscr T}_{\underline{\cal{ BC}} }\gamma^{\underline{\cal{BC}}} \Psi_{\overline{\cal A}} + \nabla_{\overline{\cal{A}} }\epsilon \label{transfegravi}\\
  \delta \rho &= \xi^{\cal M} \partial_{\cal M} \rho + \frac{1}{4} {\mathscr T}_{\underline{\cal{AB}}} \gamma^{\underline{\cal{AB}}} \rho - \gamma^{\underline{\cal A}} \nabla_{\underline{\cal{A}}} \epsilon \, ,  \label{transferho}
\end{align}
\end{subequations}
where $g^{-2}\sim\alpha'$ is a dimensionful constant, ${\mathscr T}_{\cal{AB}}$ parameterizes the local double Lorentz $\rO(9,1)_\rL\times \rO(1,9+k)_\rR$ tangent space symmetry,
\bea
\nabla_{\cal A} \epsilon \ = \ {\cal E}_{\cal A}  \epsilon  - \frac14 \omega_{\cal{A  B C}} \gamma^{\cal{B  C}} \epsilon \, ,
\label{gencovariant}
\eea
with ${\cal E}_{\cal A}=\sqrt2{\cal E}^{\cal M}{}_{\cal A}\partial_{\cal M}$, and  the identifications 
\bea
\mathcal{F}_{\cal{ABC}} &= & 3 {\cal E}_{[{\cal A}}{\cal E}^{\cal N}{}_{{\cal B}} {\cal E}_{{\cal N C}]} + g \sqrt2f_{\cal{MNP}}{\cal E}^{\cal M}{}_{{\cal A}}{\cal E}^{\cal N}{}_{{\cal B}}{\cal E}^{\cal P}{}_{{\cal C}}\, = -3 {\omega}_{[\cal{ABC}]}\, , \label{fluxidentification2}\\
  {\mathcal{F}}_{\cal A}&=& \sqrt2e^{2d}\partial_{\cal M}\left({\cal E}^{\cal M}{}_{\cal A}e^{-2d}\right)=- { \omega}_{\cal{BA}}{}^{\cal B}\, , \label{fluxidentification3}
\eea
\bea
f_{{\cal M}{\cal N}}{}^{\cal P}=\left\{\begin{matrix}f_{\alpha\beta}{}^\gamma & {\rm for} \  {\cal M},{\cal N}, {\cal P}=\alpha, \beta, \gamma \\
0 & {\rm otherwise}\end{matrix}\right.\, .\label{fidentificationhat}
\eea

Equivalent constraints to \eqref{consf} and \eqref{StrongConstraint}    must be imposed, i.e.
\begin{subequations}\label{sconst}
\begin{align}
\partial_{\cal M}\partial^{\cal M}\cdots =0\, , \qquad \partial_{\cal M}\cdots \partial^{\cal M}\cdots &=0\, , \qquad f^{\cal M}{}_{\cal {NP}}\partial_{\cal M}\cdots =0\, ,\\
f_{\cal{MNP}}=f_{[{\cal MNP}]}\, ,\quad & \quad f_{[\cal{MN}}{}^{\cal R}f_{{\cal P}]\cal R}{}^{\cal Q}=0\, .
\end{align}
\end{subequations}

The gauge fixing $\delta {\cal E}_{\alpha}{}^{\overline a}  =  0$ implies
\bea
{\mathscr T}_{\overline \alpha}{}^{\overline  b}= \Big( \partial^P \xi_\alpha {\cal E}_P{}^{\overline b} -\frac12\overline\epsilon\gamma^{\underline c} {\Psi}^{\overline b} {\cal E}_{\alpha \underline c} \Big) (\Box^{-\frac12})^{\alpha}{}_{\beta} e^\beta{}_{\overline \alpha}\, ,
\label{gaugefixing1}
\eea
and $\delta e_{\alpha}{}^{\overline \alpha} \ = \ 0$ determines   
\bea
{\mathscr T}_{\overline \alpha\overline \beta} &=& \Big( \delta(\Box^{\frac12})_{\alpha}{}^{\beta} e_{\beta[\overline{\beta}}  - {\xi}^{P} \partial_{P} {\cal E}_{\alpha[\overline{\beta}} + \partial^{P} {\xi}_{\alpha} {\cal E}_{P[\overline{\beta}}  -g{f}_{\alpha\beta }{}^\gamma {\xi}^{\beta} {\cal E}_{\gamma[\overline \beta} - \frac12 \overline{\epsilon} \gamma^{\underline b} {\Psi}_{[\overline{\beta}} {\cal E}_{\alpha \underline b} \Big)  e^{\delta}{}_{\overline{\alpha}]}(\Box^{-\frac12})^{\alpha}{}_{\delta} \, .\nn\\
\label{gaugefixing2}
\eea 
The gauge generators $\left(t^\alpha\right)_{\overline {\cal A}}{}^{\overline {\cal B}}$  implement the map
\bea
V_{\overline {\cal A}}{}^{\overline {\cal B}}=-g V_\alpha\left(t^\alpha\right)_{\overline {\cal A}}{}^{\overline {\cal B}}\, ,
\label{map}
\eea
allowing to write
\bea
-g \xi_\alpha (t^\alpha)_{\overline {\cal A}\overline {\cal B}}\equiv {\mathscr T}_{\overline {\cal A}\overline {\cal B}}\, ,\qquad\qquad
-g {\cal E}_\alpha{}^{\underline a}\left(t^\alpha\right)_{\overline {\cal C}\overline {\cal  D}}\equiv \frac1{\sqrt2}{\rm A}^{\underline a}{}_{\overline { \cal C}\overline {\cal D}} \  .
\label{Ctransf1}
\eea
 They satisfy $[t_\alpha \, , \, t_\beta] = {f}_{\alpha\beta}{}^\gamma \, t_\gamma$ and $Tr(t^\alpha t_\beta )=X_R\delta^\alpha_\beta$, where $X_R$ is the Dynkin index of the representation.

Parameterizing  $\delta {\cal E}_M{}^{\underline a}$ one gets 
\bea
\delta {\rm A}_{\underline a\overline {\cal C} \overline {\cal D}}  &=& \xi^P\partial_P {\rm A}_{\underline a\overline {\cal C} \overline {\cal D}} -  {\cal E}_{\underline a}{\mathscr T}_{\overline {\cal C} \overline { \cal D}} - 2 {\rm A}_{\underline a[\overline {\cal C}}{}^{\overline {\cal B}} {\mathscr T}_{\overline {\cal D}] \overline { \cal B}} \ -{\rm A}^{\underline b}{}_{\overline {\cal C}\overline { \cal D}}{\mathscr T}_{\underline{ab}} + \overline\epsilon \gamma_{\underline a} \Psi_{\overline{{\cal C}} \overline{{\cal D}}}  \ ,    \ \ \ \ 
\label{transfClorentz}
\eea
where 
\bea
{\Psi}_{\overline {\cal C}\overline {\cal D}}\equiv\frac g{\sqrt2} \Psi_{\overline \beta} {\cal E}_\alpha{}^{\overline \beta} (t^\alpha)_{\overline {\cal C}\overline {\cal D}}\, .
\label{gauginodft}
\eea
In order to eliminate these extra degrees of freedom, it is convenient to define
\bea
{\mathcal{F}}^*_{\underline a{\overline {\cal C}\overline { \cal D}}} \ = \ {\mathcal{F}}_{\underline a{\overline {\cal C}\overline {\cal D}}}-\frac12\overline{\Psi}_{\overline { \cal C}} \gamma_{\underline a} \Psi_{\overline { \cal D}}\, ,\label{fstar}
\eea
which allows to establish the generalized Bergshoeff-de Roo  identification between the generalized gauge and spin connections
\bea
{\rm A}_{\underline{M} {\overline { \cal C}\overline { \cal D}}} =  {\cal F}^{*}_{\underline{M} {\overline {\cal C}\overline {\cal D}}} 
\, ,\label{Fident}
\eea
 and to determine ${\Psi}_{{\overline {\cal C}\overline {\cal D}}}$ as the generalized gravitino curvature
\bea
{\Psi}_{{\overline {\cal C}\overline {\cal D}}}= \nabla_{[\overline {\cal C} } \Psi_{\overline {\cal D}]} +\frac12 \omega^{\overline {\cal B}}{}_{\overline {\cal C}\overline {\cal D}} \Psi_{\overline {\cal B}}\, ,
\label{gauginoidentf2}
\eea
since both sides of (\ref{Fident}) and (\ref{gauginoidentf2}) transform in the same way.
The main steps of the demonstration can be found in \cite{diego2018}. 

We now proceed to extract the  first order $\alpha'$-corrections to the transformation rules of the $\rO(10,10+n_g)$ generalized fields.

\subsection{Induced transformation rules on $\rO(10,10)$ multiplets}\label{sec:induced}

The  covariant transformation rules  \eqref{tranfe} induce higher derivative deformations on the transformations  \eqref{gentransf0} of the $\rO(10,10+n_g)$ fields.  In this section, we work out the first order modifications, expanding the coefficients $(\Box^{\frac12}) _{\alpha}{}^\beta$ and $(\triangle^{\frac12})_M{}^N$ in the parameterization of  the $\rO(10,10+k)$ multiplets.

To simplify the presentation, we turn off the gauge sector of the $\rO(10,10+n_g)$ multiplets, i.e. we take $n_g=0$, and obtain the induced transformation rules of the $\rO(10,10)$ fields. The gauge sector will be trivially included in the next subsection.

It is convenient to first express  the components of the generalized $\rO(10, 10+k)$ fluxes \eqref{fluxidentification2} and \eqref{fluxidentification3} in terms of the $\rO(10,10)$ fluxes \eqref{gralspinconnectionE} and \eqref{gralspinconnectiontrace}.
Keeping only the  first order terms in the expansion of the coefficients $(\Box^{\frac12}) _{\alpha}{}^\beta$ and $(\triangle^{\frac12})_M{}^N$, namely
\be
(\Box^{\frac12}) _{\alpha}{}^\beta \ \cong \ \kappa{}_{\alpha}{}^\beta  - \frac{1}{2} {\rm{A}}_{M\alpha} {\rm{A}}^{M \beta} \,  , \ \ \ \
(\triangle^{\frac12})_{MN}\cong \eta_{MN} - \frac12 {\rm A}_{M \alpha} {\rm A}_{N \beta} \kappa^{\alpha\beta} \, ,
\ee
we get the first order deformations 
\begin{subequations}\label{genfluxes}
\begin{align}
{\cal F}_{\underline {abc}}&=F_{\underline {abc}}+F^{(3)}_{\underline {abc}}\cong F_{\underline {abc}}  -\frac{3b}{4}\left(  E_{[\underline{a}}{F}^{*\overline{cd}}_{\underline{b}} - 
\frac{1}{2}F_{\underline d[\underline{ab}}F^{*\underline{d}\overline{cd}} - \frac{2}{3}F^{*\overline{c}}{}_{\overline{e}[\underline{a}}F^*_{\underline{b}}{}^{\overline{ed}}\right)F^*_{\underline{c}]\overline{cd}}\, ,  \label{efe3}\\
{\cal F}_{\overline a\underline {bc}}&=F_{\overline a\underline {bc}}+F^{(3)}_{\overline a\underline {bc}}\cong F_{\overline a\underline {bc}} -\frac{b}{4}\left(E_{\overline{a}}{F}^{*\overline{cd}}_{[\underline{b}} +{F}^{*\underline{e}\overline{cd}}{F}_{\overline{a}\underline{e}[\underline b}\right)F^{*}_{\underline{c}]\overline{cd}}\, ,  \label{efe3bar}  \\
{\cal F}_{\underline a \overline {bc}} &= {F}_{\underline a \overline {bc}} +{ F}^{(3)}_{\underline a \overline {bc}} \cong {F}_{\underline a \overline {bc}}+\frac{b}{8} F^{* }_{\underline d\overline {e f}} F^{*\overline {e f}}{}_{\underline a} F^{\underline d }{}_{\overline {bc} }\, ,\\
{\cal F}_{\overline {abc}} &= {F}_{\overline {abc}}\, , \\
{\cal F}_{\overline {ab}\underline {cd}} &\cong  { F}^{(2)}_{\overline {ab}\underline {cd}}=- 2 E_{[\underline c} {F}^*_{\underline d]\overline{ab}} +2 {F}^*_{\overline a}{}^{\overline{e}}{}_{[\underline c} {F}^*_{\underline d]\overline{eb} } +{F}_{\underline {cde}} F^{*\underline e}{}_{\overline{ab}} 
\, ,\\
{\cal F}_{\underline {a}\overline {bcd}} &\cong  { F}^{(2)}_{\underline {a}\overline {bcd}} =\frac1{\sqrt2}E_{\overline b}F^*_{\underline {a}\overline {cd}} -\frac1{\sqrt2}F_{\underline {ad}\overline {b}} F^{*\underline d}{}_{\overline{cd}}\, , \\
{\cal F}_{\underline {a}}&=F_{\underline {a}}+F^{(3)}_{\underline {a}}\cong F_{\underline {a}} +\frac b8 \left[{F}^{*\underline b}{}_{\overline{cd}} {F}_{\underline a}^{*\overline{cd}} F_{\underline b} + E_{\underline b}\left({F}^{*\underline b}{}_{\overline{cd}} {F}_{\underline a}^{*\overline{cd}}  \right) \right]\, ,\\
 {\cal F}_{\overline a}&=F_{\overline a}\, ,
\end{align}
\end{subequations}
where we used
\bea{F}^*_{\underline M\overline{bc}}\equiv P_M{}^N{F}^*_{N\overline{bc}}=\frac1{\sqrt2}E_M{}^{\underline a}F^*_{\underline a \overline {bc}}=\frac1{\sqrt2}E_M{}^{\underline a}\left(F_{\underline a \overline {bc}}-\frac12\overline\Psi_{\overline b}\gamma_{\underline a}\Psi_{\overline c}\right)\, ,\eea  $b=\frac2{(1-X_R)g^2}$, the superscripts $^{(2)}$ and $^{(3)}$ refer to the number of derivatives, and we defined 
\bea
{\cal F}_{\overline{\alpha}\underline{cd}}=\frac1{\sqrt2gX_R}{\cal F}_{\overline{\cal AB}\underline{cd}}(t_\alpha)^{\overline{\cal{AB}}}e^{\alpha}{}_{\overline\alpha}\, .
\eea

The transformation rules \eqref {tranfe} take the following form:

\noindent {\bf $-$  Vielbein}

The identification ${\cal E}_{M}{}^{\overline{a}}=E_{M}{}^{\overline{a}}$ implies $\delta {\cal E}_{M}{}^{\overline{a}} \ = \ \delta E_{M}{}^{\overline{a}}$, and from  (\ref{transfeviel}) we get
\be
\delta E_{M}{}^{\overline{a}} \ = \  \hat{\mathcal{L}}_{{\xi}} E_{M}{}^{\overline{a}} + E_{M}{}^{\overline{b}} {\mathscr T}_{\overline{b}}{}^{ \overline{a}} + {\cal E}_{M}{}^{\overline  \beta} {\mathscr T}_{\overline{\beta}}{}^{\overline{a}} + \frac12 \overline{\epsilon} \gamma^{\underline b} {\Psi}^{\overline{a}} {\cal E}_{M\underline b}\, .
\ee

Using the gauge fixing (\ref{gaugefixing1}) and the following relation
\bea
{\rm{A}}_{M}{}^{\beta} f(\Box)_{\beta}{}^{\alpha} = {\rm{A}}_{N}{}^{\alpha} f(\triangle)_{M}{}^{N}\, ,
\eea
which holds for any function $f$, one gets
\bea
\delta E_{M}{}^{\overline{a}}  = \hat{\mathcal{L}}_{{\xi}} E_{M}{}^{\overline{a}} + E_{M}{}^{\overline{b}} {\mathscr T}_{\overline{b}}{}^{\overline a}  - {\rm{A}}_{M}{}^{\beta} \partial^{P} {\xi}_{\alpha} E_{P}{}^{\overline{a}} (\Box^{-\frac12})^{\alpha}{}_{\beta} + \frac12 \overline{\epsilon} \gamma^{\underline b} {\Psi}^{\overline{a}} (\triangle^{-\frac12})_{M}{}^{N} E_{N\underline b}\, .  
\label{fullframetransf}
\eea
The second term in the r.h.s. of this expression allows to identify ${\mathscr T}_{\overline{ab}}$ with the  $\Gamma_{\overline {ab}}$ component of the Lorentz parameter  \eqref{lorpar}. The third term contains the  deformation 
\bea
\delta^{(1)}_{\Gamma} E_{M}{}^{\overline{a}}  = \frac{b}{2} \ {F}^*_{\underline M\overline{bc}}E_N{}^{\overline a}\partial^N  \Gamma^{\overline{bc}}\, ,
\label{firstLorentzframe}
\eea
which is the leading order of the  $\rO(10,10)$ covariant generalization of  the Green-Schwarz  transformation \cite{mn}. And finally, the last term in \eqref{fullframetransf} contains the first order correction to the supersymmetry transformation rule \eqref{gensusytransf0viel}, namely
\bea
\delta^{(1)}_{\epsilon} E_{M}{}^{\overline a} & =& -\frac{b}{8} \overline{\epsilon} \gamma^{\underline b} {\Psi}^{\overline{a}} { F}^*_{\underline{M} \overline{bc}} {F}^*_N{}^{\overline{bc}}E^N{}_{\underline b}\, .
\label{firstordersusyframe}
\eea

Following a similar reasoning,  one can see that the other projection transforms as 
\bea
\delta^{(1)}E_{M}{}^{\underline a}  &=& \frac{b}{2} \ {F}^{*N \overline{cd}} E_N{}^{\underline a}\big(- \partial_{\overline{M}} \Gamma_{\overline{cd}}  + \frac1{4\sqrt2} \overline{\epsilon} \gamma^{\underline b} \Psi_{\overline{b}} {F}_{ {\underline b}\overline{cd}} E_{M}{}^{\overline{b}} \big)\, ,
\label{firstorderframe2}
\eea
where we have identified 
\bea
{\mathscr T}_{\underline{ab}} = \Gamma_{\underline{ab}} + \frac{b}{4} {F}^{*}_{[\underline b}{}^{\overline{cd} } E_{\underline a]} \Gamma_{\overline{cd}} -\frac b4\ov\epsilon\gamma_{[\underline a}\Psi^{\ov{cd}}F^*_{\underline b]\ov{cd}}\, .
\label{Lambdaproyid}
\eea

\noindent {\bf $-$ Gravitino }

From \eqref{transfegravi} we get
 the first order corrections  to the transformation rules of the $\rO(10,10)$ gravitino \eqref{gentransf0gravi}, up to bilinear terms in fermions,
\be
\delta^{(1)} \Psi_{\overline{a}} \ = \frac{b}{16} \ E_{\underline b} \Gamma_{\overline{cd}} {F}_{\underline c}{}^{\overline{cd}} \gamma^{\underline{bc}} \Psi_{\overline{a}} + \frac b2 \ \Psi^{\overline{cd}} E_{\overline{a}} \Gamma_{\overline{cd}} +\frac{1}{4} {F}^{(3)}_{\overline{a} \underline{bc}} \gamma^{\underline{bc}} \epsilon ,
\label{firstordergravitino} 
\ee 
where we have kept  the leading order terms in the $\rO(10,10+k)$ gaugino identification 
\eqref{gauginoidentf2}.
Note that there are two corrections to the Lorentz transformations. The first term in the right hand side  can be interpreted as a generalized Green-Schwarz transformation  and the second one depends on the gravitino curvature, that we now define. 

\noindent {\bf $-$ Gravitino curvature}

  To leading order in \eqref{gauginoidentf2}, the induced $\rO(10,10)$ gravitino curvature is, 
\be
\Psi_{\overline{ab}}=\nabla_{[\overline a}\Psi_{\overline b]}+\frac12\omega_{\overline{cab}}\Psi^{\overline c}\, . \label{gc}
\ee
From  \eqref{transfegravi}, we find that 
 it obeys the transformation rule
\be
\delta{\Psi}_{\ov{ab}} = \xi^{M}\partial_{M}\Psi_{\ov{ab}} + 2\Psi_{\ov{c}[\ov{b}}\Gamma^{\ov{c}}{}_{\ov{a}]} + \frac{1}{4}\Gamma_{\underline{cd}}\gamma^{\underline{cd}}\Psi_{\ov{ab}} + \frac{1}{2}E_{\ov{c}}\Gamma_{\ov{ab}}\Psi^{\ov{c}} + \frac{1}{2}F^{*\underline{c}}{}_{\ov{ab}}E_{\underline{c}}\epsilon + \frac{1}{8}{\cal{F}}^{(2)}_{\underline{cd}\ov{ab}}\gamma^{\underline{cd}}\epsilon\, .
\label{Curvaturetransf}
\ee

\noindent {\bf $-$ Dilatino }

 The first order corrections to the  transformation rules  of the generalized dilatino \eqref{gentransf0rho} that are obtained from (\ref{transferho}) are
\be
\delta^{(1)} \rho \ = \frac{b}{16} E_{\underline b} \Gamma_{\overline{cd}} {F}^*_{\underline c}{}^{\overline{cd}} \gamma^{\underline {bc}} \rho - \frac{b}{8} \gamma^{\underline a} F^*_{\underline a\overline{bc}} F^{*\underline d\overline{bc}}  E_{\underline d}  \epsilon - \frac{1}{12} {F}^{(3)}_{\underline {abc}} \gamma^{\underline {abc}} \epsilon - \frac{1}{2} {F}^{(3)}_{\underline c} \gamma^{\underline c} \epsilon \, .
\label{dilatinofirstorder}
\ee

Note that the transformation rules of the dilaton \eqref{0transf} as well as the diffeomorphisms on all the fields  are not corrected.

\subsection{Including the heterotic gauge sector}

It is now trivial to include the gauge sector of the $\rO(10,10+n_g)$ formulation. We simply extend the duality group $\rO(10,10) \rightarrow \rO(10,10+n_g)$, the right Lorentz group $\rO(1,9)_\rR \rightarrow \rO(1,9 + n_g)_\rR$ and the indices $M\rightarrow M = (M,i), \bar a \rightarrow \bar A = (\bar a,\bar i)$, accordingly.  Now the generalized fluxes  and  gravitino curvature  contain the contributions of the gauge sector, and  in particular the structure constants.

A straightforward extension of the indices in equations \eqref{firstordersusyframe} - \eqref{dilatinofirstorder} gives the following transformation rules of the $\rO(10,10+n_g)$ generalized fields, up to first order,
\begin{subequations}\label{transf1}
\begin{align}
\delta E_{\mathbb M}{}^{\underline a}  &=  \hat{\cal L}_\xi E_{\mathbb M}{}^{\underline a} + E_{{\mathbb M}\underline  b} \Gamma^{\underline {ba}} - \frac12 \overline{\epsilon} \gamma^{\underline  a} \Psi^{\overline{B}} E_{{\mathbb M} \overline{B}}  \nn\\
& \ \ -\frac{b}{2}E_N{}^{\underline a} {F}^{*N}{}_{\overline{CD}}\left( \partial_{\overline{\mathbb M}} \Gamma^{\overline{CD}}  - \frac{1}{4\sqrt2} \overline{\epsilon} \gamma^{\underline b} \Psi_{\overline{B}} {F}_{\underline b}{}^{\overline{CD}}  E_{\mathbb M}{}^{\overline{B}} \right)\, , \label{transf1viel1}\\
\delta E_{\mathbb M}{}^{\overline{A}}& =   \hat{\cal L}_\xi E_{\mathbb M}{}^{\overline{A}} + E_{{\mathbb M} \overline{B}} \Gamma^{\overline{B A}} + \frac12 \overline{\epsilon} \gamma^{\underline b} \Psi^{\overline{A}} E_{{\mathbb M} \underline b}   \nn\\
&    \ \ + \frac{b}{2} \ {F}^{*}_{\underline{\mathbb M}}{}^{ \overline{CD}}\left(E_N{}^{\overline{A}} \partial^N\Gamma_{\overline{CD}} - \frac{1}{4} \overline{\epsilon} \gamma^{\underline b} \Psi^{\overline{A}}  {F}_{N \overline{CD}} E^N{}_{\underline b}\right)\, , \label{transf1viel2}\\
\delta d  &=  \xi^M\partial_{M} d-\frac12\partial_{M}\xi^{M} - \frac14 \overline{\epsilon} \rho \, , \\
\delta \Psi_{\overline{A}} &=   \hat{\cal L}_\xi \Psi_{\overline{A}} + \Psi_{\overline{B}} \Gamma^{\overline{B}}{}_{\overline{A}} + \frac{1}{4} \Gamma_{\underline {bc}} \gamma^{\underline {bc}} \Psi_{\overline{A}} + \nabla_{\overline{A}} \epsilon  \nn\\
& \ \ + \frac{b}{16} E_{\underline b} \Gamma_{\overline{CD}} {F}_{\underline c}{}^{\overline{CD}} \gamma^{\underline {bc}} \Psi_{\overline{A}}+\frac b2 \ \Psi^{\overline{DC}} E_{\overline{A}} \Gamma_{\overline{CD}} + \frac{1}{4} { F}^{(3)}_{\overline{A} \underline {bc}} \gamma^{\underline {bc} } \epsilon \, ,\\
\delta \rho & =   \hat{\cal L}_\xi \rho + \frac{1}{4} \Gamma_{\underline {bc}} \gamma^{\underline {bc}} \rho - \gamma^{\underline a} \nabla_{\underline a}\epsilon + \frac{b}{16} \ E_{\underline b}\Gamma_{\overline{CD}} {F}_{\underline c}{}^{\overline{CD}} \gamma^{\underline {bc}} \rho \nn \\ & \ \  - \frac{b}{8} \gamma^{\underline a} F_{\underline a\overline{BC}} F^{\underline d\overline{BC}}  E_{\underline d}  \epsilon - \frac{1}{12} {F}^{(3)}_{\underline {abc}} \gamma^{\underline {abc}} \epsilon- \frac{1}{2} {F}_{\underline a}^{(3)}  \gamma^{\underline a}\epsilon \, .
\end{align}
\end{subequations}

In Appendix \ref{App first} we show that the  algebra of these transformation rules  closes, up to terms with two fermions, with the following  field-dependent parameters 
\begin{subequations}\label{param1}
\begin{align}
\xi_{12}^{\mathbb M} &= [\xi_1,\xi_2]^{\mathbb M}_{C_f} - \frac{1}{\sqrt2} E^{\mathbb M}{}_{\underline a} \overline{\epsilon_1} \gamma^{\underline a} \epsilon_{2}
+b \Gamma_{[1}^{\overline{CD}} \partial^{\mathbb M} \Gamma_{2] \overline{CD}} + \frac{b}{8} {F}^{\mathbb M}{}_{\overline{CD}} {F}^*_{\underline b}{}^{\overline{CD}} \overline{\epsilon}_1 \gamma^{\underline b} \epsilon_2\, , \\
\Gamma_{12 \overline{AB}} &= 2 \xi_{[1}^P \partial_P \Gamma_{2] \overline {AB}} - 2 \Gamma_{[1 \overline A}{}^{\overline C} \Gamma_{2] \overline {CB}}+ \frac b 2 E_{\overline{B}} \Gamma_{[1}^{\overline{CD}} E_{\overline{A}} \Gamma_{2] \overline{CD}}\nn\\
& \ \ \ 
+ b \overline{\epsilon}_{[1} \gamma^{\underline b} \Psi_{[\overline{A}} E^{M}{}_{\overline{B}]} \partial_{M} \Gamma_{2]}^{\overline{CD}} {F}^*_{\underline b \overline{CD}}   + \frac{b}{4} \overline{\epsilon}_1 \gamma^{\underline b} \epsilon_2 {F}^*_{\underline b \overline{AB}}\, , \label{param}\\
\Gamma_{12\underline { ab}} &= 2 \xi_{[1}^P \partial_P \Gamma_{2] \underline {ab}} - 2 \Gamma_{[1 \underline a}{}^{\underline c} \Gamma_{2] \underline {cb}}+ \frac b 2 E_{\underline b} \Gamma_{[1}^{\overline{CD}} E_{\underline a} \Gamma_{2] \overline{CD}} \nn\\& \ \ \ + b \overline{\epsilon}_{[1} \gamma_{[\underline a} \Psi^{\overline{B}} {F}^*_{\underline b]}{}^{\overline{CD}} E^{M}{}_{\overline{B}} \partial_{M} \Gamma_{2] \overline{CD}} \, , \\
\epsilon_{12} &= \  -\frac{1}{2} \Gamma_{[1 \underline {bc}} \gamma^{\underline {bc}} \epsilon_{2]} + 2 \xi_{[1}^P \partial_{P} \epsilon_{2]}-\frac{b}{4} \gamma^{\underline {bc}} \epsilon_{[1} E^{M}{}_{\underline b} \partial_{M} \Gamma_{2] \overline{CD}} {F}^*_{\underline c}{}^{\overline{CD}}  \,  .
\end{align}
\end{subequations}

\subsection{First order corrections to ${\cal N}=1$ supersymmetric DFT}\label{sec:dftaction}
The  invariant action under the transformation rules \eqref{tranfe} is clearly of the same functional form as \eqref{DFTsusyAction} but it depends on the $\rO(10,10+k)$ multiplets, namely 
\begin{equation}
S_{\mathcal{N}=1\ {\rm DFT}} \ = \ \int d^{20+k} X\,e^{-2d}\left(\mathcal{R}(\mathcal{E},d)
  + \overline{\Psi}^{\overline{\mathcal{A}}}\gamma^{\underline{b}}\nabla_{\underline{b}}\Psi_{\overline{\mathcal{A}}}
 -\overline{\rho}\gamma^{\underline{a}}\nabla_{\underline{a}}\rho+2\overline{\Psi}^{\overline{\mathcal{A}}}\nabla_{\overline{\mathcal{A}}}\rho \right)\ . \label{Action}
\end{equation}
Hence  it  contains higher derivatives of the $\rO(10,10+n_g)$ multiplets. 

The transformation rules (\ref{tranfe})  define the following Lichnerowicz principle,
\bea
\left(\gamma^{\underline{ \cal A}}\nabla_{ \underline{\cal A}}\gamma^{ \underline{\cal B}}\nabla_{\underline{ \cal B}}-\nabla^{\overline{\cal A}}\nabla_{\overline{\cal A}}
\right)\epsilon&=&-\frac14{\cal R}\epsilon\, ,\label{lich1}\\
\left[\nabla_{\overline{\cal A}}, \gamma^{ \underline{\cal B}}\nabla_{\underline{ \cal B}}\right]\epsilon&=&\frac12\gamma^{\underline {\cal B}}{\cal R}_{{\overline{\cal A}} \underline{\cal B}}\epsilon\, ,\label{lich2}
\eea
and then  the $\rO(10,10+k)$ generalized Ricci scalar  
\bea
{\mathcal R}&=& 2{\cal E}_{\underline{\cal A}}{\cal F}^{\underline{\cal A}}+{\cal F}_{\underline{\cal A}}{\cal F}^{\underline{\cal A}}-\frac16{\cal  F}_{\underline{\cal A} \underline{\cal B} \underline{\cal C}}{ F}^{\underline{\cal A} \underline{\cal B} \underline{\cal C}}-\frac12{ \cal F}_{\overline {\cal A} \underline{\cal B} \underline{\cal C}}{\cal  F}^{\overline {\cal A} \underline{\cal B} \underline{\cal C}}\, \label{calr}
\eea
determines the corrections to the generalized Dirac operator. 

In terms of the $\rO(10,10+n_g)$ generalized fluxes, the $\rO(10,10+k)$ generalized Ricci scalar is, up to first order, 
\bea
{\cal  {R}} \ = \ {\mathbb R}+b{\mathbb R}^{(1)}&=&{\mathbb R} -{F}^{(3)}_{\overline{A}\underline{bc}} {F}^{\overline{A}\underline{bc}} - \frac13 {F}^{(3)}_{\underline{abc}}{F}^{\underline{abc}}+2{F}^{(3)}_{\underline d}{F}^{\underline d} +2E_{\underline a}F^{(3)\underline a}\nn\\
&&+\frac b4{E}_{\underline d}F^{\underline a} {F}^{*\underline d}{}_{\overline{BC}} {F}^*_{\underline a}{}^{\overline{BC}}  + \frac b8 {F}^{(2)\overline{AB}}{}_{\underline{cd}} { F}^{(2)}_{\overline{AB}}{}^{\underline{cd}} \,  ,
\label{firstorderaction}
\eea
where ${\mathbb R}$ was defined in \eqref{r}. Replacing the expressions \eqref{genfluxes} with the overlined indices extended to include the gauge sector (i.e. $\ov c, \ov d,...\rightarrow \ov C, \ov D,...$), ${\mathbb R}^{(1)}$ may be written  as
\bea
\label{R_first_order_4}
{\mathbb R}^{(1)}& = & \frac{1}{4}\left[(E_{\underline{a}}E_{\underline b}{F}^{*\underline b}{}_{\overline{CD}}) {F}^{*\underline{a}\overline{CD}} + (E_{\underline{a}}E_{\underline b}{F}^{*\underline a}{}_{\overline{CD}}) {F}^{*\underline{b}\overline{CD}}+ 2(E_{\underline a}{F}_{\underline b}^{*\overline{CD}}){F}^{*\underline a}{}_{\overline{CD}}{F}^{\underline{b}}\right.\, \nn \\
& & + (E_{\underline{a}}{F}^{*\underline{a}\overline{CD}})(E_{\underline b}{F}^{*\underline b}{}_{\overline{CD}}) + (E_{\underline a} {F}_{\underline b}^{*\overline{CD}})(E^{\underline a} {F}^{*\underline b}{}_{\overline{CD}})+ 2(E_{\underline{a}}F_{\underline b}){F}^{*\underline b}{}_{\overline{CD}}{F}^{*\underline{a}\overline{CD}}\, \nn \\ 
& & + (E_{\overline{A}}F^*_{\underline{b}\overline{CD}})F_{\underline{c}}^{*\overline{CD}}F^{*\overline{A}\underline{bc}} - (E_{\underline{a}}F^*_{\underline{b}\overline{CD}})F_{\underline{c}}^{*\overline{CD}}{F}^{\underline{abc}} + 2(E_{\underline a}{F}^{*\underline a}{}_{\overline{CD}}){F}_{\underline b}^{*\overline{CD}}{F}^{\underline{b}}\, \nn \\
& &   - 4(E_{\underline a} {F}_{\underline b}^{*\overline{CD}}){F}^{*\underline a}{}_{\overline{CE}} {F}^{*\underline b\overline{E}}{}_{\overline{D}}+ \frac{4}{3}{F}^{*\overline{E}}{}_{{\underline a}\overline{C} }{F}^*_{{\underline b}\overline{ED}}F_{\underline{c}}^{*\overline{CD}}{F}^{\underline{abc}} + {F}^{*\underline b}{}_{\overline{CD}} {F}_{\underline a}^{*\overline{CD}} F_{\underline b}{F}^{\underline{a}} \, \nn \\
& &  \left.+ {F}_{\underline{a}}^{*\overline{CE}}{F}^{*}_{\underline b\overline{E}\overline{D}}{F}^{*\underline a}{}_{\overline{CG}} {F}^{*\underline b\overline{G}\overline{D}}
 - {F}_{\underline{b}}^{*\overline{CE}}{F}^{*}_{\underline a\overline{E}\overline{D}}{F}^{*\underline a}{}_{\overline{CG}} {F}^{*\underline b\overline{G}\overline{D}} - F_{\overline{A}\underline{bd}}F^{*\underline{d}}{}_{\overline{CD}}F_{\underline{c}}^{*\overline{CD}}F^{\overline{A}\underline{bc}}\right]\, .\nn\\
\eea
Note that it depends on the generalized gravitino through ${F}^*_{\underline a\overline{BC}}$. 

Similarly, we may define
\be
\overline{\Psi}^{\overline{\mathcal{A}}}\gamma^{\underline{b}}\nabla_{\underline{b}}\Psi_{\overline{\mathcal{A}}}
 -\overline{\rho}\gamma^{\underline{a}}\nabla_{\underline{a}}\rho+2\overline{\Psi}^{\overline{\mathcal{A}}}\nabla_{\overline{\mathcal{A}}}\rho\equiv {\mathbb L}_{\rF}+{\mathbb L}_{\rF}^{(1)}\, , \label{ferdft}
\ee
where ${\mathbb L}_{\rF}$ was introduced in \eqref{lf} and the first order corrections  are given by
\bea
\label{Fermionic_2}
{\mathbb L}^{(1)}_{\rF} & = & \frac{1}{2}\left[\frac{1}{4}{\overline{\Psi}}^{\overline{A}}\gamma^{\underline{b}}E_{\underline{c}}\Psi_{\overline{A}}F^{\underline{c}{\overline{CD}}}F_{\underline{b}{\overline{CD}}} - \frac{1}{8}\overline{\Psi}^{\overline{A}}\gamma^{\underline{bcd}}\Psi_{\overline{A}}(E_{\underline{b}}{F}_{\underline{c}\overline{CD}})F_{\underline{d}}{}^{\overline{CD}}\right.\, \nn \\ 
& & + \frac{1}{16}\overline{\Psi}^{\overline{A}}\gamma^{\underline{bcd}}\Psi_{\overline{A}}F_{\underline{abc}}F^{\underline{a}}{}_{\overline{CD}}F_{\underline{d}}{}^{\overline{CD}} + \frac{1}{12}\overline{\Psi}^{\overline{A}}\gamma^{\underline{bcd}}\Psi_{\overline{A}}F_{\underline{b}\overline{C}}{}^{\overline{E}}F_{\underline{c}\overline{ED}}F_{\underline{d}}{}^{\overline{CD}}\, \nn \\ 
& & - \frac{1}{4}\bar{\Psi}^{\overline{A}}\gamma^{\underline{b}}\Psi^{\overline{C}}F_{\underline{b}}{}^{\overline{EF}}F_{\underline{d}\overline{AC}}F^{\underline{d}}{}_{\overline{EF}} + 2\overline{\Psi}^{\overline{A}}\gamma^{\underline{b}}\Psi_{\overline{CD}}(E_{\overline{A}}F_{\underline{b}}{}^{\overline{CD}}) - 2\overline{\Psi}^{\overline{A}}\gamma^{\underline{b}}\Psi_{\overline{CD}}F_{\overline{A}\underline{bc}}F^{\underline{c}\overline{CD}}\, \nn \\
& & - 2\overline{\Psi}^{\overline{CD}}\gamma^{\underline{b}}E_{\underline{b}}\Psi_{\overline{CD}} - \frac{1}{6}\overline{\Psi}^{\overline{CD}}\gamma^{\underline{bcd}}\Psi_{\overline{CD}}F_{\underline{bcd}} - 4\overline{\Psi}_{\overline{CE}}\gamma^{\underline{b}}\Psi^{\overline{E}}{}_{\overline{D}}F_{\underline{b}}{}^{\overline{CD}} - \frac{1}{4}\overline{\rho}\gamma^{\underline{a}}E_{\underline{b}}\rho F^{\underline{b}\overline{CD}}F_{\underline{a}\overline{CD}}\, \nn \\ 
& & + \frac{1}{8}\overline{\rho}\gamma^{\underline{abc}}\rho E_{\underline{a}}{F}_{\underline{b}\overline{CD}}F_{\underline{c}}{}^{\overline{CD}} - \frac{1}{16}\overline{\rho}\gamma^{\underline{abc}}\rho F_{\underline{abd}}F^{\underline{d}}{}_{\overline{CD}}F_{\underline{c}}{}^{\overline{CD}} - \frac{1}{12}\overline{\rho}\gamma^{\underline{abc}}\rho F_{\underline{a}\overline{C}}{}^{\overline{E}}F_{\underline{b}\overline{ED}}F_{\underline{c}}{}^{\overline{CD}}\, \nn \\
& & - \frac{1}{4}\overline{\Psi}^{\overline{A}}\gamma^{\underline{bc}}\rho(E_{\overline{A}}{F}_{\underline{b}}{}^{\overline{CD}})F_{\underline c\overline{CD}} + \frac{1}{4}\overline{\Psi}^{\overline{A}}\gamma^{\underline{bc}}\rho{F}_{\overline{A}\underline{bd}}{F}^{\underline{d}}{}_{\overline{CD}}F_{\underline{c}}{}^{\overline{CD}} - 2\overline{\Psi}^{\overline{CD}}F^{\underline{a}}{}_{\overline{CD}}E_{\underline{a}}\rho\, \nn \\ 
& & \left. + \overline{{\Psi}}^{\overline{CD}}\gamma^{\underline{ab}}\rho(E_{\underline{a}}F_{\underline{b}\overline{CD}}) - \overline{{\Psi}}^{\overline{CD}}\gamma^{\underline{ab}}\rho{F}_{\underline{a}\overline C}{}^{\overline E}{F}_{\underline{b}\overline{ED}} - \frac{1}{2}\overline{{\Psi}}^{\overline{CD}}\gamma^{\underline{ab}}\rho{F}_{\underline{abc}} F^{\underline{c}}{}_{\overline{CD}}\right]\, .
\eea

In conclusion, the manifestly duality covariant  first order corrections to the action of ${\cal N}=1$ supersymmetric DFT \eqref{DFTsusyAction}  in terms of  $\rO(10,10+n_g)$ multiplets are given by the addition of  ${\mathbb R}^{(1)}$ and ${\mathbb L}^{(1)}_{\rF} $, up to  bilinear terms in fermions. 
We have explicitly verified that the action
\be
S_{{\cal N}=1\ {\rm DFT}}=\int d^{20+n_g}Xe^{-2d}\left({\mathbb R}+{\mathbb R}^{(1)}+{\mathbb L}_F+{\mathbb L}^{(1)}_F\right)\, , \label{n=1}
\ee
 is invariant under the transformation rules \eqref{transf1}, up to terms with four derivatives and two fermions. The structure constants preserve a global $\rO(10,10;\mathbb R)$ symmetry.

\section{Transformation rules of the supergravity fields}\label{sec:transf}

To make contact with the heterotic string low energy effective field theory, in this section we parameterize the $\rO(10,10+n_g)$ duality multiplets in terms of  supergravity and super Yang-Mills multiplets, we analyze the deformations of the symmetry transformation rules and compare with previous proposals in the literature.

 The deformed transformation rules of the duality multiplets \eqref{transf1}  induce higher derivative corrections on the
transformation rules of the  supergravity and super Yang-Mills fields that parameterize the generalized fields \eqref{HKparam}, \eqref{gendil} and \eqref{gengrav}. We then expect an $\alpha'$-expansion of the parameterizations, that we now denote $\widetilde e_\mu{}^a,  \widetilde b_{\mu\nu},  \widetilde\phi, \widetilde A_\mu^i, \widetilde\psi_{\mu}, \widetilde\lambda, \widetilde\chi_i$, in terms of the gauge and Lorentz covariant  fields, e.g.
$ \widetilde  e_\mu{}^a \ =\ e_\mu{}^a+{\cal O}(\alpha')\ ,   \widetilde b_{\mu\nu}\ = \ b_{\mu\nu}+{\cal O}(\alpha')\, ,  
\widetilde\psi_{\mu}=\psi_{\mu}+{\cal O}(\alpha')\, ,$ etc.

To find the relations between both sets of fields, it is convenient to first work out the parameterizations  of the generalized fluxes and curvatures and their transformation rules. 
From the first order terms in the action \eqref{n=1}, we see that only the leading order expressions are necessary.   We denote  the parameterization of  $F^{*}_{\underline a \overline{CD}}$ as
\be
\hat\Omega_{aCD}=\left(\hat w^{(-)}_{acd}, \hat F_{ac}^i, \hat A_{a}{}^{ij}\right)\, , \label{gencon}
\ee
where the hats  distinguish objects that contain fermions and the collective indices of the tangent space $C=(c,i)$ include the gauge indices. In terms of  supergravity  and super Yang-Mills fields, the components are
\be\label{paramf*}
\hat {w}^{(-)}_{a{bc}}\equiv \left( w_{\mu bc}^{(-)}-\frac12 \overline{\psi}_b\gamma_\mu \psi_{{c}}\right)e^\mu{}_a\, ,
\ee
with $w_{abc}^{(\pm)}=w_{abc}\pm\frac12H_{abc}$,
\be
\hat{ F}_{a b}{}^i\equiv-\frac1{\sqrt2}\left({ F}^{ i}_{\mu\nu }-\frac12\overline{\psi}_{[\mu}\gamma_{\nu]} \chi^i\right) e^\mu{}_a  e^\nu{}_b\, , \label{hatf}
\ee
and
\be
\hat {A}_a{}^{ {ij}}\equiv-\left( A_\mu^k \ f_k{}^{ij}+\frac14\overline{ \chi}^i\gamma_\mu\chi^j\right) e^\mu{}_a\, .
\ee

The  generalized gravitino curvature $\Psi_{\ov{AB}}$ is parameterized as
\be
\widetilde\Psi_{AB}=\Psi_{AB}-\frac12\hat\Omega_{cAB}\psi^c\equiv \psi_{AB} -\frac1{2\sqrt2}\hat\Omega_{iAB}\chi^i-\frac12\hat\Omega_{cAB}\psi^c\, ,\label{gengravcur}
\ee
with
\begin{subequations}\label{gravcur}
\begin{align}
\psi_{ab}&\equiv   e^{\mu}{}_{[a}  e^{\nu}{}_{b]} D_\mu^{(+)}\psi_\nu\, ,\\
\psi_{ai}&=\frac1{2\sqrt2}\left(\partial_c\chi_i-\frac14\hat w_{abc}^{(+)}\gamma^{bc}\chi_i-\frac1{2\sqrt2}\hat F_{bci}\gamma^{bc}\psi_a\right)\, ,\\
\psi_{ij}&=\frac1{4\sqrt2}\hat F_{bc[i}\gamma^{bc}\chi_{j]}\, ,
\end{align}
\end{subequations}
and
\be
\hat\Omega_{iAB}=\left(\hat F_{abi},\hat A_{aij},\sqrt2f_{ijk}\right)\, 
\ee
is the parameterization of the generalized flux component $F_{\ov{ABi}}$.

Parameterizing the Lorentz and supersymmetry transformation rules of $ F^*_{\underline a\overline{BC}}$, namely
\bea
\delta F^*_{\underline a\overline{BC}}=-E_{\underline a}\Gamma_{\overline{BC}}+\Gamma^{\underline b}{}_{\underline a}F^*_{\underline b\overline{BC}}-2\Gamma^{\overline D}{}_{[\overline B}F^*_{\overline{C}]\ov D\underline a}+\overline\epsilon\gamma_{\underline a}\Psi_{\overline{BC}}\, ,
\eea
 we get 
\be
\delta\hat\Omega_{\mu CD}=-\partial_\mu\Lambda_{CD}+2\hat\Omega_{\mu B[D}\Lambda^B{}_{C]}+\bar\epsilon\gamma_\mu\Psi_{CD}\, , \label{transfOmega}
\ee
 where the generalized Lorentz parameters $\Gamma_{\underline{ab}}$    and $\Gamma_{\ov{AB}}$ are  parameterized as $-\wt{\Lambda}_{ab}+\ov\epsilon \gamma_{[a}\wt\psi_{b]}$ and \mbox{$\wt{\Lambda}_{AB}=(\wt{\Lambda}_{ab} , \wt{\Lambda}_{ai}, \wt{\Lambda}_{ij})$},  with $\wt{\Lambda}_{AB}=\Lambda_{AB}+{\cal O}(\ap)$, and $\Lambda_{ab}$ is the generator of $\rO(1,9)$ transformations, while $\Lambda_{ai}=\frac1{2\sqrt2}\ov\epsilon\gamma_a\chi_i$ and $ \Lambda_{ij}=f_{ijk}\xi^k$ depend on the supersymmetry and gauge parameters according to \eqref{lorgau}.

The transformation rule \eqref{transfOmega} contains, other than the standard Lorentz transformations,
the supersymmetry variation of the torsionful spin connection \cite{bdr1,Bergshoeff:1988nn}
\be
\delta_{\epsilon}\hat{w}^{(-)}_{\mu{bc}} = \overline{\epsilon} \gamma_\mu {\psi}_{{bc}}  + \frac3{4}\overline{\epsilon}  \gamma_{[\rho} \chi_{i} \hat F^i_{\mu\nu]} e^\nu{}_b e^\rho{}_c\, , \label{transfw-} 
\ee
the supersymmetry and gauge transformations of the Yang-Mills field strength,
\be
 \delta_\epsilon\hat F_{\mu ci}=\frac12\left[D_\mu\left(\bar\epsilon\gamma_c\chi_i\right)-\bar\epsilon\gamma_\mu D_c\chi_i+\frac14\bar\epsilon\gamma_\mu\left(\frac12 H_{c\nu\rho}\gamma^{\nu\rho}\chi_i-\hat F_{\nu\rho i}\gamma^{\nu\rho}\psi_c\right)\right]
\ee
and $\delta_\xi \hat F_{\mu ci}=f_{ijk}\xi^j\hat F_{\mu c}{}^k$, as well as the leading order gauge and supersymmetry transformations of the Yang-Mills connection, \eqref{transf0chi} and \eqref{gauge0} respectively.

Similarly, from the transformation rule of the generalized gravitino curvature
\be
\delta{\Psi}_{\ov{AB}} =  2\Psi_{\ov{C}[\ov{B}}\Gamma^{\ov{C}}{}_{\ov{A}]} + \frac{1}{4}\Gamma_{\underline{cd}}\gamma^{\underline{cd}}\Psi_{\ov{AB}} + \frac{1}{2}E_{\ov{C}}\Gamma_{\ov{AB}}\Psi^{\ov{C}} + \frac{1}{2}F^{*\underline{c}}{}_{\ov{AB}}E_{\underline{c}}\epsilon + \frac{1}{8}{\cal{F}}^{(2)}_{\underline{cd}\ov{AB}}\gamma^{\underline{cd}}\epsilon\, 
\label{gencurvaturetransf}
\ee
we obtain 
\be
\delta\Psi_{CD}=2\Psi_{B[D}\Lambda^B{}_{C]}+\frac18\hat {\mathscr R}_{\mu\nu CD}\gamma^{\mu\nu}\epsilon\, , \label{transfcurgravg}
\ee
where we have defined 
\be
\hat{\mathscr R}_{\mu\nu CD}  =  -2\partial_{[\mu}\hat{\Omega}_{\nu]CD} + 2\hat{\Omega}_{[\mu|C|}{}^{E}\hat{\Omega}_{\nu]ED}\, , \label{gcur}\ee
which has components
\bea
\hat {\mathscr R}_{\mu\nu cd} &=& \hat R^{(-)}_{\mu\nu cd} - \hat{F}_{\mu\tau}{}^{i}\hat{F}_{\nu\lambda i}e^{\tau}{}_{[c}e^{\lambda}{}_{d]}\, ,\\
\hat {\mathscr R}_{\mu\nu c}{}^{i} &=& \sqrt{2}\left(D^{(-)}_{[\mu}\hat{F}_{\nu]c}^i + \frac{1}{4}\ov{\chi}^{i}\gamma_{[\mu}\chi^{j}\hat{F}_{\nu]cj}\right)\, , \\
\hat {\mathscr R}_{\mu\nu}{}^{ ij} &=&  F_{\mu\nu}^{k}f^{ij}{}_{k} + \hat{F}^{i\lambda}{}_{[\mu}{}\hat{F}^j_{\nu]\lambda } + \frac{1}{2}D_{[\mu}\left(\ov{\chi}^{i}\gamma_{\nu]}\chi^{j}\right)\, .
\eea
In particular, \eqref{transfcurgravg} contains the supersymmetry transformation rule of the supergravity gravitino curvature
\bea
\delta_{\epsilon} {\psi}_{{ab}} &=&\frac{1}{8} \left( \hat R^{(-)}_{\mu\nu{ab}}+\frac32\hat{T}_{\mu\nu ab}\right) \gamma^{\mu\nu} \epsilon\, ,  \ \        \ \ \ \ \
\label{transfcurgrav}
\eea
where $\hat R^{(-)}_{\mu\nu{ab}}$ is the  two-form curvature computed from the torsionful  spin connection  $\hat{ w}^{(-)}_{\mu ab}$ and $\hat T_{\mu\nu a b}=\hat{F}^i_{[\mu\nu}\hat{ F}_{ab]i}$, in agreement with \cite{bdr1,Bergshoeff:1988nn}.

Now we turn to the parameterization of  the   elementary  fields. We start from the deformed transformation rules of the components $E_M{}^{\ov a}$ and $E_M{}^{\underline a}$  given in \eqref{transf1viel1}  and \eqref{transf1viel2}. 
Of course,  different definitions lead to supergravity multiplets that obey different transformation rules. An interesting one is the following
\bea
  \widetilde e_{\mu}{}^a&=&{ e}_{\mu}{}^ {a} - \frac{b}{8} \left(\hat{w}^{(-)}_{b{cd}} \hat{w}^{(-)a {cd}} + {2} 
\hat{T}_{b}{}^{a }  +  \hat{A}^{}_{b{ij}} \hat{A}^{a{ij}}\right) {e}_{\mu}{}^{b} \, , \label{vieltransf1}\\
\tilde\phi&=&\phi-\frac b{16}\left(\hat{w}^{(-)acd}\hat{w}^{(-)}_{a {cd}} + {2} \hat{T}+  \hat{A}^{a{ij}} \hat{A}^{}_{a{ij}}\right) 
 \, , \label{redefdil}
\eea
where  $\hat T_{ab}=\hat F_{aci}\hat F_b{}^{ci}$ and $ \hat T=\hat F_{ac}^i\hat F^{ac}_i$. The quadratic  terms in spin and gauge connections are known to be necessary in order to remove  the non-standard Lorentz transformations of the supergravity vielbein $e_\mu{}^a$
 and dilaton $\phi$  fields \cite{mn, bfmn}. Together with the gauge covariant $\hat T$ terms, these parameterizations
determine $e_\mu{}^a$ and  $\phi$  fields  that obey the  leading order  supersymmetry and Lorentz transformation rules \eqref{transf0dil} and \eqref{lor0}. To get this result, the gauge fixings $\widetilde{ e}^{\mu}{}_{\overline a}= \widetilde{e}^{\mu}{}_{\underline a}\equiv\widetilde{e}^{\mu}{}_{{a}}$, $\delta E^{i}{}_{\overline{i}}=0$ and $\delta E^{\mu}{}_{\overline{i}}=0$ are used to absorb several terms into the Lorentz parameters. As a consequence, the following parameterization is needed for the duality covariant gravitino
\bea
 \widetilde\psi_{{a}} &=&{\psi}_{{a}} -\frac b2 \hat{\Omega}_{{aCD}} {\Psi}^{{CD}}  + \frac{b}{8} \hat{\Omega}_{a}{}^{{CD}} \hat{\Omega}_{b{CD}} 
{\psi}^{b} \, . \ \ \ \ \ \ \
\label{corr4}
\eea

Interestingly, these parameterizations  induce  a deformation of the  gravitino supersymmetry variation \eqref{transf0dilino} that can be absorbed into  the torsion of the spin connection through the following 
 modification of the two-form curvature
\bea
\wt{H}_{\mu\nu\rho} & = & 3\left[\partial_{[\mu}\wt{b}_{\nu\rho]} -\zeta C_{\mu\nu\rho}^{(g)} + \frac{b}{2}\hat C_{\mu\nu\rho}^{(L)} + \frac{b}{2}\hat{F}_{[\mu}{}^{ci}D^{(-)}_{\nu}\hat{F}_{\rho]ci} +\frac{b}{8}A_{[\mu}^k\partial_\nu\left(\ov{\chi}^{i}\gamma_{\rho]}\chi^{j}\right)f_{ijk}\right.\, \nn \\ 
& & \left. \ \ \
+ \frac{b}{8}\ov{\chi}^{i}\gamma_{[\mu}\chi^{j}\left(\partial_{\nu} A_{\rho]}{}^{k}- A^l_{\nu} A_{\rho]}{}^{m}f^k{}_{lm}\right)f_{ijk} -\frac{b}{8}\ov{\chi}^{i}\gamma_{[\mu}\chi^{j}\hat F_{\nu}{}^{c i}\hat F^j_{\rho]c}\right]\, .
\label{tildeh}
\eea
The Yang-Mills Chern-Simons form $C_{\mu\nu\rho}^{(g)}$ was  defined in \eqref{CSg}, the coefficient
\be
\zeta = 1 + \frac{1}{2}b\varrho\, ,\qquad \varrho\kappa_{ij} = f_{i}{}^{kl}f_{jlk}\, , \label{kappadeform}
\ee
and $\hat C_{\mu\nu\rho}^{(L)}$ denotes  the Lorentz Chern-Simons form of the torsionful spin connection $\hat w_{\mu ab}^{(-)}$,
\be
\hat C_{\mu\nu\rho}^{(L)} = \hat{w}^{(-)}_{[\mu}{}^{cd}\partial_{\nu}\hat{w}^{(-)}_{\rho]cd} + \frac{2}{3}\hat{w}^{(-)}_{[\mu}{}^{bc}\hat{w}^{(-)}_{\nu cd}\hat{w}^{(-)}_{\rho]}{}^{d}{}_{b} \, .\label{CSL}
\ee
The gaugino bilinear terms in \eqref{tildeh} may be absorbed into the first order deformation of the Yang-Mills Chern-Simons form replacing $A_\mu^i\rightarrow\hat A_\mu{}^{jk}$, but this is not convenient for reasons that will become clear shortly.

The modified three-form $\wt H_{\mu\nu\rho}$ \eqref{tildeh} may be rewritten as the compact expression
\bea
\wt{H}_{\mu\nu\rho} & = & 3\left[\partial_{[\mu}\wt{b}_{\nu\rho]} - C_{\mu\nu\rho}^{(g)} + \frac{b}{2}\hat{\mathscr C}_{\mu\nu\rho}\right]\, ,
\label{capital_tildeh}
\eea
where
\bea
\hat{\mathscr C}_{\mu\nu\rho} & = & \partial_{[\mu}\hat{\Omega}_{\nu}{}^{CD}\hat{\Omega}_{\rho]CD} + \frac{2}{3}\hat{\Omega}_{[\mu|CD|}\hat{\Omega}_{\nu}{}^{DE}\hat{\Omega}_{\rho]E}{}^{C}\, .
\label{capital_ChernSimons}
\eea

Likewise, a parameterization of the dilatino  analogous to \eqref{corr4} 
also induces  the replacement of the lowest order $H_{\mu\nu\rho}$ by $\widetilde H_{\mu\nu\rho}$  in the  supersymmetry transformation rule  \eqref{transf0dilino}, so that the combination $\widetilde\rho=2\widetilde \lambda+\gamma^a\widetilde\psi_a$ and its supersymmetry transformation rule  are not deformed, i.e.  $\widetilde\rho=\rho$ and $\delta_\epsilon\rho=\delta^{(0)}_\epsilon\rho$.

 From $\delta E_\mu{}^{\overline i}$ and $\delta\Psi_{\overline i}$ in \eqref{transf1}, one can see that  the gauge and gaugino transformation rules are not deformed and hence it is not necessary to redefine these fields. 

Finally, from the transformation rules of the components $E_{\mu \bar{a}}$   or $E_{\mu \underline{a}}$,  and using the parameterizations defined above, we get
\be
\delta^{(1)} \tilde b_{\mu \nu} =    - \frac{b}{2} \left(\partial_{[\mu} \Lambda^{CD} \hat{\Omega}_{\nu]CD} +\bar\epsilon\gamma_{[\mu}\Psi^{CD}\hat{\Omega}_{\nu]CD}\right)\, .\label{btransf01}
\ee
This compact expression contains information about the gauge, Lorentz and supersymmetry transformations of the $\wt{b}-$field, which we now analyze separately.

Expanding the first term in \eqref{btransf01} one gets
\be
- \frac{b}{2}\partial_{[\mu} \Lambda^{CD} \hat{\Omega}_{\nu]CD} = - \frac{b}{2} \left(\partial_{[\mu} \Lambda^{cd}\hat{w}^{(-)}_{\nu]cd} + \partial_{[\mu} \xi^{k} \hat{A}_{\nu]}{}^{ij}f_{ijk} - \frac{1}{2}\partial_{[\mu}\left(\ov{\epsilon}\gamma^{c}\chi^{i}\right)\hat{F}_{\nu]ci}\right)\, .
\label{btransf_capitalGS}
\ee
The first term in the r.h.s. is the Lorentz sector of the Green-Schwarz transformation \cite{gs}, which  requires the Lorentz Chern-Simons form \eqref{CSL} in $\widetilde H_{\mu\nu\rho}$. It  cannot be eliminated through redefinitions of the $b$-field \cite{mn}. The  bilinear fermionic  terms  in $\hat{w}^{(-)}_{\nu cd}$
may be canceled   redefining $\widetilde b_{\mu\nu}=b_{\mu\nu}-\frac b2{w}_{[\mu}{}^{cd}\ov\psi{}_c\gamma_{\nu]}\psi_d$, but we choose not to do this
 because \eqref{CSL} is defined with the corresponding fermionic contribution and then  $\widetilde H_{\mu\nu\rho}$ 
is Lorentz invariant.

The bosonic piece of the second term in \eqref{btransf_capitalGS}, i.e. $\frac b2\partial_{[\mu} \xi^{k} A_{\nu]}{}^lf_l{}^{ij} f_{ijk}$,  is the first order correction to the Yang-Mills Green-Schwarz transformation in \eqref{gauge0}, reflecting the $\varrho$ deformation of the Killing metric  in \eqref{kappadeform}.  This  transformation cannot be eliminated through redefinitions of the $b$-field either. Instead,   it is convenient to cancel the fermionic terms in $\hat A_{\mu ij}$ redefining
\be
\tilde b_{\mu\nu}=b_{\mu\nu}+\frac b8A_{[\mu}^k\ov\chi^i\gamma_{\nu]}\chi^jf_{ijk}\, , \label{redefb}
\ee
in order to compare with standard results. With this redefinition \eqref{tildeh} becomes
\bea
\widetilde H_{\mu\nu\rho}=   \overline H_{\mu\nu\rho}+ \frac{3b}2\left(D^{(-)}_{[\mu}\hat F_\nu{}^{c i}\hat F_{\rho]c i} -\frac{1}{4}\ov{\chi}^{i}\gamma_{[\mu}\chi^{j}\hat F_{\nu}{}^{c i}\hat F^j_{\rho]c}
+ \frac{1}{4}\ov{\chi}^{i}\gamma_{[\mu}\chi^{j}F_{\nu\rho]}{}^{k}f_{ijk}\right)\, ,\label{nh}
\eea
where
\be
\overline H_{\mu\nu\rho}=3\left(\partial_{[\mu} b_{\nu\rho]}-\zeta C_{\mu\nu\rho}^{(g)}+\frac {b}2\hat C_{\mu\nu\rho}^{(L)}\right)\, .
\ee

 Finally the third term in \eqref{btransf_capitalGS} together with the second term in \eqref{btransf01} contain the first order deformations of the supersymmetry transformation of ${b}_{\mu\nu}$, i.e.
\be
\delta^{(1)}_\epsilon b_{\mu\nu}=\frac b2\left(\hat w^{(-)cd}_{[\mu }\delta_\epsilon\hat w^{(-)}_{\nu]cd}-\varrho A_{[\mu}^i \delta_\epsilon A_{\nu]i}+ \hat F_{[\mu }{}^{c i} \delta_\epsilon\hat F_{\nu]c i}
+D^{(-)}_{[\mu}\left(\bar \epsilon \gamma^{b} \chi^i \right)\hat {F}_{\nu]bi}  
\right) .\label{btransf2}
\ee
 The first term in \eqref{btransf2} was originally introduced in \cite{rw} to restore manifest Lorentz covariance to the supersymmetry variation of the $b$-field curvature.  It was later reobtained in \cite{bdr1} as a consequence of the assumption that the Yang-Mills  and torsionful spin connections should appear symmetrically in  ten dimensional  ${\cal N}=1$ supergravity coupled to super Yang-Mills.  The second term  in \eqref{btransf2}  reflects the $\varrho$ deformation of  the Killing metric \eqref{kappadeform} in the zeroth order supersymmetry transformation  \eqref{transf0b}.   
These two terms are the obvious analogs of the Lorentz and Yang-Mills Green-Schwarz transformations
\be
\delta_\Lambda b_{\mu\nu}= - \frac{b}{2} \partial_{[\mu} \Lambda^{cd} \hat{w}^{(-)}_{\nu]cd}\, , \qquad  \delta_\xi b_{\mu\nu}=-\zeta \partial_{[\mu} \xi^{k}  A_{\nu]}^{k}\, ,
\ee
as already noticed in \cite{rw}. Here, these transformations follow directly from the manifestly duality covariant formulation of the theory.

 Interestingly, the second term in \eqref{btransf01}
can be obtained from the  leading order transformation of the 2-form in \eqref{transf0b} with the identifications $A_\mu^i\leftrightarrow \hat\Omega_{\mu}{}^{ CD}, \chi^i\leftrightarrow\Psi^{CD}$,
i.e.  a generalization of the symmetry  $A_\mu^i\leftrightarrow \hat w_{\mu}^{(-)cd}, \chi^i\leftrightarrow\psi^{cd}$ that was used in \cite{bdr1,Bergshoeff:1988nn}  to obtain the Riemann squared superinvariant. The generalized identification plays a crucial role in the proof of supersymmetric invariance of the first order action, as we discuss in the next section and show in appendix \ref{app:susyac}.

Summing up, the definitions \eqref{vieltransf1}-\eqref{corr4} and \eqref{redefb}
 lead to supergravity and super Yang-Mills fields  that obey the leading order  transformation rules,  except for  the first order deformations in  \eqref{btransf2} and the replacement $H_{\mu\nu\rho}\rightarrow \widetilde H_{\mu\nu\rho}$ in the  supersymmetry transformations of the gravitino and dilatino, i.e.
\be
\delta_\epsilon \psi_{\mu} = \partial_{\mu} \epsilon -\frac14\widetilde w^{(+)}_{\mu ab}\gamma^{ab}\epsilon\,  , \qquad\qquad\qquad
\delta_\epsilon\lambda = \ - \frac{1}{2}\gamma^{\mu} \partial_{\mu}\phi \epsilon +\frac{1}{24}\widetilde H_{{a} {b} c}\gamma^{{a} {b} c} \epsilon \, ,  \label{transf2}
\ee
with $\widetilde w^{(+)}_{\mu ab}= w_{\mu ab}+\frac12\widetilde H_{\mu\nu\rho}e^\nu{}_ae^\rho{}_b$.
 We show in Appendix \ref{app:algehete} that these deformed transformation rules obey a closed algebra including up to three-derivative terms and bilinears in fermions.

Clearly, the transformation laws depend on the choice of parameterization. For instance, we could define
\bea
  {\tilde {e}}'_{\mu}{}^a&=&{ e}_{\mu}{}^ {a} - \frac{b}{8} \left(\hat{w}^{(-)}_{b}{}^{{cd}} \hat{w}^{(-)a}_{ {cd}}   +  \hat{A}^{}_{b{ij}} \hat{A}^{a{ij}}\right) {e}_{\mu}{}^{b} \, , \label{vieltransf2}\\
\tilde\phi'&=&\phi-\frac b{16}\left(\hat{w}^{(-)acd}\hat{w}^{(-)}_{a {cd}} +  \hat{A}^{a{ij}} \hat{A}^{}_{a{ij}}\right) 
 \, , \label{redefdil2}
\eea 
and similar ones for their superpartners,  which are related to the previous parameterizations through gauge and Lorentz covariant field redefinitions. This parameterization is known to reproduce the four-derivative terms in the bosonic sector of the heterotic string effective action when $b=\alpha'$  \cite{bfmn}. Moreover, the fields defined in this way obey the same  classical dynamics as the previous \eqref{vieltransf1} and \eqref{redefdil}
 because the corresponding  effective actions will differ by terms proportional to the leading order equations of motion. 
 However, the definitions \eqref{vieltransf2}-\eqref{redefdil2} induce  complicated first order corrections in the supersymmetry transformation rules of the supergravity fields. Hence, we prefer to keep the fields that obey transformation laws with the smallest amount of deformations.

Before turning  to the construction of the invariant action under the modified transformations, we analyze the deformations  that were proposed  in references  \cite{bdr1,Bergshoeff:1988nn}. 
In particular, we wonder if there is a parameterization of the duality covariant vielbein in terms of a gauge covariant one that transforms as proposed  in \cite{bdr1} or
\cite{Bergshoeff:1988nn}, i.e.
\be
 \delta^{(1)}{\bf e}_\mu{}^a=-\frac{3\alpha'}{32}\overline\epsilon\gamma^{\sigma\tau}\gamma_\mu\psi^\nu  T_{\lambda\nu\sigma\tau}e^{\lambda a}  \quad {\rm or}\quad \delta^{(1)}{\bf e}_\mu{}^a=\frac{3\alpha'}{16}\overline\epsilon \gamma_{[\lambda}\chi^i F_{\nu\rho]i} H_{\mu}{}^{\nu\rho}e^{\lambda a}\,
 \label{bdrtrans}
\ee
respectively, written here in our conventions. Note that we only examine the gauge dependent terms since the gravitational sectors coincide up to the order we are considering.    Specifically, we search for a quantity $E_\mu{}^a$ such that 
\be
{\bf e}_\mu{}^a={ e}_\mu{}^a+{E}_\mu{}^a
\qquad {\rm and}\qquad
\delta^{(1)}{\bf e}_\mu{}^a=\delta^{(0)}{ E}_\mu{}^a \, .
\ee
The most general expressions that can reproduce either one of \eqref{bdrtrans} can be schematically written as
\bea
E_\mu{}^a&=&a^{m}_{1}\left(\overline\psi_{..}\gamma^{...}\psi_. e\right)_\mu{}^a+a^{m}_{2}\left(\overline\psi_.\gamma^{...}\chi Fe\right)_{\mu}{}^a
\eea
or as
\bea
E_\mu{}^a&=&b^m_{1}H_{bcd}H^{acd}e_\mu{}^b+ b^m_{2}(\overline\psi_. \gamma^{...}\psi_. He)_\mu{}^a+ b^m_{3}(\overline\rho\gamma^{...}\psi_. He)_\mu{}^a+ b^m_4(\overline\chi\gamma^{...}\psi_. Fe)_\mu{}^a\nn\\
&&+ b^m_5(\overline\chi\gamma^{...}\chi Fe)_\mu{}^a+ b^m_6(\overline\rho\gamma^{...}\chi Fe)_\mu{}^a+ b^m_7(\overline\chi\gamma^{...}\chi He)_\mu{}^a\, ,
\eea
where the terms between parenthesis refer to all possible  contractions of indices and numbers of $\gamma$-matrices,  numerated by the supraindex $m$, while $ \psi_.$ and $\psi_{..}$ denote the gravitino and gravitino curvature, respectively.
We found that neither of \eqref{bdrtrans} can be reproduced. 

Indeed, the supersymmetric generalized Green-Schwarz transformation \eqref{transf1},  parameterized with the fields that reproduce the bosonic terms of the heterotic effective action,  strongly constrains the possible deformations of the theory. In particular, it does not admit the proposals \eqref{bdrtrans}.  This does not imply that  the latter are in conflict with string theory.  In order to establish the  invariance of the action that implements those supersymmetries under $\rO(n,n)$ transformations,  it  should be dimensionally reduced to $10-n$ dimensions. We stress that the deformations \eqref{btransf2} and \eqref{transf2} were obtained from the  transformation rules of the $\rO(10,10+k)$ multiplets, whose algebra  closes exactly. Hence the theory  avoids an iterative procedure which  only guarantees consistency up to a given order. Moreover, supersymmetry is manifest to all orders and  dimensional reductions will preserve the expected T-duality invariance of the theory.

\section{Heterotic string effective  action to ${\cal O}(\alpha')$ } \label{sec:hete}

In this section we  parameterize  the 
  $\rO(10,10)$ invariant ${\cal N}=1$ supersymmetric action \eqref{n=1} in terms of the supergravity and super Yang-Mills fields  that  transform under local supersymmetry according to \eqref{transf0dil}, \eqref{transf0chi}, \eqref{btransf2}  and \eqref{transf2}. We obtain  all the terms of the heterotic string effective action, up to and including four derivatives of the  fields and bilinear terms  in fermions.

 It is a straightforward though heavy exercise to parameterize  the action \eqref{n=1}.  Interestingly, using  Bianchi identities and  integrations by parts, the action of the theory to ${\cal O}(\alpha')$ may be written in the following compact form:
\be
S = \int d^{10}x\ e \ e^{-2\phi}{\mathscr L}\, , \label{faction}
\ee
with
\bea
\mathscr{L} & = & R + 4\partial_{\mu}\phi\partial^{\mu}\phi - \frac{1}{12}\wt{H}_{\mu\nu\rho}\wt{H}^{\mu\nu\rho} - \frac{1}{4}F_{\mu\nu }^iF^{\mu\nu }_i + \frac{\ap}{8}\hat{\mathscr R}_{\mu\nu AB}\hat{\mathscr R}^{\mu\nu AB}\, \nn\\
& & - \ov{\psi}{}^{\mu}\gamma^{\nu}D_{\nu}\psi_{\mu} + \ov{\rho}\gamma^{\mu}D_{\mu}\rho + 2\ov{\psi}{}^{\mu}D_{\mu}\rho - \frac{1}{2}\ov{\chi}^{i}\gamma^{\mu}D_{\mu}\chi_{i} + \ov{\chi}_{i}\left(\gamma^{\mu}\psi^{\nu} - \frac{1}{4}\gamma^{\mu\nu}\rho\right)F_{\mu\nu }^i\, \nn \\
& & + \frac{1}{24}\wt{H}_{\rho\sigma\tau}\left(\ov{\psi}{}^{\mu}\gamma^{\rho\sigma\tau}\psi_{\mu} + 12\ov{\psi}{}^{\rho}\gamma^{\sigma}\psi^{\tau} - \ov{\rho}\gamma^{\rho\sigma\tau}\rho - 6\ov{\psi}{}^{\rho}\gamma^{\sigma\tau}\rho + \frac{1}{2}\ov{\chi}^{i}\gamma^{\rho\sigma\tau}\chi_{i}\right)\, \nn \\
& & + \ap\left[\ov{\Psi}{}^{AB}\gamma^{\mu}{\mathscr D}_{\mu}(w,\hat{\Omega})\Psi_{AB} -\frac{1}{24} \ov{\Psi}{}^{AB}{\slashed H} \ov{\Psi}_{AB}- \ov{\Psi}{}^{AB}\left(\gamma^{\mu}\psi^{\nu} - \frac{1}{4}\gamma^{\mu\nu}\rho\right)\hat{\mathscr R}_{\mu\nu AB}\right]\, \nn ,
\eea
where we have taken $b=\alpha'$  and  defined ${\slashed H}=\gamma^{\mu\nu\rho}H_{\mu\nu\rho}$ and 
\bea
{\mathscr D}_{\mu}(w,\hat{\Omega})\Psi_{AB} & = & \partial_{\mu}\Psi_{AB} +2 \hat{\Omega}_{\mu [A}{}^C\Psi_{B]C}  - \frac{1}{4}w_{\mu cd}\gamma^{cd}\Psi_{AB}\, .
\eea

 As expected, the bosonic fields reproduce the expression obtained from the scattering amplitudes of the heterotic string  massless  fields up to first order in $\alpha'$ and field redefinitions \cite{higherder}, i.e.
\bea
{\cal S}|_{\rm{bos}}& =&\int d^{10}x e e^{-2\phi}\left[ R + 4\partial_{\mu}{\phi} \partial^{\mu} \phi - \frac{1}{12} \overline H^{\mu \nu \rho} \overline{H}_{\mu \nu \rho} - \frac14\zeta F^i_{\mu\nu}F_i^{\mu\nu}\right.\nn \\ &&
  \left.+\frac{\alpha'}{8} \left(R_{\mu \nu}^{(-){ab}} R^{(-)\mu \nu}{}_{{ab}} - \frac{1}{2} T_{\mu\nu} T^{\mu\nu }  - \frac{3}{2} T_{\mu\nu\rho\sigma} T^{\mu\nu\rho\sigma }  \right)+ \frac{\alpha'}{4} \ e.o.m.\right]\, ,\label{hetebos}
\eea
where we have included only the terms involving purely  bosonic fields (recall that the hatted expressions contain fermions) and $e.o.m.$ refers to the leading order  equations of motion $\Delta g_{\mu\nu}, \Delta\phi$, $\Delta A_{\nu i} $ and $\Delta b_{\rho\nu}$ that are given in Appendix \ref{app:eom}, namely
\bea
e.o.m.=\frac12\Delta e_{\mu a}T^{\mu a} - \left(\frac{1}{4}\Delta\phi T_{\mu\nu} +\Delta(Ab)_{\nu }^i \Delta A_{i\mu}+ A^i_\lambda A_{i\rho }\Delta b^{\lambda}{}_{\mu}\Delta b^{\rho}{}_{\nu}\right)g^{\mu\nu}\, ,\ \  \ \label{eom}
\eea
with $\Delta(Ab)_{\nu }^i=\left(\Delta A_{\nu }^i - 2A^{i}_{\lambda }\Delta b^{\lambda}{}_{\nu}\right)$. 
The first order correction to the Killing metric  included in the coefficient $\zeta$  and all the terms in $e.o.m.$ may be eliminated through gauge  covariant field redefinitions.  However,  as we argued in the previous section,
 the redefined fields would obey more complicated  supersymmetry transformation rules.  Reversing the argument, we can think that by adding terms proportional to the equations of motion in the action, the deformations of the supersymmetry transformation rules can be minimized.

The apparent simplicity of the first order corrections   that involve  bilinears in fermions  in \eqref{faction} is due to the definitions  \eqref{gencon}, \eqref{gengravcur} and \eqref{gcur}.
The terms that are independent of the super Yang-Mills fields  (i.e. those in which all the collective indices $A, B,...$ take the values $a, b,...$) exactly agree with  equation  (2.11) of \cite{{Bergshoeff:1988nn}}. The latter was obtained replacing $A_\mu^i\rightarrow \hat w_\mu^{(-)cd}$ and $ \chi^i\rightarrow \psi^{cd}$ in the  leading order Lagrangian  \eqref{BdRAction}.   Actually,  one can recover the Lagrangian  ${\cal L}(R^{2})$ of \cite{Bergshoeff:1988nn}  replacing
$$\Psi_{AB}\rightarrow\psi_{ab}\, , \qquad \hat{{\mathscr R}}_{\mu\nu AB} \rightarrow R^{(-)}_{\mu\nu ab}\, , \qquad  \wt{H}_{\mu\nu\rho}\rightarrow \ov H_{\mu\nu\rho}\, $$ 
in
 \eqref{faction}.  However  the structures with collective tangent space indices $A,B,...$ contain super Yang-Mills fields in addition to the supergravity fields. Note that $\wt H_{\mu\nu\rho}$ involves  the generalization of the Lorentz Chern-Simons form \eqref{CSL} defined in \eqref{capital_ChernSimons}.
As expected, the  terms  in which the collective indices take the values $i,j,...$ do not agree with the corresponding expressions ${\mathscr L}(RF^2)+{\mathscr L}(F^4)$ in \cite{Bergshoeff:1988nn}, since the supersymmetry transformation rules of the fields differ by Yang-Mills field-dependent terms.

The supersymmetric invariance of
 the action \eqref{faction} is shown in appendix \ref{app:susyac}. It simply results from the observation that both the action and the  transformation rules of the fields  have the same structure as the corresponding ones in \cite{Bergshoeff:1988nn}, albeit  with collective  indices, except for the  terms contained in the parameter $\Lambda_{ci}=\frac1{2\sqrt2}\bar\epsilon\gamma_c\chi_i$, which cancel in the variation of the action.

\section{Outlook and final remarks} \label{sec:conclu}

In this paper we have obtained the first order corrections to  ${\cal N}=1$ supersymmetric DFT performing a perturbative expansion of the  exact  supersymmetric and duality covariant framework  introduced in \cite{diego2018}.
The  action  has the same functional form as the leading order one constructed in  \cite{Hohm:2012}, but it is expressed in terms of \mbox{$\rO(10,10+k)$} multiplets, where $k$ is the dimension of the $\rO(1,9+k)$ group. 
Decomposing the \mbox{$\rO(10,10+k)$} duality group  in terms of  $\rO(10,10+n_g)$  multiplets, the theory contains higher derivative terms to all orders. We kept all the terms with up to and including four derivatives of the fields and  bilinears   in fermions.

The transformation rules of the $\rO(10,10+k)$ multiplets obey a closed algebra and  induce higher-derivative deformations on those of the $\rO(10,10+n_g)$  fields. In particular, they produce a supersymmetric generalization of the duality covariant  Green-Schwarz transformation that was found in \cite{mn}.  We showed that the algebra of  deformations closes  up to first order and constructed the invariant action with up to and including four derivatives of the $\rO(10,10+n_g)$  multiplets and bilinears in fermions.

To make contact with the heterotic string low energy effective field theory, we parameterized the duality covariant multiplets in terms of supergravity and super Yang-Mills fields. The inclusion of higher-derivative terms requires unconventional non-covariant field redefinitions in the parameterizations of the duality covariant structures. The definitions that reproduce the four-derivative interactions of the  bosonic fields of the heterotic string effective action were found    in \cite{mn,bfmn}.  Here, we worked with a  set of fields related to the latter
 through gauge covariant redefinitions. Except for the two-form, the fields defined in section \ref{sec:transf}  obey the leading order  transformation rules with a 
modification of the two-form curvature in the supersymmetry variations.
  The  Lorentz and non-abelian gauge transformations  of the two-form   are deformed by the standard Green-Schwarz mechanism, as expected, and its supersymmetry transformations are deformed by Green-Schwarz-like terms plus some extra Yang-Mills dependent higher-derivative terms. 

The deformed transformations obey a closed algebra, which guarantees the existence of  an invariant action. We constructed
such action in section \ref{sec:hete},  by parameterizing the manifestly duality covariant expression \eqref{n=1}  in terms of the fields that obey   supersymmetry transformation rules with the minimal set of deformations.  As expected, the interactions of the  bosonic  fields agree with the  results obtained from the heterotic string scattering amplitudes  \cite{higherder}, up to terms proportional to the leading order equations of motion. 
To our knowledge, the three-derivative low energy  interactions involving fermions have not been constructed directly from string theory. The action and transformation rules that we have obtained follow from an exact supersymmetric and duality covariant formalism. Hence the theory  avoids an iterative  procedure which  only guarantees consistency up to a given order. Moreover, supersymmetry is manifest to all orders and dimensional reductions will preserve the expected T-duality symmetry of the theory.

 Supersymmetric extensions of the Yang-Mills and Lorentz Chern-Simons forms have been constructed using the Noether method. In particular, a supersymmetric \mbox{${\mathscr L}(R)+{\mathscr  L}(R^2)$} invariant   was obtained in \cite{bdr1,Bergshoeff:1988nn} from the leading order action \eqref{BdRAction}, using the symmetry between the gauge and torsionful spin connections. The three-derivative  terms   that are  independent of the Yang-Mills  fields   in the action \eqref{faction} coincide with those results.
But not surprisingly, the Yang-Mills field-dependent terms disagree with the corresponding expressions of the ${\mathscr  L}(RF^2)+{\mathscr  L}(F^4)$ invariants proposed in those references, since the  deformations of the transformation rules differ by Yang-Mills field-dependent terms. The supersymmetric and T-duality covariant generalized Green-Schwarz transformation  strongly restricts the
modifications to the leading order supersymmetry transformation rules, and in particular, it does  not allow the proposals of \cite{bdr1,Bergshoeff:1988nn}. As argued in section \ref{sec:transf} this does not imply that  the latter are in conflict with string theory.  In order to establish if they are compatible with the required T-duality symmetry, the corresponding invariant action should be dimensionally reduced.

The effort employed in the construction of the higher-derivative fermionic sector of the 
heterotic string effective field theory is justified
for various reasons. First of all,   an intriguing consequence of the duality covariant formalism is the natural appearance of the generalized collective tangent space indices $C, D,...$, which allows to include the higher-derivative Yang-Mills field-dependent terms into {\it gravitational} structures such as $\hat{\mathscr R}_{\mu\nu CD},\hat\Omega_{\mu CD}$ or $ \Psi_{CD}$.  In particular, it  leads to  relatively mild modifications of the leading order supersymmetry transformation rules of the fields, which permits the use of 
the  leading order  Killing spinor equations to obtain classical solutions containing higher-derivative corrections  \cite{ort}. These features not only simplify  the  construction of new supersymmetric solutions but also allow to easily extend the
  known solutions for the gravitational sector  to the Yang-Mills sector.

The fermionic contributions to the action are also relevant for applications to four-dimensional physics. 
Both the superpotential and D-terms can be more easily computed
from the fermionic couplings \cite{louis} and the higher derivative corrections to
these terms as well as to
the Yukawa couplings could also have interesting  consequences for string phenomenology and moduli fixing.

 An obvious natural extension of our work  would be to determine further interactions  beyond the first order. The quartic  interactions of the Yang-Mills fields that we have reproduced  are mirrored by corresponding quartic Riemann curvature  terms \cite{higherder}.  Consequently, we  expect that the higher orders of  perturbation will reproduce these higher-derivative  corrections.  It would be interesting to see if the generalized structures with capital indices
 persist to higher orders. If they do, the formulation would contain information about higher than four-point functions in the string scattering amplitudes. 

Nevertheless, there is  another  quartic Riemann curvature structure  that has no analog in the Yang-Mills sector \cite{higherder}.  At tree level, these terms are proportional to the transcendental coefficient $\zeta(3)$.  
The analysis of the higher-derivative terms is technically more challenging but also more interesting, since further duality covariant   structures, or even a more drastic change of scheme, seem to be necessary as advocated in \cite{wulff}.

Performing a generalized Scherk-Schwarz compactification of the sub-leading corrections to ${\cal N}=1$ supersymmetric  DFT would be another promising line of research, as this  would produce  higher-derivative  corrections  to  lower dimensional gauged supergravities \cite{abmn,bfmn}. We hope to return to these and related questions in the future.

\subsection*{Acknowledgements}
We sincerely thank W. Baron and D. Marqu\'es for kindly sharing their project \cite{diego2018} and for important insight and comments. C.N.  would also like to acknowledge correspondence with N. Berkovits, J. Gates and H. Gomez and  thank the A. S. ICTP for hospitality during the initial stages of this work. This research was partially supported
by PIP-CONICET- 11220150100559CO, UBACyT  and  ANPCyT- PICT-2016-1358.
\appendix

\section{Conventions and definitions}\label{app:conventions}
In this appendix we introduce the conventions and definitions used throughout the paper. Space-time and tangent space Lorentz indices are denoted $\mu, \nu, \dots $ and $a, b, \dots$, respectively. 

The covariant derivative  acting on a gauge tensor $G_\mu{}^{ci}$ and on a spinor $\epsilon$ is, respectively,
\bea
D_\mu^{(\pm)} G_{\nu c}^{ i}&=&\partial_\mu G_ {\nu c}^{ i}-\Gamma_{\mu\nu}^\rho G_{\rho c}^{ i}- w_{\mu c}^{(\pm)d} G_{\nu d}^{ i}-A_\mu^j G_{\nu c}^{k}f^i{}_{jk} ,\,\\
D^{(\pm)}_\mu\epsilon&=& \partial_\mu\epsilon-\frac14 w^{\pm}_{\mu ab}\gamma^{ab}\epsilon \, ,
\eea
with
\bea
\Gamma^{\sigma}_{\mu\nu} &= & \frac{1}{2}g^{\sigma\rho}\left(\partial_{\mu}g_{\nu\rho} + \partial_{\nu}g_{\mu\rho} - \partial_{\rho}g_{\mu\nu}\right)\, ,
\eea
and the torsionful spin connection 
\be
 {w}^{(\pm)}_{a{bc}}\equiv \left( w_{abc} \pm\frac12 H_{abc}\right)\, ,
\ee
where
\be
w_{\mu bc}=e_\mu{}^a\left(-e^\mu{}_{[a}{} e^\nu{}_{b]}{}\partial_\mu e_{\nu c} +e^\mu{}_{[a}{} e^\nu{}_{c]}{}\partial_\mu e_{\nu b}
+e^\mu{}_{[b}{} e^\nu{}_{c]}{}\partial_\mu e_{\nu a}\right)\, .
\ee
The identity $D_\mu e_\nu{}^a=\partial_\mu e_\nu{}^a-\Gamma_{\mu\nu}^\rho e_\rho{}^a -w_{\mu}{}^{a}{}_be_\nu{}^b=0$ implies
\be
w_{\mu a}{}^b=-e^\nu{}_a\partial_\mu e_\nu{}^b+\Gamma^\sigma_{\mu\nu}e_\sigma{}^be^\nu{}_a\, .
\ee

The commutator of covariant derivatives acting on gauge tensors and spinors is
\bea
\left[D^{(\pm)}_{\mu},D^{(\pm)}_{\nu}\right]{F}_{\rho ci} &=& - R^{\sigma}{}_{\rho\mu\nu}{F}_{\sigma ci} + R^{(\pm)}_{\mu\nu c}{}^{d}{F}_{\rho di} - F_{\mu\nu}{}^{j}{F}_{\rho c}{}^{k}f_{ijk}\\
\left[D^{(\pm)}_{\mu},D^{(\pm)}_{\nu}\right]\epsilon &=& \frac14 R^{(\pm)}{}_{\mu\nu ab}\gamma^{ab}\epsilon\, ,
\eea
where the Riemann tensor is defined as
\bea
R^\rho{}_{\sigma\mu\nu}&=&\partial_{\mu}\Gamma^\rho_{\nu\sigma}-\partial_{\nu}\Gamma^\rho_{\mu\sigma}+
\Gamma^\rho_{\mu\kappa}\Gamma^\kappa_{\nu\sigma}-\Gamma^\rho_{\nu\kappa}\Gamma^\kappa_{\mu\sigma}\nn\\
&=&e^{\rho a}e_{\sigma }{}^bR_{\mu\nu ab}=e^{\rho a} e_{\sigma }{}^b\left(-2\partial_{[\mu} w_{\nu]ab} + w_{\mu a}{}^{{c}} w_{\nu cb} - w_{\nu a}{}^{{c}} w_{\mu cb} \right)\, ,
\eea
and the Yang-Mills field strength is
\be
F_{\mu\nu}^i=2\partial_{[\mu}A_{\nu]}^i-f^i{}_{jk}A_\mu^jA_\nu^k\, .
\ee
The Ricci tensor and scalar are
\be
R_{\mu\nu}=R^\rho{}_{\mu\rho\nu}\, , \qquad R=g^{\mu\nu}R_{\mu\nu}=R_{\mu\nu}{}^{ab}e^\mu{}_ae^\nu{}_b\, .
\ee

\subsection{Some useful gamma function identities}\label{app:gammaids}

To distinguish $\rO(1,9)_\rR$ and $\rO(9,1)_\rL$ tangent space indices in DFT we use $\ov a, \ov b, \dots$ and $\underline a, \underline b, \dots$, respectively.
The Clifford algebra $\{\gamma _{\underline a} ,\gamma_{\underline b}\}=-2P_{\underline{ab}}$ determines the following identities for the $\rO(9,1)_\rL$ gamma matrices 
 \begin{subequations}\label{iden}
\begin{align}
\gamma_{\underline a} \gamma_{\underline b} &= \gamma_{\underline {ab}} - P_{\underline {ab}} \; , \\
\gamma_{\underline {ab}} \gamma_{\underline c} &=  \gamma_{\underline {ab c}} - 2 \gamma_{[\underline a} P_{\underline b]\underline c} \; ,\\
\gamma_{\underline a}\gamma_{\underline {bc}}&=\gamma_{\underline {abc}}-2P_{\underline a[\underline b}\gamma_{\underline c]}\; , \\ 
  \gamma_{\underline {ab}} \gamma^{\underline {c d}} & =  \gamma_{\underline {ab}}{}^{\underline {c d}} -4 \gamma_{[\underline a}{}^{[\underline d} P_{\underline b]}{}^{\underline c]} 
+ 2 P_{[\underline b}{}^{[\underline c}\, P_{\underline a]}{}^{\underline d]} \; ,\\
 \gamma_{\underline {ab}} \gamma^{\underline {c d e}} & = \gamma_{\underline {ab}}{}^{\underline {c d e}} -6 \gamma_{[\underline a}{}^{[\underline {d e}} P_{\underline b]}{}^{\underline c]}  
+ 6 \gamma^{[\underline e} P_{[\underline b}{}^{\underline c}\, P_{\underline a]}{}^{\underline d]} \; ,\\
  \gamma_{\underline {ab c}} \gamma^{\underline {d e}} & =  \gamma_{\underline {ab c}}{}^{\underline {d e}} -6 \gamma_{[\underline {ab}}{}^{[\underline e} P_{\underline c]}{}^{\underline d]} 
+ 6 \gamma_{[\underline a} P_{\underline c}{}^{[\underline d}\, P_{\underline b]}{}^{\underline e]} \; ,\\
\gamma_{\underline {ab c}} \gamma^{\underline {d e f}} & =  \gamma_{\underline {ab c}}{}^{\underline {d e f}} - 9 \gamma_{[\underline {ab}}{}^{[\underline {e f}} P_{\underline c]}{}^{\underline d]} + 18 \gamma_{[\underline a}{}^{[\underline f} P_{\underline c}{}^{\underline d} P_{\underline b]}{}^{\underline e]} -6 P_{[\underline c}{}^{[\underline d}\, P_{\underline b}{}^{\underline e}\, P_{\underline a]}{}^{\underline f]} \; ,\\
 C\gamma^{\underline a} C^{-1}&=-(\gamma^{\underline a})^t\, , \qquad  C^{-1}\gamma_{\underline {ab}}C=-(\gamma_{\underline {ab}})^t\, ,
\end{align}
 \end{subequations}
where $C^{-1}=C^t=-C$ and $\underline a, \underline b=0,\dots,9$. 
\subsection{Leading order components of the generalized fluxes}
\label{app:fluxes}

Using the parameterizations introduced in section 2 and solving the strong constraint in the supergravity frame, the non-vanishing determined components of the generalized spin connection  are, to leading order,
 \begin{subequations}\label{fluxes}
\begin{align}
F_{\overline a\underline{ bc}}&=-\left(w_{abc}+\frac12 H_{abc}\right)\equiv -
w^{(+)}_{ a bc}\, ,\\
F_{\underline  a \overline{bc}}& = \left(w_{abc}-\frac12 H_{abc}\right)\equiv w^{(-)}_{ a bc}\, ,\\
F_{\overline{a bc}}&=3\left(w_{[{abc}]}-\frac16H_{{abc}}\right)
\, ,\\
F_{\underline {a bc}}& =-3\left(w_{[abc]}+\frac16H_{abc}\right)\, ,\\
F_{\overline i \underline{ab}}&=
F_{\underline a\overline {bi}}=F_{\overline {abi}}=- \frac1{\sqrt2} e^ \mu{}_a e^\nu{}_b e{}_{i\overline i}F^i_{\mu\nu}\, ,\\
F_{\underline a \overline i \overline j}& = -e^i{}_{\overline i} e^j{}_{\overline j} e^\mu{}_a A_{\mu}{}^{k} f_{ijk}\, ,\\
F_{\overline{ijk}}&=\sqrt2e^i_{\overline i }e^j_{\overline j}e^k_{\overline k }f_{ijk}\, ,\\
F_{\underline a}&=F_{\overline a}\ =\left(\partial_\mu e_a^\mu+e_a^\mu e_b^\nu\partial_\mu e^b_\nu-2e^\mu{}_{ a}\partial_\mu\phi\right)\, ,
\end{align}
 \end{subequations}
where 
\bea
H_{abc}&=&e_{[a}^\mu e_b^\nu e_{c]}^\rho H_{\mu\nu\rho}=3e_{a}^\mu e_b^\nu e_{c}^\rho\left(\partial_{[\mu}b_{\nu\rho]}-A_{[\mu}^i\partial_\nu A_{\rho]i}+\frac13
f_{ijk}A_{\mu}^i A_\nu^j A_\rho^k\right)\, ,\label{hache}
\eea 
and $f_{ijk}$ are the structure constants of the $\hog$ or $\mre_8\times \mre_8$ gauge groups.

\subsection{The leading order action and equations of motion}
\label{app:eom}
Here we rewrite the zeroth order action \eqref{BdRAction}  in terms of the  dilatino $\lambda$ of the supergravity multiplet and compare with the corresponding expression in \cite{Bergshoeff:1988nn}. We also  list the leading order equations of motion of all the massless fields derived  from it.

Rewriting  the generalized dilatino $\rho=2\lambda+\gamma^\mu\psi_\mu$ in  terms of $\lambda$ and $\psi$ and integrating by parts, the action \eqref{BdRAction} takes the form

\bea
S &=& \int d^{10}x \ e \ e^{-2\phi}\left[R(w(e)) -\frac{1}{12}H_{\mu \nu \rho}H^{\mu \nu \rho} + 4\partial_{\mu}\phi\partial^{\mu}\phi-\frac14 F^i_{\mu\nu}F_i^{\mu\nu} \right. \nn\\ 
&  & \ \   - \bar{\psi}_{\mu}\gamma^{\mu \nu \rho}{D}_{\nu}\psi_{\rho}  + 4\bar{\lambda}\gamma^{\mu \nu}{D}_{\mu}\psi_{\nu} + 4\bar{\lambda}\gamma^{\mu}{D}_{\mu}\lambda -\frac12\bar\chi^i{\slashed{ D}}\chi_i\nn\\
&&\ \ + \ 4\bar{\psi}_{\mu}\gamma^{\nu}\gamma^{\mu}\lambda \partial_{\nu}\phi - 2\bar{\psi}_{\mu}\gamma^{\mu}\psi^{\nu}\partial_{\nu}\phi -\frac14\bar\chi_i\gamma^\mu\gamma^{\nu\rho}F_{\nu\rho}^i\left(\psi_\mu+\frac13\gamma_\mu\lambda\right)\nn \\
& & \left. \ \ + \frac{1}{24}H_{\rho \sigma \tau}\left(\bar{\psi}^{\mu}\gamma_{[\mu}\gamma^{\rho \sigma \tau}\gamma_{\nu ]}\psi^{\nu} + 4\bar{\psi}_{\mu}\gamma^{\mu \rho \sigma \tau}\lambda - 4\bar{\lambda}\gamma^{\rho \sigma \tau}\lambda +\frac12\bar\chi^i\gamma^{\rho\sigma\tau}\chi_i\right) \right] \, .
\label{BdRAction2}
\eea
It matches the corresponding expression in \cite{Bergshoeff:1988nn} with the following field redefinitions:  $\phi^{-3}\rightarrow e^{-2\phi}$, $R\rightarrow -R$, $H_{\mu\nu\lambda}\rightarrow\frac{1}{3\sqrt{2}}H_{\mu\nu\lambda}$, $B_{\mu\nu}\rightarrow\frac{1}{\sqrt{2}}b_{\mu\nu}$, $\lambda\rightarrow\frac{1}{\sqrt{2}}\lambda$, $A_\mu\rightarrow \frac{1}{\sqrt{2}}A_\mu$, $\chi\rightarrow \frac{1}{\sqrt{2}}\chi$. 

The leading order equations of motion of all the massless fields, written in terms of $\rho$, are
\bea
\Delta e_{\mu}{}^{a} & =&  \frac{1}{2}e_{\mu}{}^a\Delta\phi +2 R_{\mu}{}^{a} + 8D_{\mu}\phi D^{a}\phi - \frac{1}{2}H_{\mu\lambda\sigma}H^{a\lambda\sigma} -F_{\mu\lambda i}F^{a\lambda i}\, \nn \\
&&   - 2\ov{\psi}_{\mu}\gamma^{\lambda}D_{\lambda}\psi^a -2 \ov{\psi}{}^{\lambda}\gamma_{\mu}e^{\nu a}D_{\nu}\psi_{\lambda} +2 \ov{\rho}\gamma_{\mu}D^a\rho + 4\ov{\psi}_{\mu}D^a\rho - \ov{\chi}^{i}\gamma_{\mu}D^a\chi_{i}\, \nn \\
&&  + \frac{1}{4}\ov{\psi}{}^{\lambda}\gamma_{\mu}{}^{\sigma\tau}\psi_{\lambda}H^a{}_{\sigma\tau} - \frac{1}{4}\ov{\rho}\gamma_{\mu}{}^{\sigma\tau}\rho H^a{}_{\sigma\tau} + \frac{1}{8}\ov{\chi}^{i}\gamma_{\mu}{}^{\sigma\tau}\chi_{i}H^a{}_{\sigma\tau} + \ov{\psi}{}^{\sigma}\gamma_{\mu}{}^{\tau}\rho H^a{}_{\sigma\tau}\, \nn \\
&&  - \frac{1}{2}\ov{\psi}{}_{\mu}\gamma_{\sigma\tau}\rho H^{a\sigma\tau} + 2\ov{\psi}{}_{\mu}\gamma^{\sigma}\psi^{\tau}H^a{}_{\sigma\tau} - \ov{\psi}{}^{\sigma}\gamma_{\mu}\psi^{\tau}H^a{}_{\sigma\tau} + \frac{1}{12}\ov{\psi}{}_{\mu}\gamma^{\rho\sigma\tau}\psi^aH_{\rho\sigma\tau}\, \nn \\
&&  + 2\ov{\chi}_{i}\gamma_{\mu}\psi_{\lambda}F^{a\lambda i} - 2\ov{\chi}_{i}\gamma_{\lambda}\psi_{\mu}F^{a\lambda i} - \ov{\chi}_{i}\gamma_{\mu\lambda}\rho F^{a\lambda i}\, ,\label{eomviel}\\
\Delta\phi & = & -2\mathcal{L} \, ,\label{eomphi}\\
\Delta b_{\nu\rho} & =&  \frac{1}{2}D^{\mu}H_{\mu\nu\rho} - D^{\mu}\phi H_{\mu\nu\rho}\,\nn \\
&&   - \frac{1}{8}D^{\mu}\left(\ov{\psi}{}^{\lambda}\gamma_{\mu\nu\rho}\psi_{\lambda} + 12\ov{\psi}{}_{[\mu}\gamma_{\nu}\psi_{\rho]} - \ov{\rho}\gamma_{\mu\nu\rho}\rho - 6\ov{\psi}{}_{[\mu}\gamma_{\nu\rho]}\rho + \frac{1}{2}\ov{\chi}^{i}\gamma_{\mu\nu\rho}\chi_{i}\right)\, \nn \\
&& + \frac{1}{4}\left(\ov{\psi}{}^{\lambda}\gamma_{\mu\nu\rho}\psi_{\lambda} + 12\ov{\psi}{}_{[\mu}\gamma_{\nu}\psi_{\rho]} - \ov{\rho}\gamma_{\mu\nu\rho}\rho - 6\ov{\psi}{}_{[\mu}\gamma_{\nu\rho]}\rho + \frac{1}{2}\ov{\chi}^{i}\gamma_{\mu\nu\rho}\chi_{i}\right)D^{\mu}\phi\, , \ \ \ \ \ \label{eomb}\\
\Delta A_{\mu}{}^{i} & = & \frac{1}{2}H_{\mu\nu\rho}F^{\nu\rho i} + A_{\rho}{}^{i}\Delta b^{\rho}{}_{\mu} - D^{\nu}F_{\mu\nu}{}^{i} + 2F_{\mu\nu}{}^{i}D^{\nu}\phi - \frac{1}{2}\ov{\chi}^{j}\gamma_{\mu}\chi^{k}f^{i}{}_{jk}\, \nn \\
&&  - \frac{1}{8}F^{\nu\rho i}\left(\ov{\psi}{}^{\lambda}\gamma_{\mu\nu\rho}\psi_{\lambda} + 12\ov{\psi}{}_{[\mu}\gamma_{\nu}\psi_{\rho]} - \ov{\rho}\gamma_{\mu\nu\rho}\rho - 6\ov{\psi}{}_{[\mu}\gamma_{\nu\rho]}\rho + \frac{1}{2}\ov{\chi}^{j}\gamma_{\mu\nu\rho}\chi_{j}\right)\, \nn \\
&&  + 2D^{\nu}\ov{\chi}^{i}\left(\gamma_{[\mu}\psi_{\nu]} - \frac{1}{4}\gamma_{\mu\nu}\rho\right)  + 2\ov{\chi}^{i}D^{\nu}\left(\gamma_{[\mu}\psi_{\nu]} - \frac{1}{4}\gamma_{\mu\nu}\rho\right) \nn\\
&&  - 4\ov{\chi}^{i}\left(\gamma_{[\mu}\psi_{\nu]} - \frac{1}{4}\gamma_{\mu\nu}\rho\right)D^{\nu}\phi\, ,\label{eoma} \\
\Delta\psi_{\mu}& = &2D_{\nu}\ov{\psi}_{\mu}\gamma^{\nu} - 2\ov{\psi}_{\mu}\gamma^{\nu}\partial_{\nu}\phi + 2D_{\mu}\ov{\rho} + \frac{1}{12}\ov{\psi}{}_{\mu}\gamma^{\rho\sigma\tau}H_{\rho\sigma\tau} - \frac{1}{4}H_{\mu\nu\rho}\left(4\ov{\psi}^{\rho}\gamma^{\nu} - \ov{\rho}\gamma^{\nu\rho}\right)\, ,\nn\\
&&  - F_{\mu\nu}{}^{i}\ov{\chi}_{i}\gamma^{\nu}\label{eompsi}\\
\Delta\rho & = &- 2D_{\mu}\ov{\rho}\gamma^{\mu} + 2\ov{\rho}\gamma^{\mu}\partial_{\mu}\phi - 2D_{\mu}\ov{\psi}^{\mu} + 4\ov{\psi}^{\mu}\partial_{\mu}\phi - \frac{1}{12}H_{\rho\sigma\tau}\left(\ov{\rho}\gamma^{\rho\sigma\tau} + 3\ov{\psi}^{\rho}\gamma^{\sigma\tau}\right)\, ,\nn\\
&&  - \frac{1}{4}F_{\mu\nu}{}^{i}\ov{\chi}_{i}\gamma^{\mu\nu}\label{eomrho}\\
\Delta\chi_{i} & =& D_{\mu}\ov{\chi}_{i}\gamma^{\mu} - \ov{\chi}_{i}\gamma^{\mu}\partial_{\mu}\phi + \ov{\chi}_{i}\frac{1}{24}\gamma^{\rho\sigma\tau}H_{\rho\sigma\tau} - \left(\ov{\psi}^{\nu}\gamma^{\mu} - \frac{1}{4}\ov{\rho}\gamma^{\mu\nu}\right)F_{\mu\nu i}\, .\label{eomchi}
\eea

\section{Algebra of transformations of $\rO(10,10+n_g)$ fields }\label{app:closure}
In this appendix we show that the algebra of transformation rules  closes, up to terms with two fermions. We first review the algebra of zeroth order transformations \eqref{gentransf0} and in \ref{App first} we  include the first order corrections. We define $[\delta_1,\delta_2]=-\delta_{12}$.
\subsection{Leading order algebra }\label{app:closure1}

We  focus on the  algebra determined by the leading order transformations (\ref{gentransf0}) and show that it closes with the parameters \eqref{par0}.  
We split the algebra of transformations on the generalized fields into the following commutators:
\begin{itemize}
\item Supersymmetry transformations of the dilaton 
\begin{eqnarray}
\label{closured1}
\big[\delta_{\epsilon_1},\delta_{\epsilon_2}\big]d & = &  \frac{1}{2}\overline{\epsilon}_{[2}\left(\gamma^{\underline a}\sqrt2 E_{\underline a}{}^{M}\partial_{M}\epsilon_{1]}-\frac{1}{4}\gamma^{\underline a}\omega_{\underline {abc}}\gamma^{\underline {bc}}\epsilon_{1]}\right)\nn\\
&=&-\xi'^{M}_{12}\partial_{M}d+\frac{1}{2}\partial_{M}\xi'^{M}_{12}=-\delta_{\xi'_{12}}d\, ,
\end{eqnarray}
where  we have used $\bar{\epsilon}_{1}\gamma^{\underline a}\epsilon_{2}=-\bar{\epsilon}_{2}\gamma^{\underline a}\epsilon_{1}$ and $\overline{\epsilon}_{1}\gamma^{\underline {abc}}\epsilon_{2}=\overline{\epsilon}_{2}\gamma^{\underline {abc}}\epsilon_{1}$, and defined
\begin{equation}
\label{diffParameter1}
\xi'^{M}_{12}=-\frac{1}{\sqrt2}E^M{}_{\underline c}\left(\bar{\epsilon}_{1}\gamma^{\underline c}\epsilon_{2}\right)\, .
\end{equation}
\item Diffeomorphisms on the dilaton
\begin{eqnarray}
\big[\delta_{\xi_1},\delta_{\xi_2}\big]d 
& = & -\xi''^{M}_{12}\partial_M d+\frac12\partial_M\xi''^{M}_{12}=-\delta_{\xi''_{12}}d\, ,
\end{eqnarray}
with 
\begin{equation}
\label{diffParameter2}
\xi''^{M}_{12}=2\xi_{[1}^N\partial_N\xi_{2]}^M\, .
\end{equation}

\item Mixed supersymmetry and double Lorentz transformations on the dilaton
\begin{eqnarray}
\delta_{[\Gamma,\epsilon]}d & = &- \frac1{8}\overline\epsilon_{[2}\Gamma_{1]\underline {bc}}\gamma^{\underline {bc}}\rho =-\delta_{\epsilon'_{12}}d\, ,
\end{eqnarray}
where we have defined $\delta_{[\Gamma,\epsilon]}=[\delta_{\Gamma_1},\delta_{\epsilon_2}]+[\delta_{\epsilon_1},\delta_{\Gamma_2}]$ and
\begin{equation}
\label{LorParameter1}
\overline\epsilon'_{12}=-\frac12\Gamma_{[1\underline {ab}}\overline\epsilon_{2]}\gamma^{\underline {ab}}\, .
\end{equation}

\item Mixed diffeomorphisms and supersymmetry variations on the dilaton
\begin{eqnarray}
\delta_{[\epsilon,\xi]}d 
& = & \frac12\xi_{[1}^M\partial_M\overline\epsilon_{2]}\rho=-\delta_{\epsilon''_{12}} d\, ,
\eea
with 
\be
\overline\epsilon''_{12}=2\xi_{[1}^M\partial_M\overline\epsilon_{2]}\, .\label{susyParameter1}
\ee

\item Supersymmetry variations of the frame
\begin{eqnarray}
\big[\delta_{\epsilon_1},\delta_{\epsilon_2}\big]E_{{\mathbb M}{\underline a}}
 & = & \frac1{\sqrt2}E^{N}{}_{\overline B}\partial_N\left(\overline{\epsilon}_{[1}\gamma_{\underline a}\epsilon_{2]}\right)E_{\mathbb M}{}^{\overline B}-\frac12\left(\overline{\epsilon}_{[1}\gamma_{\underline c}\epsilon_{2]}\right)\omega_{\overline B \underline a}{}^{\underline c}E_{\mathbb M}{}^{\overline B}\, .
\end{eqnarray}
 Projecting   with $E^{\mathbb M}{}_{\overline C}$, we get 
\begin{eqnarray}
E^{\mathbb M}{}_{\overline C}\big[\delta_{\epsilon_1},\delta_{\epsilon_2}\big]E_{{\mathbb M} {\underline a}}& = & -
E^{\mathbb N}{}_{\overline C}\delta_{\xi'_{12}} E_{{\mathbb N}\underline a}
\end{eqnarray}
where we have used (\ref{gralspinconnectionE}) and $\xi'^{\mathbb M}_{12}$ is  the generalization of (\ref{diffParameter1}), i.e.  
\begin{equation}
\label{diffParametergen1}
\xi'^{\mathbb M}_{12}=-\frac{1}{\sqrt2}E^{\mathbb M}{}_{\underline c}\left(\bar{\epsilon}_{1}\gamma^{\underline c}\epsilon_{2}\right)\, .
\end{equation}
 Projecting   with $E^{\mathbb M}{}_{\underline c}$ we find
\begin{eqnarray}
E^{\mathbb M}{}_{\underline c}\big[\delta_{\epsilon_1},\delta_{\epsilon_2}\big]E_{{\mathbb M} {\underline a}}& = & -
E^{\mathbb N}{}_{\underline c}\delta_{\Gamma'_{12}} E_{{\mathbb N}\underline a}
\end{eqnarray}
with
\bea
\Gamma'^{\underline{ab}}_{12}=E^{[\underline a}\left(\overline\epsilon_1\gamma^{\underline c]}\epsilon_2\right)-\frac1{2}\left(\overline\epsilon_1\gamma^{\underline c}\epsilon_2\right)F^{\underline{ab}}{}_{\underline c}\, .
\eea

Following similar steps, we get
\bea
E^{\mathbb M}{}_{\underline  c}\big[\delta_{\epsilon_1},\delta_{\epsilon_2}\big]E_{{\mathbb M}{\overline A}}=-E^{\mathbb M}{}_{\underline c}\delta_{\xi'_{12}} E_{{\mathbb M}\overline A}\, ,\quad E^{\mathbb  M}{}_{{\overline B}}\big[\delta_{\epsilon_1},\delta_{\epsilon_2}\big]E_{{\mathbb M}{ \overline A}}= -
E^{\mathbb M}{}_{\overline C}\delta_{\Gamma'_{12}} E_{{\mathbb M}\overline A}\nn\, ,
\eea
with
\bea
\Gamma'^{\overline{AB}}_{12}=-\frac1{2}\left(\overline\epsilon_1\gamma^{\underline c}\epsilon_2\right)F^{\overline{AB}}{}_{\underline c}\, .
\eea

\item Diffeomorphisms and double Lorentz variations of the frame 
\bea
\delta_{[\Gamma,\xi]} E^{\mathbb M} {}_{\mathbb A}
&=&-\left(\delta_{\Gamma''_{12}}+\delta_{\xi''_{12}}\right)E^{\mathbb M}{}_{\mathbb A}\, , 
\eea
where
\bea
\Gamma''_{12\mathbb{AB}}&=&2\xi_{[1}^M\partial_M\Gamma_{2]\mathbb{AB}}-2\Gamma_{[1\mathbb A}{}^{\mathbb C}\Gamma_{2]\mathbb{CB}}\\
\xi''^{\mathbb M}_{12}&=&2\xi^{P}_{[1}\partial_{P}\xi_{2]}^{\mathbb M}-\xi_{[1}^{\mathbb N}\partial^{\mathbb  M}\xi_{2]\mathbb N}+f_{\mathbb PQ}{}^{\mathbb M} \xi_{1}^{\mathbb P} \xi_2^{\mathbb Q}\,  .\label{diffParameter3}
\eea
Note that $\xi''^M_{12}$  in (\ref{diffParameter2}) does not contain the second and third terms  in the r.h.s. of this expression, due to the strong constraint.

\item Mixed diffeomorphisms and supersymmetry variations of the frame 
\bea
\delta_{[\epsilon, \xi]} E^{\mathbb M} {}_{\underline a}
&=& \xi_{[1}^N\partial_N\overline\epsilon_{2]}\gamma_{\underline a}\Psi_{\overline B}E^{{\mathbb M}\overline B}
=-\delta_{\epsilon''_{12}}E^{\mathbb M}{}_{\underline a}\, ,
\eea
where $\epsilon''_{12}$ is defined in \eqref{susyParameter1}. A similar result is obtained  for $E^{\mathbb M}{}_{\overline A}$.

\item Mixed double Lorentz and supersymmetry variations of the frame 
\bea
\delta_{[\Gamma,\epsilon ]}  E^{\mathbb M} {}_{\underline a}
&=&\frac14\overline\epsilon_{[1}\Gamma_{2]\underline {bc}}\gamma^{\underline {bc}}\gamma_{\underline a}\psi_{\overline B}E^{{\mathbb M}\overline B}=-\delta_{\epsilon'_{12}}E^{\mathbb M}{}_{\underline a}\, ,
\eea
where $\epsilon'_{12}$ is defined in \eqref{LorParameter1}. A similar result is obtained for $E^{\mathbb M}{}_{\overline A}$.

\item Mixed diffeomorphisms and supersymmetry transformations of the gravitino
\bea
\delta_{[\epsilon, \xi]}  \Psi_{\overline{A}} &=& E_{\overline A}(2\xi_{[2}^M\partial_M\epsilon_{1]})-\frac12\omega_{\overline A\underline {bc}}\gamma^{\underline {bc}}\xi_{[2}^M\partial_M\epsilon_{1]}=-\nabla_{\overline A}\epsilon''_{12}=-\delta_{\epsilon''_{12}}\Psi_{\overline A}\, .
\eea

\item Mixed  supersymmetry and double Lorentz transformations of the gravitino 
\bea
\delta_{[\Gamma,\epsilon]}  \Psi_{\overline{A}} &=&\frac12\nabla_{\overline A}\left(\Gamma_{[2\underline {bc}}\gamma^{\underline {bc}}\epsilon_{1]}\right)\equiv -\nabla_{\overline A}\epsilon'_{12}=-\delta_{\epsilon'_{12}}\Psi_{\overline A}
\eea

\item Diffeomorphisms and double Lorentz transformations of the gravitino
\bea
\delta_{[\Gamma, \xi]}  \Psi_{\overline{A}} 
&=&-\left(\delta_{\Gamma''_{12}}+\delta_{\xi''_{12}}\right)\Psi_{\overline A}\, .
\eea

 \item Mixed supersymmetry and double Lorentz transformations of the dilatino 
\bea
\delta_{[\Gamma,\epsilon]}  \rho =-\frac12\gamma^a\nabla_{ \underline a}(\Gamma_{[2\underline {bc}}\gamma^{\underline {bc}}\epsilon_{1]})= \gamma^{\underline a}\nabla_{\underline  a}\epsilon'_{12}=-\delta_{\epsilon'_{12}}\rho\, .
\eea

\item Diffeomorphisms and double Lorentz transformations of the dilatino
\bea
\delta_{[\Gamma, \xi]}  \rho =-(\delta_{\Gamma''_{12}}+\delta_{\xi''_{12}})\rho\, .\, .
\eea

 \item Mixed diffeomorphisms and supersymmetry  transformations of the dilatino
\bea
\delta_{[\xi, \epsilon]} \rho = \gamma^{\underline a}\nabla_{\underline  a}\epsilon''_{12}=-\delta_{\epsilon''_{12}}\rho\, .\, .
\eea

\end{itemize}

Summarizing we have found, up to  bi-linear  terms in  fermions,
\begin{subequations}\label{zeroalgebra}
\begin{align}
E^{\mathbb M}{}_{\overline C}\big [\delta_1,\delta_2\big ]E_{{\mathbb M}\underline a}&=-E^{\mathbb M}{}_{\overline C}\left(\delta_{\xi_{12}}+\delta_{\Gamma_{12}}+\delta_{\epsilon_{12}}\right)E_{{\mathbb M}\underline a}\, , \\
 E^{\mathbb M}{}_{\underline c}\big [\delta_1,\delta_2\big ]E_{{\mathbb M}\overline A}&=-E^{\mathbb M}{}_{\underline c}\left(\delta_{\xi_{12}}+\delta_{\Gamma_{12}}+\delta_{\epsilon_{12}}\right)E_{{\mathbb M}\overline A}\, ,\\
E^{\mathbb  M}{}_{{\overline B}}\big[\delta_{1},\delta_{2}\big]E_{{\mathbb M}{ \overline A}}&= -
E^{\mathbb M}{}_{\overline C}\delta_{\Gamma_{12}} E_{{\mathbb M}\overline A}\, ,\\
E^{\mathbb M}{}_{\underline c}\big[\delta_{1},\delta_{2}\big]E_{{\mathbb M} {\underline a}}& =  -
E^{\mathbb N}{}_{\underline c}\delta_{\Gamma_{12}} E_{{\mathbb N}\underline a}\, ,\\
\big [\delta_1,\delta_2\big ]d&=-\left(\delta_{\xi_{12}}+\delta_{\epsilon_{12}}\right)d\, ,\\
\big [\delta_1,\delta_2\big ]\Psi_{\overline A}&=-\left(\delta_{\xi''_{12}}+\delta_{\Gamma_{12}}+\delta_{\epsilon_{12}}\right)\Psi_{\overline A}\, ,\\
\big [\delta_1,\delta_2\big ]\rho&=-\left(\delta_{\xi''_{12}}+\delta_{\Gamma_{12}}+\delta_{\epsilon_{12}}\right)\rho\, ,
\end{align}
\end{subequations}
where $\delta_1=\delta_{\xi_1}+\delta_{\epsilon_1}+\delta_{\Gamma_1}$ and $\xi^{\mathbb M}_{12}=\xi'^{\mathbb M}_{12}+\xi''^{\mathbb M}_{12}$, $\Gamma_{12{\mathbb AB}}=\Gamma'_{12{\mathbb AB}}+\Gamma''_{12{\mathbb AB}}$, $\epsilon_{12}=\epsilon'_{12}+\epsilon''_{12}$. The commutator of supersymmetry variations on the gravitino and  dilatino  as well as the missing terms $\delta_{\xi'_{12}}\rho$ and $\delta_{\xi'_{12}}\Psi_{\overline A}$ are not included as they are of higher order in fermions.

\subsection{First order algebra }
\label{App first}
We now work out the algebra of  first order transformations \eqref{transf1}  and show that it closes with the parameters \eqref{param1}, up to terms with two fermions.
Here we denote $\delta\equiv\delta^{(0)}+\delta^{(1)}$ and  $[\delta_1,\delta_2]=\delta^{(1)}_1\delta^{(0)}_2+\delta^{(0)}_1\delta^{(1)}_2-(1\leftrightarrow 2)=-\delta_{12}^{(1)}$.
We split the algebra as in the previous section.

\noindent $-$  Double Lorentz transformations on the generalized frame 
\bea
\big [\delta_{\Lambda_1},\delta_{\Lambda_2}\big] E_{\mathbb M}{}^{\overline{A}} 
&=&  \frac b 2 \left[ \delta_{\Lambda_1}\left(\mathcal{F}_{\underline{\mathbb M}}^{* {} \ \overline{C} \overline{D}}\right) E_{ N}{}^{\overline{A}} \partial^{N} \Lambda_{2 \overline{C} \overline{D}} - \delta_{\Lambda_2}\left(\mathcal{F}_{\underline{\mathbb M}}^{* {} \ \overline{C} \overline{D}}\right) E_{ N}{}^{\bar{A}}\partial^{ N} \Lambda_{1 \overline{C} \overline{D}} \right] \, . 
\eea
Rewriting
\bea
\delta_{\Lambda_1}\left(\mathcal{F}_{\underline{\mathbb M}}^{* {} \ \overline{C} \overline{D}}\right) E_{ N}{}^{\overline{A}} \partial^{ N} \Lambda_{2 \overline{C} \overline{D}} &=& \big(-\partial_{\underline{\mathbb M}} \Lambda_{1}^{\overline{C} \overline{D}} + 2 \mathcal{F}^*_{\underline{\mathbb M}}{}^{\overline{B} \overline{D}} \Lambda_{1 \overline{B}}{}^{\overline{C}} \big) E_{ N}{}^{\overline{A}} \partial^{ N} \Lambda_{2 \overline{C} \overline{D}} \, ,
\eea
with $\partial_{\mathbb M}=\partial_{\underline{\mathbb M}}+\partial_{\overline {\mathbb M}}$ and 
\be
-2 E^{{\mathbb P} \overline{A}} \partial_{\mathbb M} \Lambda_{[1}{}^{\overline{CD}} \partial_{\mathbb P}\Lambda_{2] \overline{CD}} \ = \ E^{{\mathbb P} \overline{A}} \left[\partial_{\mathbb M}\left(-\Lambda_{1}{}^{\overline{C} \overline{D}} \partial_{\mathbb P}\Lambda_{2 \overline{C} \overline{D}}\right) + \partial_{\mathbb P} \left(\Lambda_{1}{}^{\overline{C} \overline{D}} \partial_{\mathbb M}\Lambda_{2 \overline{C} \overline{D}}\right)  \right]\, ,
\ee
we get
\bea
\big [\delta_{\Lambda_1},\delta_{\Lambda_2}\big] E_{\mathbb M}{}^{\overline{A}} =-\left(\delta_{\Lambda^{(1)'}_{12}}+\delta_{\xi^{(1)'}_{12}}\right)E_{\mathbb M}{}^{\overline{A}}\, , 
\eea
where
\be
\xi^{(1)'}_{12 M} \ = \ b \Lambda_{[1}^{\overline{C} \overline{D}} \partial_{M} \Lambda_{2] \overline{C} \overline{D}}\, , \quad
\Lambda^{(1)'}_{12 \overline{A} \overline{B}} \ = \ \frac{b}{2} E_{\overline{B}} \Lambda_{[1}^{\overline{C} \overline{D}} E_{\overline{A}} \Lambda_{2] \overline{C} \overline{D}}\, . \label{xi12a}
\ee

Repeating the procedure for $E_{\mathbb M}{}^{\underline a}$, we  find  
\bea
\big [\delta_{\Lambda_1},\delta_{\Lambda_2}\big] E_{\mathbb M} {}^{{\underline a}}=-\left(\delta_{\Lambda^{(1)'}_{12}}+
\delta_{\xi^{(1)'}_{12}}\right)E_{\mathbb M} {}^{\underline a}\, , 
\eea
with $\xi^{(1)'\mathbb M}_{12}$  defined in \eqref{xi12a} and 
\be
\Lambda^{(1)'}_{12 \underline {ab}} \ = \ \frac{b}{2} E_{\underline {b}} \Lambda_{[1}^{\overline{C} \overline{D}} E_{\underline {a}} \Lambda_{2] \overline{C} \overline{D}}\, .
\ee

\noindent $-$ Mixed supersymmetry and double Lorentz transformations on the generalized frame 

Using 
\be
\delta^{(0)}_{\epsilon_1} \mathcal{F}^{*}_{\underline{\mathbb M}}{}^{\overline{C} \overline{D}} \ = \ - \overline{\epsilon}_{1} \gamma^{b} \big( \frac{1}{2} \Psi^{\overline{A}} E_{{\mathbb M} \overline{A}} \mathcal{F}_{\underline b}{}^{\overline{CD}} + E_{{\mathbb M}\underline  b} {\nabla}^{[\overline D} \Psi^{\overline{C}]} + \frac12 E_{{\mathbb M}\underline  b} \Psi^{\overline{A}} \mathcal{F}^{\overline{CD}}{}_{\overline{A}} \big), 
\ee
 we get the first order contribution to  the mixed transformation rules of $E_{\mathbb M}{}^{\overline{A}}$ 
\bea
\delta_{[\epsilon,\Lambda]} E_{\mathbb M}{}^{\overline{A}}&=& \frac b 2 \left[ -\frac12 \overline{\epsilon}_1 \gamma^{\underline b} \Psi^{\overline{B}} E_{{\mathbb M} \overline{B}} \mathcal{F}_{\underline b}{}^{\overline{C} \overline{D}} E_{{ N}}{}^{\bar{A}} \partial^{ N} \Lambda_{2 \overline{C} \overline{D}} - \frac12 \overline{\epsilon}_2 \gamma^{\underline b} \Psi^{\overline{A}} \partial_{\overline{\mathbb M}} \Lambda_{1}^{\overline{C} \overline{D}} \mathcal{F}_{\underline b \overline{C} \overline{D}} \right. \notag \\ && + \frac{1}{16} \overline{\epsilon}_2 \gamma^{\underline b} \Lambda_{1\underline {cd}} \gamma^{\underline {cd}} \Psi^{\overline{A}} \mathcal{F}_{\underline{\mathbb M}}{}^{\overline{C} \overline{D}} \mathcal{F}_{\underline b \overline{C} \overline{D}} - \frac14 \overline{\epsilon}_2 \gamma^{\underline b} \Psi^{\overline{A}} {\cal F}_{\underline{\mathbb M}}{}^{\overline{C} \overline{D}} {\cal F}_{\underline a \overline{C} \overline{D}} \Lambda_{1}^{\underline a}{}_{\underline b} \notag \\ && + \frac14 \mathcal{F}_{\underline{\mathbb M}}{}^{\overline{C} \overline{D}} \overline{\epsilon}_1 \gamma^{\underline b} \Psi^{\overline{A}} E^{ N}{}_{\underline b} \partial_{ N} \Lambda_{2 \overline{C} \overline{D}} \notag + \frac14 \overline{\epsilon}_2 \gamma^{\underline b} \Psi^{\overline{A}} \partial_{\underline{\mathbb M}} \Lambda_{1}^{\overline{C} \overline{D}} \mathcal{F}_{\underline b \overline{C} \overline{D}} \\ && \left. - \frac18 \overline{\epsilon}_2 \gamma^{\underline b} E^{{N}}{}_{\underline a} \partial_{ N} \Lambda_{1 \overline{CD}} \mathcal{F}_{\underline c}{}^{\overline{CD}} \gamma^{\underline {ac}} \Psi^{\overline{A}} E_{{\mathbb M}\underline b}  - (1 \leftrightarrow 2) \right] \, .
\label{inter2}
\eea
The first two terms  are  a Lorentz transformation with parameter 
\be
\Lambda_{12\overline{A} \overline{B}}^{(1)''} \ = \ b \ \overline{\epsilon}_{[1} \gamma^{\underline b} \Psi_{[\overline{A}} E^{N}_{\overline{B}]} \partial_{ N} \Lambda_{2]}^{\overline{C} \overline{D}} \mathcal{F}_{\underline b \overline{C} \overline{D}}\, . 
\ee
From the second line,  only one term survives after commuting the gamma matrices, which corresponds to a first order supersymmetric variation with zeroth order parameter
$
\overline{\epsilon}'_{12} \ = \ - \frac12 \overline{\epsilon}_{[1} \gamma^{\underline {cd}} \Lambda_{2]\underline {cd}}$.

In the same way, from the remaining terms  we find a first-order supersymmetry parameter
\be
\overline{\epsilon}^{(1)'}_{12} \ = \ \frac{b}{4}  \overline{\epsilon}_{[1} E^{M}{}_{\underline a} \partial_{M} \Lambda_{2] \overline{C} \overline{D}} \mathcal{F}_{\underline c}{}^{\overline{C} \overline{D}} \gamma^{\underline {ac}}\, .\label{firstep}
\ee

Consider now  the component $E_{\mathbb M}{}^{\underline a}$
 \bea
\delta_{[\epsilon,\Lambda]} E_{\mathbb M}{}^{{\underline a}}
&=& \frac{b}{2} \left[ - \frac12 \overline{\epsilon}_1 \gamma^{\underline c} \Psi^{\overline{B}} E_{{\mathbb M}\underline c} E^{N}{}_{\overline{B}} \partial_{ N} \Lambda_{2 \overline{C} \overline{D}} \mathcal{F}^{\underline a \overline{C} \overline{D}} - \frac12 \overline{\epsilon}_2 \gamma^{\underline a} \Psi^{\overline{B}} \mathcal{F}_{\underline{\mathbb M}}{}^{\overline{C} \overline{D}} E^{ N}{}_{\overline{B}} \partial_{N} \Lambda_{1 \overline{C} \overline{D}} \right. \notag \\ && + \frac14 \overline{\epsilon}_2 \gamma^{\underline b}(-\frac14 \Lambda_{1\underline {cd}} \gamma^{\underline {cd}} \Psi_{\overline{B}}) \mathcal{F}_{\underline b \overline{C} \overline{D}} \mathcal{F}^{\underline a \overline{C} \overline{D}} E_{\mathbb M}{}^{\overline{B}} - \frac14 \overline{\epsilon}_2 \gamma^{\underline b} \Psi_{\overline{B}} \mathcal{F}_{\underline b \overline{C} \overline{D}} E_{N}{}^{\underline a} \partial^{N} \Lambda_{1}^{\overline{C} \overline{D}} E_{\mathbb M}{}^{\overline{B}} \notag \\ &&+  \frac12 E_{\mathbb M}{}^{\overline{B}} \overline{\epsilon}_2 \gamma^{\underline c} \Psi_{\overline{B}} E^{ N}{}_{\underline c} \partial_{ N} \Lambda_{1 \overline{C}}{}^{\overline{D}} \mathcal{F}^{\underline a \overline{C}}{}_{\overline{D}} - \frac12 \overline{\epsilon}_2 \gamma^{\underline a} E_{{\mathbb M} \overline{B}}(-\frac14 E^{N}{}_{\underline b} \partial_{N} \Lambda_{1 \overline{C} \overline{D}} \mathcal{F}_{\underline c}{}^{\overline{C} \overline{D}} \gamma^{\underline {bc}} \Psi^{\overline{B}}) \notag \\ && \left. + \frac14 \overline{\epsilon}_2 \gamma^{\underline b} \Psi_{\overline{B}}(- E^{N}{}_{\underline b}\partial_{N} \Lambda_{1 \overline{C} \overline{D}} + \mathcal{F}_{\underline c \overline{CD}} \Lambda_{1}^{\underline c}{}_{\underline b}) \mathcal{F}^{\underline a \overline{C} \overline{D}} E_{\mathbb M}{}^{\overline{B}} - (1 \leftrightarrow 2) \right] \, .
\label{inter5}
\eea
 The  first line is a zeroth order Lorentz transformation with parameter    
\be 
\Lambda_{12\underline {ab}}^{(1)''} \ = \  b \ \overline{\epsilon}_{[1} \gamma_{[\underline a} \Psi^{\overline{B}} \mathcal{F}_{\underline b] }{}^{\overline{C} \overline{D}} E^{M}{}_{\overline{B}} \partial_{ M} \Lambda_{2] \overline{C} \overline{D}}\, .
\ee
 Commuting the gamma matrices  in the first term of the second line, the second contribution in the  fourth line is canceled and we get again a supersymmetry transformation with zeroth order parameter
$
\overline{\epsilon}''_{12}  =  - \frac12 \overline{\epsilon}_{[1} \gamma^{\underline {cd}} \Lambda_{2]\underline {cd}}\, .
$
Finally, commuting the gamma matrices  in the second term of the third line,   various cancellations  leave a supersymmetry transformation with first order parameter
\eqref{firstep}.

\noindent $-$  Supersymmetry variations  on the generalized frame
\bea
E^{\mathbb M}{}_{\underline c} \big[ \delta_{\epsilon_1} , \delta_{\epsilon_2}  \big] E_{{\mathbb M} \overline{A}} & = & \frac{b}{2} \left[  -\frac12 \overline{\epsilon}_2 \big( E^{\mathbb N}{}_{[\underline d} E^{ P}{}_{\overline{A}} \partial_{P} \mathcal{F}_{\underline{\mathbb N}}{}^{\overline{CD}} \mathcal{F}_{\underline c] \overline{CD}} + E^{\mathbb N}{}_{\overline{A}} E^{ P}{}_{[\underline c} \partial_{P} \mathcal{F}_{\mathbb N}{}^{\overline{CD}} \mathcal{F}_{\underline d] \overline{CD}} \notag \right. \\ && + \mathcal{F}^{\underline{N}}{}_{\overline{CD}} \mathcal{F}_{[\underline c}{}^{\overline{CD}} \partial_{ N} E^{\mathbb P}{}_{\underline d]} E_{{\mathbb P} \overline{A}} \big) \gamma^{\underline d} \epsilon_1 + \frac14 \overline{\epsilon}_2 \mathcal{F}_{\overline{A}}{}^{\underline b}{}_{\underline d} \mathcal{F}_{\underline c}{}^{\overline{C} \overline{D}} \mathcal{F}_{\underline b \overline{C} \overline{D}} \gamma^{\underline d} \epsilon_1 \notag \\ && \left. + \frac14 \overline{\epsilon}_2 E^{P}{}_{\overline{A}} \partial_{P} \mathcal{F}_{\underline d}{}^{\overline{CD}} \mathcal{F}_{\underline c}{}^{\overline{CD}} \gamma^{\underline d} \epsilon_1 - \frac14 E^{P}{}_{\overline{A}} \partial_{P} \big( \overline{\epsilon}_2 \gamma^{\underline b} \epsilon_1 \mathcal{F}_{\underline b \overline{CD}} \big) \mathcal{F}_{\underline c}{}^{\overline{CD}}-(1\leftrightarrow 2) \right] \, . \nn
\eea

The first and last terms of the r.h.s. combine into a  Lorentz transformation with parameter
\be
\Lambda^{(1)'''}_{12 \overline{A} \overline{B}} \ = \  \frac{b}{4} \overline{\epsilon}_1 \gamma^{\underline c} \epsilon_2 \mathcal{F}_{\underline c \overline{A} \overline{B}}\, , 
\ee
while the other terms  form a diffeomorphism with  first order parameter
\be
\xi^{(1)''}_{12{\mathbb M}} \ = \ \frac{b}{8} \mathcal{F}_{{\mathbb M} \overline{C} \overline{D}} \mathcal{F}_{\underline b}{}^{\overline{C} \overline{D}} \overline{\epsilon}_1 \gamma^{\underline b} \epsilon_2 \, .
\ee

The same result holds for  $E^{\mathbb M}{}_{\overline{C}} \big[ \delta_{\epsilon_1} , \delta_{\epsilon_2} \big] E_{{\mathbb M}\underline a}$, while
\be
E^{\mathbb M}{}_{\overline{C}} \big[ \delta_{\epsilon_1} , \delta_{\epsilon_2} \big] E_{{\mathbb M} \overline{A}} \ = \ 0
\, , \qquad
E^{\mathbb M}{}_{\underline c} \big[ \delta_{\epsilon_1} , \delta_{\epsilon_2} \big] E_{{\mathbb M}\underline a} \ = \ 0.
\ee

\noindent $-$ Mixed diffeomorphism and Lorentz variations of the generalized frame

Recalling that diffeomorphisms are not deformed, we get to first order 
\be
\delta_{[\Lambda,\xi]}
E_{{\mathbb M}\mathbb A} \ = \ b \ E^{P}{}_{\mathbb A} \partial_{ [P} (2 \xi_{[1}^{ {N}} \partial_{N} \Lambda_{2 ]\overline{CD}}) \mathcal{F}^*_{\underline{\mathbb M]}}{}^{\overline{CD}} 
\, ,
\ee
which is a first-order Lorentz transformation with a zeroth order parameter.
We use the convention $A_{[\overline{A}} B_{\underline {b}]} \ = \ \frac12 A_{\overline{A}} B_{\underline {b}} - \frac12 A_{\overline{B}} B_{\underline {a}}$ to interchange projected indices.

\noindent $-$ Mixed diffeomorphism and supersymmetry variations  on the generalized frame

This case is similar to the previous one. We start with 
\be
\delta_{[\epsilon,\xi]}
E_{{\mathbb M}\overline A}  = \frac{b}{2} \left(- \frac14 \xi_2^P \partial_{P} \overline{\epsilon}_1 \right) \gamma^{\underline b} \Psi_{\overline{A}} \mathcal{F}_{\underline{\mathbb M}}{}^{\overline{CD}} \mathcal{F}_{\underline b \overline{CD}} - (1 \leftrightarrow 2) \, ,
\ee
which is a first order supersymmetry transformation with a zeroth order parameter. It is straightforward to see that the same result holds for $E_{{\mathbb M}\underline a}$.

\noindent $-$ Double Lorentz variations on the generalized gravitino 
\bea
 \left[ \delta_{\Lambda_1} , \delta_{\Lambda_2} \right] \Psi_{\overline{A}}  &=& \frac{b}{2} \left[ \Lambda_2^{\overline{B}}{}_{\overline{A}} \left(\delta_{\Lambda_1}^{(1)} \Psi_{\overline{B}}\right)- \frac14 \Lambda_{2\underline {bc}} \gamma^{\underline {bc}} \left(\delta_{\Lambda_1}^{(1)} \Psi_{\overline{A}}\right) \right.\notag \\ && - \delta_{\Lambda_1}^{(0)}\left(\frac14 \left(E^{M}{}_{\underline b} \partial_{M} \Lambda_{2 \overline{CD}} \mathcal{F}_{\underline c}{}^{\overline{CD}} \gamma^{\underline {bc}} \Psi_{\overline{A}}\right) +\left(2 \nabla^{\overline{D}} \Psi^{\overline{C}} - \omega_{\overline{E}}{}^{\overline{DC}} \Psi^{\overline{E}}\right) E^{ M}{}_{\overline{A}} \partial_{M} \Lambda_{2 \overline{CD}}\right) \notag \\ && \left.\ \ - \left(2 {\nabla}^{\overline{D}} \Psi^{\overline{C}} - \omega_{\overline{E}}{}^{\overline{DC}} \Psi^{\overline{E}}\right) \delta^{(0)}_{\Lambda_1} \left(E^{ M}{}_{\overline{A}} \partial_{M} \Lambda_{2 \overline{CD}}\right) \right]- (1\leftrightarrow2)
\, .
\eea
After some straightforward manipulations, we finally obtain  Lorentz transformations with the following parameters
\bea
\Lambda_{12 \overline{AB}} = - 2 \Lambda_{[1 \overline{A}}{}^{\overline{C}} \Lambda_{2] \overline{CB}}\, , \ \ \  \Lambda^{(1)'}_{12 \overline{AB}} \ = \ \frac b 2 E_{\overline{B}} \Lambda_{[1}{}^{\overline{CD}} E_{\overline{A}} \Lambda_{2] \overline{CD}}\, \ \
{\rm and} \  \ 
\Lambda^{(1)'}_{12\underline {ab}} \ = \ \frac b 2 E_{\underline b} \Lambda_{[1 \overline{CD}} E_{\underline a} \Lambda_{2]}{}^{\overline{CD}}\nn
\eea

\noindent $-$ {Mixed Lorentz and supersymmetry transformations on the generalized gravitino}

\bea
\delta_{[\Lambda,\epsilon]} \Psi_{\overline A} &=& \frac{b}{2} \left[ \Lambda^{\overline B}_{2\overline A}\delta_{\epsilon_1}^{(1)}\Psi_{\overline B}+\frac1{16}\Lambda_{2\underline {ab}}\gamma^{\underline {ab}} {\cal F}^{(3)}_{\overline A\underline { cd}}\gamma^{\underline {cd}}\epsilon_1
+\delta_{\Lambda_1}^{(1)}E^M{}_{\overline A}\partial_M\epsilon_2 \right. \nn\\
&& \ -\frac14 E^{M}{}_{\underline b} \partial_{M} \Lambda_{2\overline {CD}}{\cal F}_{\underline c}{}^{\overline {CD}}\gamma^{\underline {bc}}\nabla_{\overline A}\epsilon_1 -\frac14\delta_{\Lambda_1}^{(1)}{\cal F}_{\overline {A}\underline {bc}} \gamma^{\underline {bc}}\epsilon_2 \notag \\ && \ -2\delta_{\epsilon_1}^{(0)}(\nabla^{\overline D}\Psi^{\overline C}) E^{ M}{}_{\overline{A}} \partial_{M} \Lambda_{2\overline{CD}} + {\omega}_{\overline B}{}^{\overline {DC}}\delta_{\epsilon_1}^{(0)}\Psi^{\overline B} E^{M}{}_{\overline{A}} \partial_{ M}\Lambda_{2\overline {CD}} \notag \\ && \ \left. -\frac14\delta_{\Lambda_1}^{(0)} {\cal F}^{(3)}_{\overline {A}\underline {bc}}\gamma^{\underline {bc}}\epsilon_2-(1\leftrightarrow 2) \right] \, .
\eea
Commuting the gamma matrices in the second term of the r.h.s, and combining it with the corresponding term in the ($1\leftrightarrow 2$) operation, we recognize a supersymmetry transformation with zeroth order parameter $\epsilon'_{12}=-\frac12 \bar{\epsilon}_{[1} \gamma^{\underline {ab}} \Lambda_{2]\underline {ab}}$.

The first term in the second line together with  the corresponding term in the ($1\leftrightarrow 2$) operation, gives a zeroth order supersymmetry transformation with first order parameter $\epsilon^{(1)'}_{12}=\frac{b}{4}\gamma^{\underline {bc}}\epsilon_{[1} E^{M}{}_{\underline b}\partial_{ M}\Lambda_{2]\overline {CD}}{\cal F}_{\underline c}{}^{\overline {CD}}$. The remaining terms cancel and then we get
\bea
\delta_{[\Lambda,\epsilon]}\Psi_{\overline A}&=&-\left(\delta^{(0)}_{\epsilon^{(1)'}_{12}}+\delta_{\epsilon'_{12}}^{(1)}\right)\Psi_{\overline A}
\eea
up to terms with two fermions.

\noindent $-$ {Mixed diffeomorphisms and supersymmetry transformations on the generalized gravitino}
\bea
\delta_{[\xi,\epsilon]} \Psi_{\overline A}&=&\frac b 4 \xi_{[1}^{ M} {\cal F}^{(3)}_{\overline A \underline {bc}} \gamma^{\underline {bc}}\partial_{M} \epsilon_{2]}= -\delta_{\epsilon_{12}}^{(1)}\Psi_{\overline A}\, ,
\eea
with $\epsilon_{12}=2\xi_{[1}^{M} \partial_{M}\epsilon_{2]}$.

\noindent $-$  Double Lorentz variations on the generalized dilatino 
\bea
\big[ \delta_{\Lambda_1} , \delta_{\Lambda_2} \big] \rho &=& - \frac14 \Lambda_{2\underline {ab}} \gamma^{\underline {ab}} \big(-\frac14 E^{M}{}_{\underline b} \partial_{ M} \Lambda_{1 \overline{CD}} \mathcal{F}_{\underline d}{}^{\overline{CD}} \gamma^{\underline {bd}} \rho \big)  - \frac14 \Lambda_1{}^{\underline c}{}_{\underline b} E^{ M}{}_{\underline c} \partial_{ M} \Lambda_{2 \overline{CD}} \mathcal{F}_{\underline d}{}^{\overline{CD}} \gamma^{\underline {bd}} \rho \notag \\ && - \frac14 E^{ M}{}_{\underline b} \partial_{ M} \Lambda_{2 \overline{CD}} \big( - E^{N}{}_{\underline c} \partial_{N} \Lambda_1{}^{\overline{CD}} + \mathcal{F}_{\underline a}{}^{\overline{CD}} \Lambda_1{}^{\underline a}{}_{\underline c} + 2 \mathcal{F}_{\underline c}{}^{\overline{BD}} \Lambda_{1 \overline{B}}{}^{\overline{C}} \big) \gamma^{\underline {bc}} \rho \notag \\ && - \frac14 E^{M}{}_{\underline b} \partial_{M} \Lambda_{2 \overline{CD}} \mathcal{F}_{\underline c}{}^{\overline{CD}} \gamma^{\underline {bc}} (-\frac14 \Lambda_{1\underline {ad}} \gamma^{\underline {ad}} \rho)-(1\leftrightarrow 2)\, .
\eea
In the second line (adding the ($1\leftrightarrow 2$) operation) we recognize a Lorentz transformation with  first and zeroth order parameters
\be
\Lambda^{(1)'}_{12\underline {ab}} \ = \ \frac b 2 E_{\underline a} \Lambda_{[1 \overline{CD}} E_{\underline b} \Lambda_{2]}{}^{\overline{CD}}\, \quad {\rm and} \quad
\Lambda_{12 \overline{AB}} \ = \ -2 \Lambda_{[2\overline{CB}} \Lambda_{1] \overline{A}}{}^{\overline{C}}.
\ee
Commuting the gamma matrices of the third line, it is straightforward to see that the remaining terms cancel.

\noindent $-$ {Mixed Lorentz and supersymmetry transformations on the generalized dilatino}

This computation is similar to the one associated to the gravitino. We find the following supersymmetry parameters 
\bea
\epsilon'_{12}=-\frac12 \bar{\epsilon}_{[1} \gamma^{\underline {ab}} \Lambda_{2]\underline {ab}} \quad {\rm and}\quad\epsilon^{(1)'}_{12}=\frac{b}{4}\gamma^{\underline {bc}}\epsilon_{[2} E^{M}{}_{\underline b}\partial_{M} \Lambda_{1]\overline {CD}}{\cal F}_{\underline c}{}^{\overline {CD}}\, , 
\eea
so that finally
\bea
\delta_{[\epsilon,\Lambda]}\rho=-\delta^{(1)}_{\epsilon_{12}}\rho\, .
\eea

\noindent $-$ {Mixed diffeomorphism and supersymmetry transformations on the generalized dilatino}

\bea
\delta_{[\xi,\epsilon]} \rho &=& \xi_2^M\partial_M\left(-\frac1{12} {\cal F}^{(3)}_{\underline {abc}}\gamma^{\underline {abc}} \epsilon_1   -\frac14\left(\omega_{\underline {cd}}{}^{\underline c} {\cal F}^{\underline d}{}_{\overline{CD}}{\cal F}_{\underline a}{}^{\overline {CD}}+ E^{N}{}_{\underline d} \partial_{N}({\cal F}^{\underline d}{}_{\overline {CD}}{\cal F}_{\underline a}{}^{\overline{CD}})\right)\gamma^{\underline a}\epsilon_1 \right)\nn\\
&&-(1\leftrightarrow 2) \nn \\ &=& -\delta_{\epsilon_{12}}^{(1)}\rho\, .
\eea

In equations \eqref{param1} of the main text we collect the parameters that appear in this algebra of first order transformation rules.

\section{Supersymmetry of heterotic string effective action}\label{app:susyac}
In the first part of this appendix we prove that the higher-derivative deformations of the transformation rules of the supergravity fields satisfy a closed algebra up to ${\cal O}(\alpha')$ and up to terms with two fermions. In the second part,  we show that the action \eqref{faction} is invariant under these  supersymmetry transformations.

\subsection{Supersymmetry algebra}\label{app:algehete}

It is well known that the algebra of leading order transformations of  supergravity and super Yang-Mills fields closes. Moreover,
the replacement $H_{\mu\nu\rho}\rightarrow\wt H_{\mu\nu\rho}$ in the supersymmetry transformations of the gravitino and dilatino does not affect the  leading order closure on any field except for  the $b$-field.  Hence we focus on the algebra of first order transformation rules on $b_{\mu\nu}$.

 It is convenient to first look at the brackets acting on  ${\wt b}_{\mu\nu}=b_{\mu\nu}+\frac b8A_{[\mu}^k\ov\chi^i\gamma_{\nu]}\chi^jf_{ijk}$. Up to first order and bilinear terms in fermions, we need the following transformation rules:
\begin{subequations}\label{tranfru}
\begin{align}
 \delta\psi_a & =  \psi_b\Lambda^b{}_a-\frac14\gamma^{bc}\Lambda_{bc}\psi_a +\partial_{\mu} \epsilon -\frac14\widetilde w^{(+)}_{\mu ab}\gamma^{ab}\epsilon \, , \\
\delta A_\mu^i &= \partial_\mu\xi^i+f^i{}_{jk}\xi^jA_\mu^k+\frac12\bar\epsilon\gamma_\mu\chi^i\, ,   \\
\delta\chi& =f^i{}_{jk}\xi^j\chi^k -\frac14\Lambda_{bc}\gamma^{bc}\chi-\frac14 F_{\mu\nu}^i\gamma^{\mu\nu}\epsilon \, ,  \\
\delta \wt b_{\mu \nu} & = 2\partial_{[\mu}\xi_{\nu]}-\zeta\partial_{[\mu}\xi^i A_{\nu]i}  - \frac{b}{2} \left(\partial_{[\mu} \Lambda^{CD} \hat{\Omega}_{\nu]CD} +\bar\epsilon\gamma_{[\mu}\Psi^{CD}\hat{\Omega}_{\nu]CD}\right)\, ,\\
\delta\hat{\Omega}_{\mu CD} & = - \partial_{\mu}\Lambda_{CD} + 2\hat{\Omega}_{\mu E[D}\Lambda^{E}{}_{C]} + \ov{\epsilon}\gamma_{\mu}\Psi_{CD}   = - {\mathscr D}_{\mu}\Lambda_{CD} + \ov{\epsilon}\gamma_{\mu}\Psi_{CD}\, ,\\
\delta\hat{{\mathscr R}}_{\mu\nu CD} & =  2\hat{{\mathscr R}}_{\mu\nu E[D}\Lambda^{E}{}_{C]} - 2{\mathscr D}_{[\mu}\left(\ov{\epsilon}\gamma_{\nu}\Psi_{CD}\right)\, ,\\
\delta\Psi_{CD} &= 2\Psi_{E[D}\Lambda^{E}{}_{C]} + \frac{1}{8}\hat{{\mathscr R}}_{\mu\nu CD}\gamma^{\mu\nu}\epsilon\, .\label{grav_curv_transf}
\end{align}
\end{subequations}

We exclude the diffeomorphisms since it is trivial to see that all the transformation rules of $b_{\mu\nu}$ (i.e. Lorentz, supersymmetry, abelian and non-abelian gauge transformations) transform as tensors under  diffeomorphisms and hence their commutators  are trivial. Therefore, we  compute the brackets
\bea
\label{clausura}
\left([\delta_{1},\delta_{2}]\wt{b}_{\mu\nu}\right)^{(1)} & = &\left(\delta^{(1)}_{1}\delta^{(0)}_{2} - \delta^{(1)}_{2}\delta^{(0)}_{1}\right)\wt{b}_{\mu\nu} + \left(\delta^{(0)}_{1}\delta^{(1)}_{2} - \delta^{(0)}_{2}\delta^{(1)}_{1}\right)\wt{b}_{\mu\nu}\, .
\eea The first term in the r.h.s. gives \bea
\label{primero}
\delta^{(1)}_{1}\delta^{(0)}_{2}\wt{b}_{\mu\nu} - (1\leftrightarrow 2) 
& = & \frac{3\ap}{4}\ov{\epsilon}_{2}\gamma^{\lambda}\epsilon_{1}\hat{\mathscr{C}}_{\mu\nu\lambda}\, ,
\eea
and the second one can be written as
\bea
\label{segundo}
\delta^{(0)}_{1}\delta^{(1)}_{2}\wt{b}_{\mu\nu} - (1\leftrightarrow 2) & = & \ap\partial_{[\mu}\left(\Lambda_{2}^{CD}\partial_{\nu]}\Lambda_{1CD}\right) + \ap\partial_{[\mu}\left(\Lambda_{1}^{CD}\Lambda_{2}^{E}{}_{C}\right)\hat{\Omega}_{\nu]ED} \nn \\
& & + \frac{\ap}{4}\ov{\epsilon}_{2}\gamma^{\lambda}\epsilon_{1}\hat{\Omega}_{[\mu}{}^{CD}\hat{{\mathscr R}}_{\nu]\lambda CD}-\frac {\ap}{2}\partial_{[\mu}\left(\ov\epsilon_2\gamma^\sigma\epsilon_1\hat\Omega_{c\sigma}^i \right) \hat{\Omega}_{\nu]ci}\, .
\eea
Adding both contributions, we get
\be
\left([\delta_{1},\delta_{2}]\wt{b}_{\mu\nu}\right)^{(1)} = 2\partial_{[\mu}\xi_{12\nu]} - \frac{\ap}{2}\partial_{[\mu}\Lambda_{12}^{CD}\hat{\Omega}_{\nu]CD}-\frac {\ap}{2}\partial_{[\mu}\left(\ov\epsilon_2\gamma^\sigma\epsilon_1\hat\Omega_{c\sigma}^i \right) \hat{\Omega}_{\nu]ci}\, ,\label{suma}
\ee
with \be
\xi_{12\nu} = \frac{\ap}{2}\left[\Lambda_{2}^{CD}\partial_{\nu}\Lambda_{1CD} + \frac{1}{4}\ov{\epsilon}_{2}\gamma^{\lambda}\epsilon_{1}\hat{\Omega}_{\nu}{}^{CD}\hat{\Omega}_{\lambda CD}\right]\, .
\ee
and 
\be
\Lambda_{12}^{CD} = 2\Lambda_{1}^{CE}\Lambda_{2E}{}^{D} + \frac{1}{2}\ov{\epsilon}_{2}\gamma^{\lambda}\epsilon_{1}\hat{\Omega}_{\lambda}{}^{CD}\, .
\ee

To see the algebra of  transformations on $b_{\mu\nu}$, note that
\be
\left([\delta_{1},\delta_{2}]\right)^{(1)}b_{\mu\nu} = \left([\delta_{1},\delta_{2}]\wt{b}_{\mu\nu}\right)^{(1)} - \frac{\ap}{8}\left([\delta_{1},\delta_{2}]\right)^{(0)}\left(A_{[\mu}{}^{k}\ov{\chi}^{i}\gamma_{\nu]}\chi^{j}f_{ijk}\right)\, ,
\ee
and it is easy to see that the second term in the r.h.s. vanishes. Rewriting \eqref{suma} in terms of supergravity and super Yang-Mills fields, the brackets that mix supersymmetry with Lorentz and abelian gauge transformations vanish, while the supersymmetry algebra gives
\bea
\left([\delta_{\epsilon_1},\delta_{\epsilon_2}]\right)^{(1)}b_{\mu\nu}=\partial_{[\mu}(\xi_{12})_{\nu]}-\ap\partial_{[\mu}\Lambda_{12}^{cd}\hat w_{\nu]cd}-\frac\ap2\varrho\partial_{[\mu}\xi_{12}^iA_{\nu]i}\, ,
\eea
with
\bea
(\xi_{12})_{\nu}&=&\frac\ap4\bar\epsilon_2\gamma^\lambda\epsilon_1\hat\Omega_{\nu}{}^{CD}\hat\Omega_{\lambda CD}\, ,\nn\\
\Lambda_{12}^{cd}&=&\frac14\bar\epsilon_2\gamma^\lambda\epsilon_1\hat w_\lambda^{(-)cd}\, ,\nn\\
\xi_{12}^i&=&-\frac12\bar\epsilon_2\gamma^\lambda\epsilon_1A_\lambda^i\, .
\eea

\subsection{Invariance of the action}

Here we prove the supersymmetric invariance of the action
\be
S = \int d^{10}x\ e \ e^{-2\phi}{\mathscr L}\, ,
\ee
with
\bea
\mathscr{L} & = & R + 4\partial_{\mu}\phi\partial^{\mu}\phi - \frac{1}{12}\wt{H}_{\mu\nu\rho}\wt{H}^{\mu\nu\rho} - \frac{1}{4}F_{\mu\nu }^iF^{\mu\nu }_i + \frac{\ap}{8}\hat{\mathscr R}_{\mu\nu CD}\hat{\mathscr R}^{\mu\nu CD}\, \nn\\
& & - \ov{\psi}{}^{\mu}\gamma^{\nu}D_{\nu}\psi_{\mu} + \ov{\rho}\gamma^{\mu}D_{\mu}\rho + 2\ov{\psi}{}^{\mu}D_{\mu}\rho - \frac{1}{2}\ov{\chi}^{i}\gamma^{\mu}D_{\mu}\chi_{i} + \ov{\chi}_{i}\left(\gamma^{\mu}\psi^{\nu} - \frac{1}{4}\gamma^{\mu\nu}\rho\right)F_{\mu\nu }^i\, \nn \\
& & + \frac{1}{24}\wt{H}_{\rho\sigma\tau}\left(\ov{\psi}{}^{\mu}\gamma^{\rho\sigma\tau}\psi_{\mu} + 12\ov{\psi}{}^{\rho}\gamma^{\sigma}\psi^{\tau} - \ov{\rho}\gamma^{\rho\sigma\tau}\rho - 6\ov{\psi}{}^{\rho}\gamma^{\sigma\tau}\rho + \frac{1}{2}\ov{\chi}^{i}\gamma^{\rho\sigma\tau}\chi_{i}\right)\, \nn \\
& & + \ap\left(\ov{\Psi}{}^{CD}\gamma^{\mu}D_{\mu}(w,\hat{\Omega})\Psi_{CD} - \frac{1}{24}\ov{\Psi}{}^{CD}{\slashed H} \ov{\Psi}_{CD}- \ov{\Psi}{}^{CD}\left(\gamma^{\mu}\psi^{\nu} - \frac{1}{4}\gamma^{\mu\nu}\rho\right)\hat{\mathscr R}_{\mu\nu CD}\right)\,  \ \ \ \ \ \ \ \label{flagr}
\eea

Since the leading order action is known to be invariant \cite{Chapline:1982ww}, we analyze the  ${\cal O}(\alpha')$ variation, namely
\bea
\label{var_ap}
\left(\delta S\right)^{(1)} & = & \int d^{10}xee^{-2\phi}\left[- e_{\mu}{}^{a}\delta^{(0)}e^{\mu}{}_{a}{\mathscr L}^{(1)} - 2\delta^{(0)}\phi{\mathscr L}^{(1)} + \delta^{(0)}{\mathscr L}^{(1)} + \delta^{(1)}{\mathscr L}^{(0)}\right]\, .
\eea
Using the transformation rules \eqref{tranfru}  we get
\bea
&&\left(\delta S\right)^{(1)}  =  - \frac{\ap}{8}\ov{\epsilon}\rho\left(H_{\mu\nu\rho}\hat{\mathscr{C}}^{\mu\nu\rho} - \frac{1}{2}\hat{\mathscr R}_{\mu\nu CD}\hat{\mathscr R}^{\mu\nu CD}\right) + \frac{3\ap}{8}\ov{\epsilon}\gamma^{(\mu}\psi^{\lambda)}H_{\mu\nu\rho}\hat{\mathscr{C}}_{\lambda}{}^{\nu\rho}\, \nn \\
& & - \frac{3\ap}{2}\ov{\epsilon}\gamma^{\mu}\psi^{\nu}\left(\partial_{\rho}\phi\hat{\mathscr{C}}_{\mu\nu}{}^{\rho} - \frac{1}{2}{\mathscr{D}}_{\rho}\hat{\mathscr{C}}_{\mu\nu}{}^{\rho}+\frac1{12}\hat{\mathscr R}_{\mu\rho CD}\hat{\mathscr R}_{\nu}{}^{\rho CD}\right) + \frac{3\ap}{8}\ov{\epsilon}\gamma_{\mu}\chi^{i}F_{\nu\rho i}\hat{\mathscr{C}}^{\mu\nu\rho} \, \nn \\
& & +\frac{\ap}{2}\delta^{(0)}\hat{\Omega}_{\mu}{}^{CD}\left(\Delta b^{\mu\nu}\hat{\Omega}_{\nu CD} + \frac{1}{2}H^{\mu\nu\rho}\hat{\mathscr{R}}_{\nu\rho CD} + 2\partial_{\nu}\phi\hat{\mathscr R}^{\mu\nu CD} - {\mathscr{D}}_{\nu}\hat{\mathscr R}^{\mu\nu CD}\right)\, \nn \\
& & + \frac{\ap}{8}\delta^{(0)}\ov{\psi}{}^{\mu}\left(\gamma^{\rho\sigma\tau}\psi_{\mu}\hat{\mathscr{C}}_{\rho\sigma\tau} + 12\gamma^{\sigma}\psi^{\tau}\hat{\mathscr{C}}_{\mu\sigma\tau} - 3\gamma^{\sigma\tau}\rho\hat{\mathscr{C}}_{\mu\sigma\tau} + 8\gamma^{\nu}\Psi^{CD}\hat{\mathscr R}_{\mu\nu CD}\right)\, \nn \\
& & - \frac{\ap}{8}\delta^{(0)}\ov{\rho}\left(\left(\gamma^{\rho\sigma\tau}\rho + 3\gamma^{\sigma\tau}\psi^{\rho}\right)\hat{\mathscr{C}}_{\rho\sigma\tau} - 2\gamma^{\mu\nu}\Psi^{CD}\hat{\mathscr R}_{\mu\nu CD}\right) + \frac{\ap}{16}\delta^{(0)}\ov{\chi}^{i}\gamma^{\rho\sigma\tau}\chi_{i}\hat{\mathscr{C}}_{\rho\sigma\tau}\, \nn \\
& & + 2\ap\delta^{(0)}\ov{\Psi}{}^{CD}\left({\slashed D}(w,\hat{\Omega})\Psi_{CD} -\left({\slashed\partial}\phi + \frac{1}{24}{\slashed H}\right)\Psi_{CD} + \frac{1}{2}\left(\gamma^{\mu}\psi^{\nu} - \frac{1}{4}\gamma^{\mu\nu}\rho\right)\hat{\mathscr R}_{\mu\nu CD}\right)\, \nn \\
& & + \delta^{(1)}\wt{b}_{\mu\nu}\Delta b^{\mu\nu} - 2\delta^{(1)}\ov{\psi}{}^{\mu}\Delta\psi_\mu + 2\delta^{(1)}\ov{\rho}\Delta\rho\, . \ \ \ \ \label{varia}
\eea
The variations \eqref{tranfru} depend on the supersymmetry parameter   explicitly and  through $\Lambda_{ci}=\frac1{2\sqrt2}\bar\epsilon\gamma_c\chi_i$. The explicit dependence  has the same structure as the corresponding transformations in \cite{bdr1,{Bergshoeff:1988nn}}, replacing the collective indices $C,D,...$ by $c,d,...$. Since the corresponding actions also have the same structure, we can assure that those terms cancel in \eqref{varia}. The $\Lambda_{ci}$-dependent terms are contained in $\delta^{(0)}\hat\Omega_{\mu CD}, \delta^{(1)}b_{\mu\nu}$ and $\delta^{(0)}\Psi_{CD}$. We can disregard the latter as they are higher than bilinear in fermions. The former two may be written as
\bea
\left(\delta S\right)^{(1)} & = &  \frac{\ap}{2}D_{\rho}\left[\partial_{\mu}\Lambda^{CD}\hat{\Omega}_{\nu CD} - \hat{\Omega}_{\mu}{}^{ED}\hat{\Omega}_{\nu}{}^{C}{}_{D}\Lambda_{EC}\right]H^{\mu\nu\rho}\, \nn \\
& & + \frac{\ap}{2}{\mathscr{D}}_{\nu}{\mathscr{D}}_{\mu}\Lambda^{CD}\hat{\mathscr R}^{\mu\nu}{}_{CD} - \frac{\ap}{4}\mathscr{D}_{\mu}\Lambda^{CD}H^{\mu\nu\rho}\hat{\mathscr{R}}_{\nu\rho CD}\, \, \nn ,
\eea
which can be easily shown to vanish after performing some integrations by parts.

\end{document}